\documentclass[12pt]{iopart}
\usepackage{braket}
\usepackage{graphicx}
\begin{document}

\pdfminorversion=4

\title[]{Dispersion-engineered $\chi^{(2)}$ nanophotonics: a flexible tool for nonclassical light}

\author{Marc Jankowski$^{1,2}$, Jatadhari Mishra$^{2}$, and M. M. Fejer$^{2}$}

\address{$^1$NTT Research Inc. Physics and Informatics Labs, 940 Stewart Drive, Sunnyvale, California}
\address{$^2$Edward L. Ginzton Laboratory, Stanford University, Stanford, CA.}
\ead{marc.jankowski@ntt-research.com}
\vspace{10pt}
\begin{indented}
\item[]December 2020
\end{indented}

\begin{abstract}
This article reviews recent progress in quasi-phasematched $\chi^{(2)}$ nonlinear nanophotonics, with a particular focus on dispersion-engineered nonlinear interactions. Throughout this article, we establish design rules for the bandwidth and interaction lengths of various nonlinear processes, and provide examples for how these processes can be engineered in nanophotonic devices. In particular, we apply these rules towards the design of sources of non-classical light and show that dispersion-engineered devices can outperform their conventional counterparts. Examples include ultra-broadband optical parametric amplification as a resource for measurement-based quantum computation, dispersion-engineered spontaneous parametric downconversion as a source of separable biphotons, and synchronously pumped nonlinear resonators as a potential route towards single-photon nonlinearities.\end{abstract}

%

\section*{Introduction}

Crystals with quadratic ($\chi^{(2)}$) nonlinearities form the backbone of many modern optical systems, where they can be used for second-harmonic generation (SHG), sum- and difference-frequency generation (SFG and DFG), optical parametric amplification (OPA), and spontaneous parametric down-conversion (SPDC). In the context of quantum optics, these devices can be used for the generation~\cite{URen2005GenerationOP,Yokoyama2013}, manipulation~\cite{Wakui:07}, transmission~\cite{DeGreve2012, Liao2017}, and detection~\cite{Pelc:11,Shaked2018} of quantum light. The recent development of quasi-phasematched (QPM) interactions in nanophotonic waveguides with $\chi^{(2)}$ nonlinearities has made possible an entirely new class of nonlinear devices, where the linear dispersion and nonlinear optical properties of the waveguide can be co-engineered by lithographically patterning both the waveguide geometry and the $\chi^{(2)}$ coefficient associated with the nonlinear medium. Until recently, efficient nonlinear interactions have been achieved in state-of-the-art platforms using either modal phase-matching in nanowaveguides, or quasi-phasematching in weakly-guiding diffused waveguides. QPM interactions in weakly-guiding waveguides rely on a periodic poling of $\chi^{(2)}$ to correct for phase drifts between the interacting waves, e.g. $k_{2\omega}-2k_\omega = 2\pi/\Lambda_G$ for second-harmonic generation (SHG), where $\Lambda_G$ is the period of the modulation~\cite{Armstrong1962,Franken1963,Langrock2006}. In these systems the phase-matching bandwidths (and hence useful lengths for pulsed interactions) have ultimately been limited by the material dispersion that dominates over geometrical dispersion in weakly-guiding waveguides. In contrast, the sub-wavelength confinement found in direct-etched nanophotonic waveguides can be used to achieve phase-matching using the geometry dependence of the phase-velocity of TE and TM modes~\cite{Chen2018,Luo2019,Bruch2018,Bruch2019,Chang2018,Chiles2019,Stanton2020, Guidry:20}. In these systems geometrical dispersion can dominate over material dispersion, but the design of these waveguides is constrained to geometries that achieve phase-velocity matching between the interacting waves, e.g. $n_\omega=n_{2\omega}$ for SHG. Recent work has focused on quasi-phasematched interactions in nanophotonic waveguides, which overcome both of these limitations~\cite{Wang2018,Chang2016,Timurdogan2017,Billat2017,Porcel2017,XLu2020,Chen2020,Nagy2019,Nagy2020}. Quasi-phasematching can be achieved for almost any waveguide geometry of interest, which frees up the geometric dispersion as a design parameter. This freedom enables a new set of design rules where multiple dispersion orders, such as the group velocities and group-velocity dispersion of the interacting waves, can be simultaneously engineered to achieve favorable characteristics across a wide range of wavelengths~\cite{Hickstein2019,Singh2020,Jankowski2020}.

\begin{figure}
    \centering
    \includegraphics[width=\columnwidth]{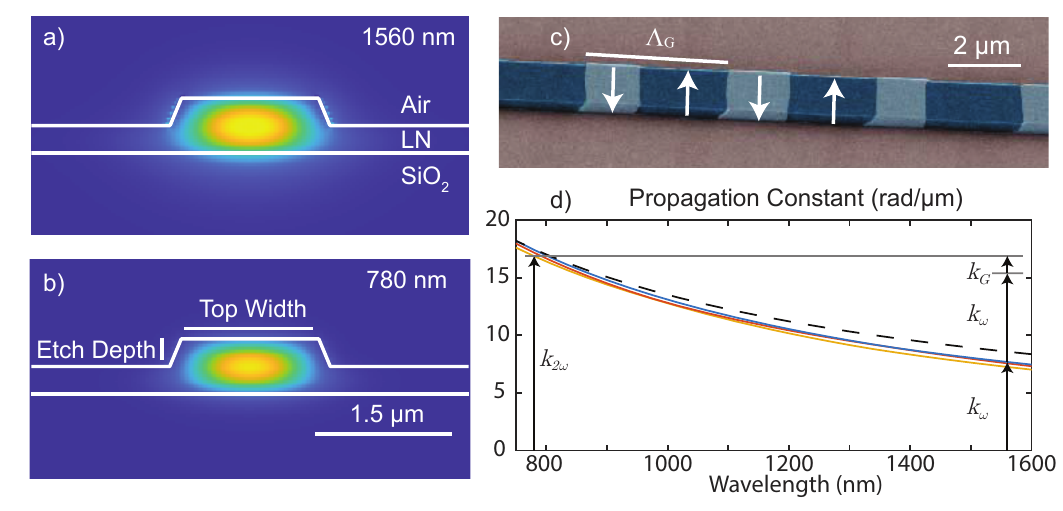}
    \caption{a,b) Schematic of the waveguide cross section, showing the electric field distribution, $E_x$, associated with the TE$_{00}$ modes of the fundamental and second harmonic, respectively. c) False color SEM image of a fabricated PPLN waveguide, showing ferroelectric domains with period $\Lambda_G$. d) The dispersion relations of the waveguide. The propagation constant at long wavelengths differs substantially from bulk (dashed line). The phase-mismath between a TE$_{00}$ fundamental (blue line) and a TE$_{00}$ second harmonic (orange line) can be compensated with a suitable choice of the grating period, $k_G=2\pi/\Lambda_G$. Panel (c) is reproduced from Wang \textit{et al}., Optica \textbf{5}, 1438 (2018). Copyright 2018 Authors, licensed under the terms of the OSA Open Access Publishing Agreement}
    \label{fig:Fig1}
\end{figure}

The purpose of this review is two-fold. First, we provide design rules for dispersion-engineered QPM devices, with particular focus on how to engineer the bandwidths of nonlinear interactions. Second, we apply these rules to the design of nonlinear components that can be used to generate and manipulate quantum light. Examples include broadband optical parametric amplification for the generation and detection of squeezed light, high-purity separable biphotons for heralding, and microcavities with efficient few-photon nonlinear interactions. When combined with low-loss linear photonic circuits and efficient integrated detectors, the nonlinear components discussed here can be used to enable a number of emerging platforms for integrated quantum photonics~\cite{OBrien2007,OBrien2009}. We note here that a number of excellent reviews have discussed recent developments in thin-film lithium niobate (TFLN) nanophotonics~\cite{zhu2021integrated, Wang2020Q, Honardoost2020, Qi2020, Boes2018, Sun2020}, as well as progress in platforms for integrated quantum photonics~\cite{Moody2020, Lukin2019, Lukin2020, Wang2019quantum}. This review complements these works by clarifying the role of dispersion engineering in the design of nonlinear photonic devices, which has not been comprehensively discussed in the literature. While the examples studied here predominantly consider TE (Z-polarized) modes in X-cut TFLN ridge waveguides, these design rules are applicable to any tightly-confining QPM device.

This work proceeds in eight sections. \Sref{sec:fab} briefly discusses the fabrication of nanophotonic devices in periodically-poled lithium niobate (PPLN) thin films. In \sref{sec:CW} we review continuous-wave (CW) interactions in nonlinear nanowaveguides. This section establishes the theoretical framework and figures of merit used throughout this review. \Sref{sec:BW} discusses the bandwidths associated with nonlinear interactions, and provides an example design of dispersion-engineered SHG. \Sref{sec:pulsednlo} extends the theoretical framework introduced in \sref{sec:CW} to pulsed interactions in dispersion-engineered QPM devices. We also introduce quasi-static nonlinear photonic devices. In these devices several of the dominant dispersion orders are eliminated simultaneously, thereby increasing the interaction lengths of short pulses by orders of magnitude. \Sref{sec:OPA} presents the design of ultra-broadband optical parametric amplifiers (OPAs) to produce cluster states for measurement-based quantum computation. \Sref{sec:SPDC} discusses the generation of separable biphotons by combining dispersion engineering with a non-uniform QPM grating. \Sref{sec:SPND} discusses quasi-static interactions in nonlinear resonators driven by short pulses and shows that such devices provide a viable route towards efficient single-photon nonlinear interactions. We also compare the relative nonlinearity of several emerging photonics platforms as a guide for researchers interested in developing highly nonlinear devices. \Sref{sec:conclusion} summarizes this work and discusses many of the opportunities available for future work.

\section{Fabrication of PPLN Nanowaveguides}\label{sec:fab}

\begin{figure}[t]
  \centering
  \includegraphics[width=\columnwidth]{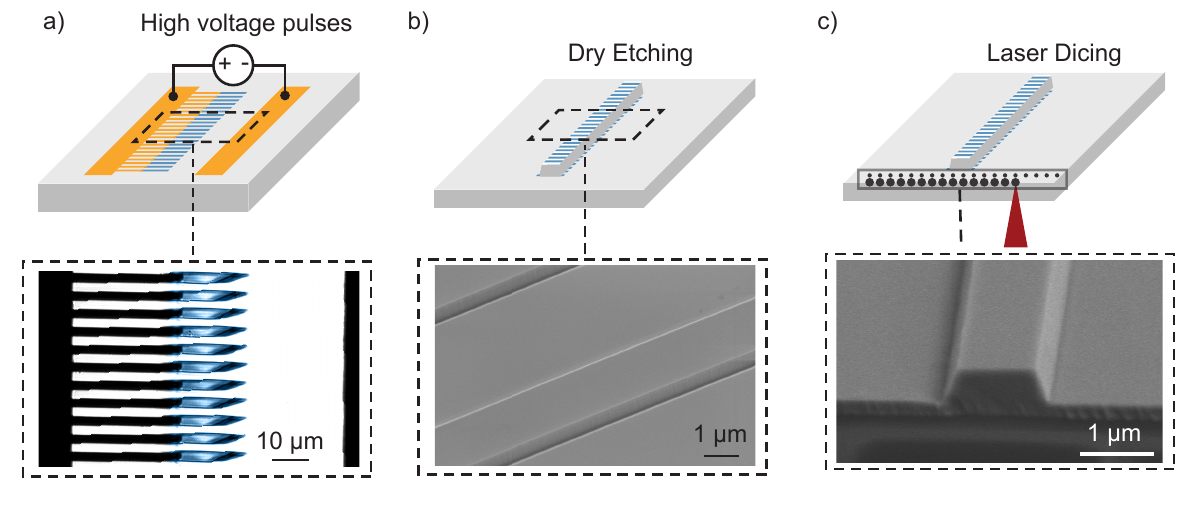}
\caption{\label{fig:fab}a) Schematic of the poling process, resulting in high fidelity domain inversion with a $\sim$50\% duty cycle. b) Waveguides are patterned using an Ar$^+$ assisted dry etch, resulting in smooth sidewalls. c) Facet preparation is performed using laser dicing, resulting in optical-quality end-facets. Figure adapted from Jankowski \textit{et al}., Optica \textbf{7}, 40 (2020). Copyright 2020 Authors, licensed under the terms of the OSA Open Access Publishing Agreement.}
\end{figure}

Throughout this review, we focus on nonlinear interactions in nanophotonic PPLN waveguides. Here we briefly describe the fabrication process for the devices used in~\cite{Wang2018,Jankowski2020,mckenna2021ultralowpower}, with similar approaches described in~\cite{Chang2016,Chen2020,Nagy2019,Nagy2020,Zhao2020,Rao2019}. These waveguides are fabricated in three steps. First, we periodically pole an X-cut magnesium-oxide- (MgO-) doped lithium niobate thin film (Figure \ref{fig:fab}(a)) using the methods described in~\cite{Wang2018}. The metal electrodes consist of a 15-nm-thick Cr adhesion layer and a 150-nm-thick Au layer, deposited by electron-beam evaporation. We perform the periodic domain inversion by applying several 580 V, 5-ms-long pulses at room temperature with the sample submerged in oil, which corresponds to a poling electric field of $\sim$7.6 kV/mm. The inset shows a colorized 2-photon microscope image of the resulting inverted domains with a duty cycle of $\sim$50\%. The poled region typically has a width of 10-25 $\mu$m and a length of 4-6 mm. After periodic poling, we remove the electrodes using metal etchant. The second step is to pattern and etch the waveguides using the process described in~\cite{Zhang2017}. Here, aligned electron-beam lithography is used to create waveguide patterns inside the poled region. Each poled region can accommodate multiple ridge waveguides (two to three in our case) without cross-talk due to the strong optical confinement, allowing for dense device integration. The patterns are then transferred to the LN device layer using an optimized Ar$^+$-based dry etching process to form ridge waveguides~\cite{Zhang2017}. This yields low-loss ($\sim 0.03$ dB/cm) ridge waveguides (Figure \ref{fig:fab}(b)). The inset shows a scanning electron microscope (SEM) image of the ridge waveguides, showing smooth sidewalls. Finally, facet preparation is done using a DISCO DFL7340 laser saw (Figure \ref{fig:fab}(c)). Here, $\sim$10-$\mu$J pulses are focused into the substrate to create a periodic array of damage spots, which act as nucleation sites for crack propagation. The sample is then cleaved. The inset shows an SEM image of the resulting end-facets, which exhibit $<$10-nm facet roughness. We note here that the propagation direction (z) is along the crystalline Y-axis, such that light polarized in the plane of the thin film ($E_x$) is oriented along the crystalline Z-axis.

\section{Continuous-Wave $\chi^{(2)}$ Interactions in Nanophotonic Devices}\label{sec:CW}

We now consider continuous-wave (CW) interactions in nonlinear nanophotonic devices and establish the theoretical framework that will be used throughout the remainder of this review. The analysis presented here relies on solutions to the coupled-wave equations (CWEs) for SHG and three-wave mixing (TWM). These equations, along with their associated nonlinear coupling, are derived in \ref{sec:CWEs}.

\subsection{Second-Harmonic Generation}

We first consider SHG between a pair of modes at frequency $\omega$ and $2\omega$. The evolution of the complex field envelopes, $A_\omega$, is given by the CWEs for SHG
\begin{eqnarray}
\partial_z A_\omega(z) &= -i\kappa A_{2\omega}(z)A_\omega^*(z) \exp(-i\Delta k z)\label{eqn:CWE01}\\
\partial_z A_{2\omega}(z) &= -i\kappa A_\omega^2(z) \exp(i\Delta k z),\label{eqn:CWE02}
\end{eqnarray}
where $A_\omega$ is normalized to have units of W$^{-1/2}$, such that $|A_\omega|^2=\mathrm{P}_\omega$ is the power contained in the fundamental. The phase-mismatch is given by $\Delta k = k_{2\omega} - 2k_\omega - 2\pi/\Lambda_G$, where $\Lambda_G$ is the period of the QPM grating, and the nonlinear coupling, $\kappa$, is given by
\begin{equation}
\kappa = \frac{\sqrt{2 Z_0} \omega d_\mathrm{eff}}{c n_\omega \sqrt{n_{2\omega}A_{\mathrm{eff}}}},\label{eqn:kappa}
\end{equation}
where $d_\mathrm{eff}=$ is the effective nonlinear coefficient,  $d_\mathrm{eff}=2d_{33}/\pi$ for a PPLN grating with a $50\%$ duty cycle, $Z_0$ is the impedance of free space, and $n_\omega$ is the effective refractive index of the mode at frequency $\omega$. The effective area, $A_\mathrm{eff}$, measures the relative strength of the nonlinear interaction due to tight confinement. Typical values of the effective area for doubling wavelengths around 2-$\mu$m are $A_\mathrm{eff}\sim$1 $\mu$m$^2$ in nanophotonic devices, which is a $\sim$50-fold improvement when compared to weakly-guiding diffused waveguides. For guided wave devices the effective area scales as $\lambda^{-2}$, resulting in a quadratic scaling of $\kappa$ as a given device is scaled to operate at shorter wavelengths. For bulk nonlinear devices driven by confocally-focused Gaussian beams, $A_\mathrm{eff}\sim$100's - 1000's of $\mu$m$^2$.

\begin{figure}[t]
    \centering
    \includegraphics[width=\columnwidth]{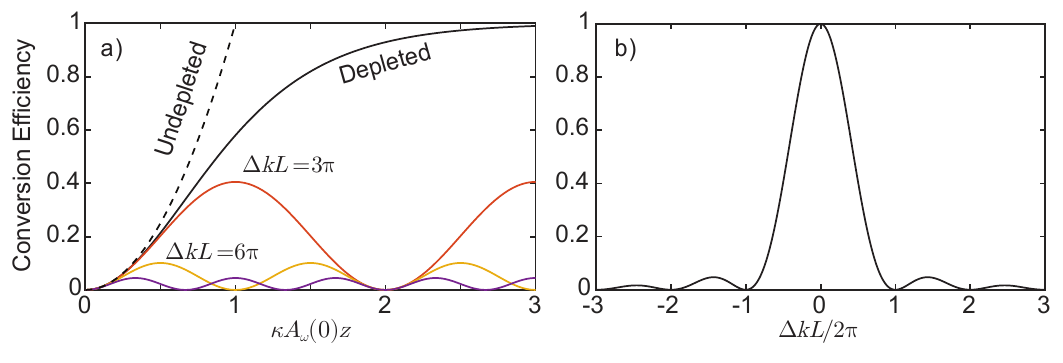}
    \caption{a) The experimentally relevant operating regimes of SHG: undepleted (dashed line), depleted (solid black), and phase-mismatched ($\Delta kL=3\pi, 6\pi, 9\pi$). b) The transfer function for undepleted SHG, $\mathrm{sinc}^2(\Delta k L/2)$.}
    \label{fig:Fig3}
\end{figure}

In the absence of pump depletion the fundamental envelope remains unchanged, $A_\omega(z)\sim A_\omega(0)$, in which case the equation of motion for the second harmonic is readily integrated to find 
\begin{equation}
    A_{2\omega}(z) = -i\kappa A_\omega^2(0)z\exp(i\Delta k z/2)\mathrm{sinc}(\Delta k z/2)\label{eqn:SHG-undepleted}.
\end{equation}
For $\Delta k=0$, (\ref{eqn:SHG-undepleted}) gives rise to the familiar quadratic scaling of output power with input power, $\mathrm{P}_{2\omega} = \eta_0 \mathrm{P}_\omega^2 z^2$ (Figure \ref{fig:Fig3}(a)), associated with SHG. $\eta_0 = \kappa^2$, quoted in $\%$/W-cm$^2$, is the normalized efficiency of SHG and is the typical figure of merit for nonlinear waveguides; waveguides with larger $\eta_0$ can achieve efficient frequency conversion with either less power or shorter devices. For $\Delta k \neq 0$ the second harmonic envelope oscillates sinusoidally in $z$, with an amplitude given by $2\kappa A_\omega^2(0)/\Delta k$ (Figure \ref{fig:Fig3}(a)). In terms of conversion efficiency, $\eta=\mathrm{P}_{2\omega}(L)/\mathrm{P}_{\omega}(0)$, (\ref{eqn:SHG-undepleted}) takes the form
\begin{equation}
\eta = \eta_0 \mathrm{P}_{\omega}(0)L^2\mathrm{sinc}^2(\Delta k L/2).\label{SHG-TF}
\end{equation}
The $\mathrm{sinc}^2(\Delta k L/2)$ factor in (\ref{SHG-TF}) is referred to as the SHG transfer function, and is shown in Figure \ref{fig:Fig3}(b). Along with $\eta_0$, the transfer function characterizes the performance of a nonlinear device. Device inhomogeneities and loss mechanisms cause the transfer function to deviate from an ideal $\mathrm{sinc}^2$ shape~\cite{Fejer1992, Bortz1994, Hum2007}. Phase errors due to inhomogeneities generally broaden transfer functions, with the total area of the transfer function conserved. Loss mechanisms give rise to a wide variety of behaviors: an overall reduction of total device efficiency, broadened transfer functions, either an artificial suppression or an enhancement of the normalized efficiency inferred from the power of each harmonic output from the waveguide. In practice, the SHG transfer function is typically measured by detuning the wavelength of the fundamental by $\Omega$ to vary the phase-mismatch, $\Delta k(\Omega) = k_{2\omega}(2\omega+2\Omega)-2k_\omega(\omega+\Omega)-2\pi/\Lambda_G$.

For large $\eta_0 P_\omega(0)L^2$ the conversion efficiency can become sufficiently large to deplete the fundamental. In this case, (\ref{eqn:CWE01}-\ref{eqn:CWE02}) can be integrated, assuming $\Delta k=0$, to find the evolution of the field envelopes accounting for pump depletion,
\begin{eqnarray}
A_{2\omega}(z) = -i A_\omega(0)\mathrm{tanh}(\kappa A_\omega(0) z),\\
A_\omega(z) = A_\omega(0)\mathrm{sech}(\kappa A_\omega(0) z).
\end{eqnarray}
Saturated SHG is a useful diagnostic tool: the generated second harmonic power is background-free, and therefore can be used to determine the input power $P_\omega$ and the normalized efficiency $\eta_0$ by fitting the $\mathrm{tanh}^2$ dependence of the second harmonic. In the undepleted limit, measuring $\eta_0$ requires an accurate measurement of both the fundamental and second harmonic power.

\subsection{Experimental demonstrations of SHG in PPLN nanowaveguides}

\begin{figure}[t]
    \centering
    \includegraphics[width=\columnwidth]{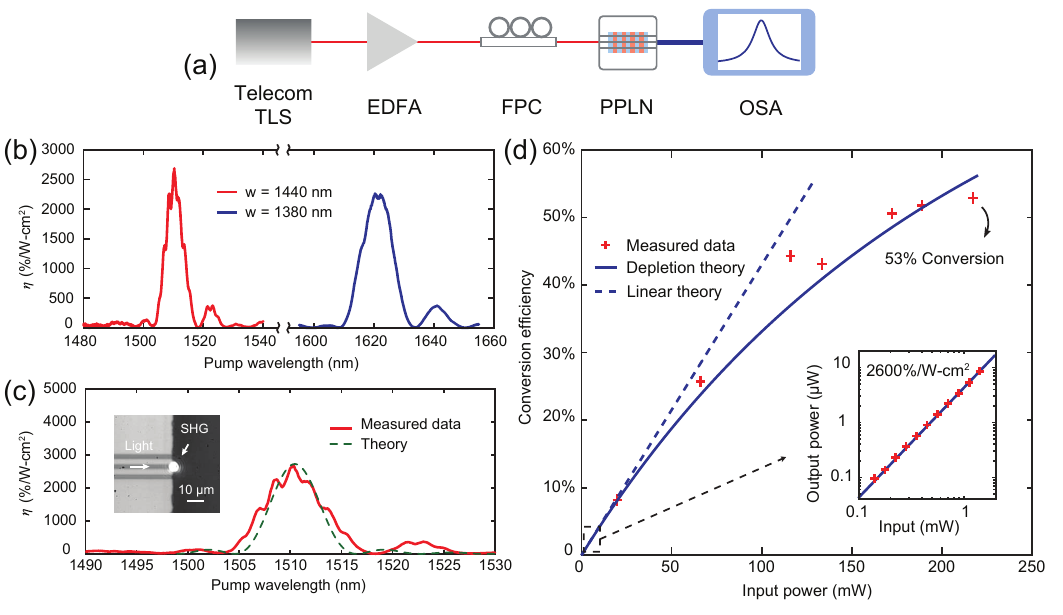}
    \caption{a) Schematic of the characterization setup. Light from a tunable laser (TLS) is amplified using an erbium-doped fiber amplifier (EDFA), and the polarization state is set using a fiber polarization controller (FPC). The light output from a PPLN waveguide is collected with an optical spectrum analyzer (OSA). b) Measured SHG transfer functions for two waveguides with the same poling period but different top widths; the phase-matched wavelength is seen to tune 1.8 nm per nm change in the top width. c) Comparison of the measured (solid red) and simulated (dashed green) SHG transfer function for the 1440-nm-wide device (solid curve). Inset: CCD camera image of the scattered second harmonic at the output waveguide facet. d) The recorded conversion efficiency as a function of in-coupled pump power. Inset: the undepleted limit, showing a normalized efficiency of 2600\%/W-cm$^2$. Figure adapted from Wang \textit{et al}., Optica \textbf{5}, 1438 (2018). Copyright 2018 Authors, licensed under the terms of the OSA Open Access Publishing Agreement.}
    \label{fig:Fig4}
\end{figure}

Having established the relevant operating regimes of SHG, we now review a number of the early experimental demonstrations of SHG in periodically poled TFLN waveguides. These demonstrations focused on characterizing $\eta_0$ and the SHG transfer function to verify that these devices could achieve performance comparable to theoretical predictions. The first demonstration of QPM interactions used heterogeneously integrated SiNx waveguides on periodically poled TFLN~\cite{Chang2016}. These devices achieved a normalized efficiency of 160$\%$/W-cm$^2$, which was an order of magnitude below theory (1600$\%$/W-cm$^2$), and exhibited transfer functions three times wider than theory. These effects are consistent with a lossy second harmonic, and it was later determined that these discrepancies were due to lateral leakage of the second harmonic~\cite{Boes2019}.

The first demonstrations using direct-etched PPLN waveguides avoided these problems and achieved normalized efficiencies around 2600$\%$/W-cm$^2$~\cite{Wang2018}. These results are shown in Figure \ref{fig:Fig4}. The samples were characterized using end-fire coupling (Figure \ref{fig:Fig4}(a)) and the transfer functions are shown in Figure \ref{fig:Fig4}(b-c). We note here that the measured transfer functions exhibit fringes every 2-3 nm due to a weak Fabry-Perot cavity formed between the fiber and the end-facet of the waveguide. These measured transfer functions are in reasonable agreement with theoretical predictions (Figure \ref{fig:Fig4}(c)), which suggest that the devices do not exhibit strong loss or inhomogeneities. We note here that while the authors of~\cite{Wang2018} quoted a theoretical value in excess of 4000\%/W-cm$^2$, these values were calculated using $d_{33}=25$ pm/V. Using a Miller's delta scaling of the parameters in~\cite{Shoji1997, Choy1976}, a more accurate estimate is $d_{33}=21.7$ pm/V for doubling of 1560-nm light, resulting in a theoretical normalized efficiency of 3000\%/W-cm$^2$. Further reductions of the normalized efficiency due to the measured 40\% duty cycle yield a theoretical value of 2600\%/W-cm$^2$, in good agreement with the experimental results. Later work has focused on achieving duty cycles closer to 50\% and tighter mode confinement, and there have now been multiple demonstrations of normalized efficiencies in excess of 4000\%/W-cm$^2$ at 1.5 $\mu$m~\cite{Zhao2020,Rao2019}.

\subsection{Three-wave mixing}

Having established the nonlinear coupling and transfer function for SHG, we now generalize these results to three-wave mixing, which encompass processes such as sum- and difference-frequency generation (SFG and DFG), optical parametric amplification (OPA), and spontaneous parametric downconversion (SPDC). The CWEs for three-wave mixing are given by
\begin{eqnarray}
\partial_z A_{\omega_3}(z) &= -i\kappa_3 A_{\omega_2}(z)A_{\omega_1}(z) \exp(i\Delta k z),\label{eqn:TWM03}\\
\partial_z A_{\omega_2}(z) &= -i\kappa_2 A_{\omega_3}(z)A_{\omega_1}^*(z) \exp(-i\Delta k z),\label{eqn:TWM02}\\
\partial_z A_{\omega_1}(z) &= -i\kappa_1 A_{\omega_3}(z)A_{\omega_2}^*(z) \exp(-i\Delta k z),\label{eqn:TWM01}
\end{eqnarray}
where $\omega_3=\omega_2+\omega_1$ and $\omega_2 > \omega_1$. The phase-mismatch is given by $\Delta k = k_{\omega_3} - k_{\omega_2} - k_{\omega_1}-2\pi/\Lambda_G$, and the nonlinear coupling is given by
\begin{eqnarray}
&\kappa_j = \frac{\sqrt{2 Z_0} \omega_j d_\mathrm{eff}}{c \sqrt{n_{\omega,1}n_{\omega,2}n_{\omega,3}A_{\mathrm{eff}}}}.\label{eqn:kappaTWM}
\end{eqnarray}
We note here that (\ref{eqn:TWM03}-\ref{eqn:TWM01}) satisfy the Manley-Rowe relations, $\partial_z \mathrm{P}_{\omega_3}/\omega_3=-\partial_z \mathrm{P}_{\omega_2}/\omega_2=-\partial_z \mathrm{P}_{\omega_1}/\omega_1$,

In the undepleted limit, the solution to the CWEs for SFG and DFG are essentially identical to that of SHG, e.g. $A_{\omega_3}(z) = -i\kappa_3 A_{\omega_1}(0)A_{\omega_2}(0)z\exp(i\Delta k z/2)\mathrm{sinc}(\Delta k z/2)$ for undepleted SFG, and therefore the previous analysis in terms of normalized efficiency and transfer function is sufficient to study these interactions. In contrast, OPA exhibits solutions that grow exponentially with $z$. For convenience, we introduce flux amplitudes $a_s = A_{\omega_2}/\sqrt{\hbar\omega_2}$, $a_i = A_{\omega_1}/\sqrt{\hbar\omega_1}$, and adopt the usual pump-signal-idler nomenclature used for OPA ($\omega_3 = \omega_p$, $\omega_2=\omega_s$, and $\omega_1=\omega_i$). With this notation the coupled-wave equations for the signal and idler have symmetric nonlinear couplings,
\begin{eqnarray}
\partial_z a_{s}(z) &= \gamma a_{i}^*(z) \exp(-i\Delta k z),\label{eqn:dz_OPA01}\\
\partial_z a_{i}^*(z) &= \gamma^* a_{s}(z) \exp(i\Delta k z),\label{eqn:dz_OPA02}
\end{eqnarray}
where $\gamma=-i\sqrt{\kappa_1\kappa_2}A_{\omega_3}(0)$. These equations are solved to find

\begin{eqnarray}
    \left[\begin{array}{c}
         \tilde{a}_s(z)  \\
         \tilde{a}_i^*(z) 
    \end{array}\right]=
    \left[\begin{array}{cc}
         \mu & \nu \\
         \nu^* & \mu^*
    \end{array}\right]
    \left[\begin{array}{c}
         a_s(0)  \\
         a_i^*(0) 
    \end{array}\right]\label{eqn:OPA01},
\end{eqnarray}
where $\mu(z) = \left[\mathrm{cosh}(gz)+\frac{i\Delta k}{2g}\mathrm{sinh}(gz)\right]$, $\nu(z) = \frac{\gamma}{g}\mathrm{sinh}(gz)$, and $\tilde{a}=a\exp(i\Delta k z/2)$. The field gain coefficient is given by $g=|\gamma|\sqrt{1-(\Delta k/2|\gamma|)^2}$. 

We may assume $\gamma$ is real without loss of generality since only the relative phase $\phi_\gamma - \phi_s - \phi_i$ contributes to the nonlinear dynamics. For a phase-matched interaction, exponential growth occurs for the quadratrure $x(z)=(a_s + a_i^*)/2=\exp(\gamma z)x(0)$, and deamplification occurs for $y(z)=(a_s - a_i^*)/(2i)=\exp(-\gamma z)y(0)$. In the general case, these quadratures are given by $x(z)=(\tilde{a}_s + \tilde{a}_i^*)/2$ and $y(z)=(\tilde{a}_s - \tilde{a}_i^*)/(2i)$, and evolve as
\begin{eqnarray}
    \left[\begin{array}{c}
         x(z)  \\
         y(z) 
    \end{array}\right]=
    \left[\begin{array}{cc}
         \bar{\mu}_{+} & \bar{\nu} \\
         -\bar{\nu} & \bar{\mu}_{-}
    \end{array}\right]
    \left[\begin{array}{c}
         x(0)  \\
         y(0) 
    \end{array}\right]\label{eqn:OPA02},
\end{eqnarray}
where $\bar{\mu}_\pm=\mathrm{cosh}(gz)\pm\frac{\gamma}{g}\mathrm{sinh}(gz)$, and $\bar{\nu}=\frac{\Delta k}{2g}\mathrm{sinh}(gz)$. Phase-mismatch has three effects: the quadratures are phase-shifted relative to the field envelopes, the amplification bandwidth is restricted to signal and idler frequencies that satisfy $|\Delta k|<2|\gamma|$, and the quadratures are coupled together. Under most circumstances, the phase-mismatch induced coupling can be ignored and the quadratures exhibit a power gain, $G_\pm = |\mathrm{c}(gz)\pm \gamma \mathrm{s}(gz)/g|^2$, where $+$ and $-$ correspond to the amplified and deamplified quadratures, respectively.

The last three-wave interaction we consider here is SPDC, which occurs in the limit of low gain ($|\gamma L|^2\sim 0.1$). In this case, (\ref{eqn:OPA01}) becomes
\begin{eqnarray}
a_{s}(z) = &a_{s}(0) + \gamma z a_{i}^*(0)\exp(-i\Delta k z/2)\mathrm{sinc}(\Delta k z/2)\label{eqn:OPA03},\\
a_{i}(z) = &a_{i}(0) + \gamma z a_{s}^*(0)\exp(-i\Delta k z/2)\mathrm{sinc}(\Delta k z/2).\label{eqn:OPA04}
\end{eqnarray}
The generated signal and idler are limited to frequencies within the bandwidth set by the DFG transfer function, $\mathrm{sinc}(\Delta k L/2)$, which provides a lower bound for the OPA bandwidth. We note here that while a more detailed treatment of SPDC is given in \sref{sec:SPDC}, we may already gain some insight about this process using (\ref{eqn:OPA03}-\ref{eqn:OPA04}). SPDC occurs in low-gain optical parametric amplifiers in the absence of any input signal and idler. Instead of $a_{s}(0)=a_{i}(0)=0$ we may take $a_{s}(0)$ and $a_{i}(0)$ to be noise fields that correspond to semi-classical vacuum fluctuations. Under these conditions, we see that the pump field amplifies these vacuum fluctuations to produce signal and idler photons in a range of frequencies that satisfy $|\Delta k L| < 2 \pi$.

\subsection{Quantum nonlinear optics}\label{sec:QNLO}

Throughout this review, we will consider a number of waveguide designs that have been engineered to generate non-classical light, and therefore a more complete description of the generated fields is given in terms of the evolution of the field annhiliation operator $\hat{a}$, and creation operator $\hat{a}^\dagger$, respectively. In many cases, such as in traveling-wave OPA and SPDC, the solutions of the coupled-wave equations are sufficient, with the classical c-number fields $a_s$ and $a_i^*$ in (\ref{eqn:OPA01}) replaced by the operators $\hat{a}_s$ and $\hat{a}_i^\dagger$, respectively~\cite{Caves1987,Crouch1988,Harris2007}. For example, in traveling-wave OPA $\hat{a}_s$ and $\hat{a}_i^\dagger$ take on the familiar form,
\begin{eqnarray}
    \left[\begin{array}{c}
         \hat{a}_s(L)\exp(i\Delta k L/2)  \\
         \hat{a}_i^\dagger(L)\exp(-i\Delta k L/2)
    \end{array}\right]=
    \left[\begin{array}{cc}
         \mu & \nu \\
         \nu^* & \mu^*
    \end{array}\right]
    \left[\begin{array}{c}
         \hat{a}_s(0)  \\
         \hat{a}_i^\dagger(0) 
    \end{array}\right]\label{eqn:QOPA00},
\end{eqnarray}
where $\mu$ and $\nu$ are defined below (\ref{eqn:OPA01}).

In general, a description of the system dynamics is given by the interaction Hamiltonian for three-wave interactions, as obtained from the electric dipole Hamiltonian,
\begin{equation}
    \hat{H}_\mathrm{int}/\hbar = g_H(\hat{a}_{\omega_3}\hat{a}_{\omega_2}^{\dagger }\hat{a}_{\omega_1}^{\dagger} + h.c.) 
    \label{eqn:H},
\end{equation}
where $\hat{a}_{\omega_3}\hat{a}_{\omega_2}^{\dagger }\hat{a}_{\omega_1}^{\dagger}$ describes the annihilation of a photon at $\omega_3$ and the creation of a photon at $\omega_2$ and $\omega_1$, and $g_H$ is the coupling rate. Here we have assumed a single mode at each wavelength. In the following sections we will rely on a phenomenological approach where $g_H$ is calculated using physically measurable parameters such as the the transfer functions associated with nonlinear interactions, $\mathrm{sinc}(\Delta k L/2)$, the normalized efficiency, $\eta_0$, and the parametric gain, $\gamma$. In this approach, we may establish a correspondence between the interaction Hamiltonian (\ref{eqn:H}) and the CWEs by taking all three waves to be coherent states with mean photon number $|\alpha_{\omega_j}|^2$, such that $\hat{a}_{\omega_j}\rightarrow \alpha_{\omega_j}$ becomes a complex number. As an example, for optical parametric amplification (in pump-signal-idler nomenclature) we have
\begin{eqnarray}
\partial_t \alpha_{s} = -i g_H \alpha_{p}\alpha_{i}^*,\label{eqn:QOPA01}\\
\partial_t \alpha_{i} = -i g_H \alpha_{p}\alpha_{s}^*.\label{eqn:QOPA02}
\end{eqnarray}
In the undepleted limit, $\alpha_{p}(t)=\alpha_{p}(0)$, the evolution of the signal is given by $\partial_t^2 \alpha_{s} = |g_H|^2 |\alpha_{p}|^2\alpha_{s}$, which again yields solutions that grow exponentially in time. In the limit of low gain, (\ref{eqn:QOPA01}-\ref{eqn:QOPA02}) become
\begin{eqnarray}
\alpha_{s}(T) = \alpha_{s}(0)+\gamma_H T \alpha_{i}^*(0),\label{eqn:QOPA03}\\
\alpha_{i}(T) = \alpha_{i}(0)+\gamma_H T \alpha_{s}^*(0).\label{eqn:QOPA04}
\end{eqnarray}
where $\gamma_H = -i g_H\alpha_{p}(0)$. Comparing (\ref{eqn:QOPA03}-\ref{eqn:QOPA04}) with (\ref{eqn:OPA03}-\ref{eqn:OPA04}) and assuming an interaction time set by the length of the device, $T=L/v_g$, we find $\gamma_H = \gamma v_g \mathrm{sinc}(\Delta k L/2)\exp(-i \Delta k L/2)$.

Having established this correspondence, this review will largely focus on how to engineer the waveguide properties such as the normalized efficiency and the bandwidth associated with nonlinear interactions, and in many cases the classical CWEs are sufficient to gain the necessary insight and design rules for a given nonlinear process. 

\section{Dispersion-engineered nonlinear interactions}\label{sec:BW}

The frequency dependence of the phase mismatch, $\Delta k$, ultimately determines the bandwidth that can be generated by SHG, SFG, DFG, and OPA, and in most cases of interest the response of the generated harmonics to a driving nonlinear polarization is filtered by $\mathrm{sinc}(\Delta k L/2)$. In this section we derive the bandwidth of nonlinear interactions in terms of the dispersion orders associated with modes in a nonlinear waveguide. We first derive the bandwidth associated with SHG, and provide examples of dispersion-engineered nonlinear waveguides that enhance this bandwidth by orders of magnitude. We then generalize this treatment for arbitrary three-wave interactions.

\subsection{The bandwidth of second-harmonic generation}

\begin{figure}
\centering
\includegraphics[width=\columnwidth]{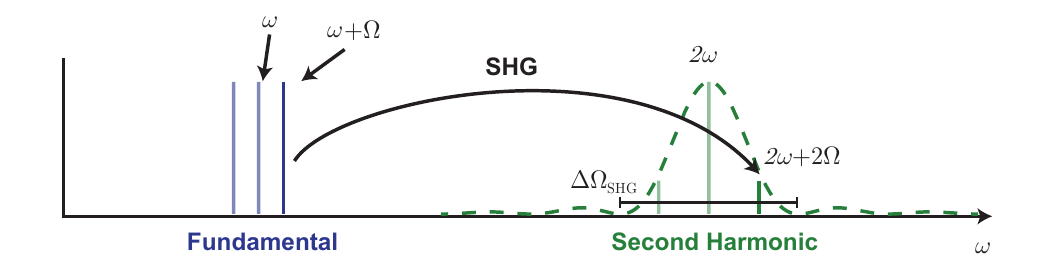}
\caption{\label{fig:Fig5}The SHG bandwidth, $\Delta\Omega_\mathrm{SHG}$, is determined by the temporal walk-off $\Delta k' L$. The height of the $\mathrm{sinc}^2$ function at $2\omega+2\Omega$ denotes the reduction in SHG due to filtering by the SHG transfer function.}
\end{figure}

We begin by considering the the amount of angular frequency detuning, $\Omega$, that we can impart on the fundamental wave before the first zero in the generated second harmonic power is reached at $2\omega+2\Omega$ (Figure \ref{fig:Fig5}). The dispersion of the refractive index as $\Omega$ is varied gives rise to a variation of the phase-mismatch,
\begin{equation}
\Delta k(\Omega) = k(2\omega+2\Omega)-2k(\omega+\Omega) - 2\pi/\Lambda_G.
\end{equation}
Given an arbitrary $\Delta k(\Omega)$, we define the SHG bandwidth, $\Delta \Omega_\mathrm{SHG}$ as the full width between the first zeroes of the $\mathrm{sinc}^2(\Delta k(\Omega) L/2)$ transfer function, which occur at $\Delta k(\Omega)L = \pm 2 \pi$. In conventional devices the bandwidth of the SHG transfer function is determined, to first order, by the mismatch of the group velocities of the interacting waves. Taylor series expanding $\Delta k$ with respect to $\Omega$, we find
\begin{equation}
\Delta k(\Omega) = \Delta k_0 + 2\Delta k'\Omega + \left(2k_{2\omega}''-k_\omega''\right)\Omega^2,\label{eqn:dkSHG}
\end{equation}
where $\Delta k_0 = \Delta k(0)$ is the phase-mismatch at $\Omega=0$, $\Delta k' = v_{g,2\omega}^{-1}-v_{g,\omega}^{-1}$ represents the group-velocity-mismatch between the interacting waves, and $k_\omega''$ represents the group velocity dispersion (GVD) at frequency $\omega$. For an interaction phase-matched at $\Omega=0$, and neglecting terms of order $\mathcal{O}(\Omega^2)$, we find that $\Delta k L = \pm 2 \pi$ when $2\Delta k'\Omega L=\pm 2\pi$. We note here that there is an intuitive time domain description of this phenomenon in terms of temporal walk-off. If we define the accumulated group delay between the fundamental and second harmonic due to temporal walk-off as $\tau_\mathrm{walk-off} = \Delta k' L$, then the full width between the zeros of the SHG transfer function, $\Delta\Omega_\mathrm{SHG}$, is given by
\begin{equation}
\Delta \Omega_\mathrm{SHG} = \frac{2\pi}{\tau_\mathrm{walk-off}}.\label{eqn:SHGBW}
\end{equation}

\Eref{eqn:SHGBW} implies that the bandwidth of an SHG device is determined by the total amount of temporal walk-off that would be accumulated between an interacting fundamental and second harmonic, and that $\Delta \Omega_\mathrm{SHG}$ decreases linearly with increasing device length. While we treat pulsed interactions in \sref{sec:pulsednlo}, \eref{eqn:SHGBW} already allows us to develop some intuition about pulsed nonlinear processes. We see here that as long as the pulses used in a nonlinear interaction are long compared to the accumulated group delay, $\tau\gg\tau_\mathrm{walk-off}$, or alternatively $\Delta\Omega \ll \Delta\Omega_\mathrm{SHG}$, the generated second harmonic bandwidth will not be filtered by the SHG transfer function. This constraint limits pulsed interactions to either short devices or long pulses.

For the special case of a group-velocity-matched interaction, $\Delta k'=0$, we find that $\Delta k L = \pm 2 \pi$ when $\left(2k_{2\omega}''-k_\omega''\right)\Omega^2 L=\pm 2\pi$. In this limit, the scaling of $\Delta \Omega_\mathrm{SHG}$ with respect to device length is no longer linear, $\Delta \Omega_\mathrm{SHG}\propto L^{-1/2}$. Group velocity matching greatly enhances the bandwidth of SHG. As an example, for fields polarized along the extraordinary axis in bulk lithium niobate, $\Delta k'=100$~fs/mm for doubling of 2-$\mu$m light. Therefore, in a 1-mm-long bulk crystal we expect the SHG bandwidth to be $10$~THz. In a dispersion-engineered medium with the same GVD parameters as the bulk media ($k_\omega'' = -50$~fs$^2$/mm, and $k_{2\omega}'' = 250$~fs$^2$/mm), but with $\Delta k'=0$, we expect the bandwidth to be $34$~THz. Given the $L^{-1/2}$ scaling of the SHG bandwidth and the numbers used here, one could increase $L$ by an order of magnitude ($L$=1 cm) before the SHG bandwidth becomes comparable to 10~THz, thereby reducing the power requirements for SHG by two orders of magnitude while retaining a broadband transfer function.

\subsection{Example design: ultra-broadband second-harmonic generation}\label{sec:exampleSHG}

\begin{figure}[t]
    \centering
    \includegraphics[width=\columnwidth]{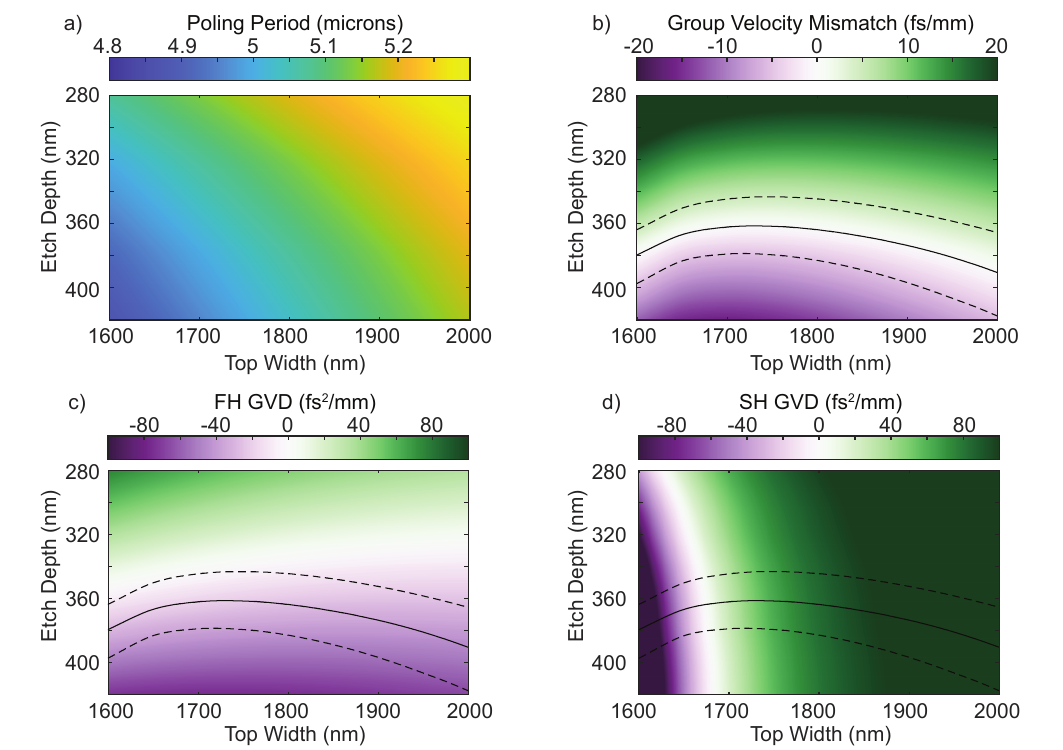}
    \caption{Variation in the phase-mismatch and dispersion orders for SHG as a function of waveguide geometry at 2060-nm for a 700 nm thin film. a) The poling period needed to achieve quasi-phasematching, $2\pi/\Delta k$, b) The group velocity mismatch, $\Delta k'$, c,d) The group velocity dispersion of the fundamental (FH; $k_\omega''$), and second harmonic (SH; $k_{2\omega}''$).  Solid black lines correspond to $\Delta k' = 0$, and dashed black lines correspond to $\Delta k' = \pm 5$ fs/mm.}
    \label{fig:SHGDesign}
\end{figure}

In nanophotonic PPLN waveguides, the propagation constants of the interacting harmonics may become strongly modified by both the tight confinement of the fundamental and avoided mode crossings near the second harmonic. As expected, this renders $\Delta k'$, $k_\omega''$, $k_{2\omega}''$, and higher dispersion orders functions of the waveguide geometry, which therefore enables engineering of $\Delta k(\Omega)$. In many cases even a simple ridge waveguide has sufficiently many degrees of freedom to achieve multiple favorable dispersion orders simultaneously (e.g. $\Delta k'=0$ and $2k_{2\omega}''=k_\omega''$) at a desired wavelength. In contrast, in bulk media the interacting wavelengths can typically be chosen to achieve one favorable dispersion order (e.g. $\Delta k'=0$ or $k_\omega''=0$). In this section we consider a design example where the geometry of a nonlinear waveguide is chosen for ultra-broadband SHG of wavelengths around 2060 nm.

Figure \ref{fig:SHGDesign}(a-d) shows the variation of the poling period and dispersion orders as a function of waveguide geometry for a 700-nm thin film. Group velocity matching occurs for etch depths around 360 nm (Figure \ref{fig:SHGDesign}(b)), anomalous dispersion at the fundamental occurs for etch depths greater than 340 nm \ref{fig:SHGDesign}(c)), and the second harmonic switches from normal to anomalous dispersion for top widths narrower than 1700 nm \ref{fig:SHGDesign}(d)).

\begin{figure}[t]
    \centering
    \includegraphics[width=\columnwidth]{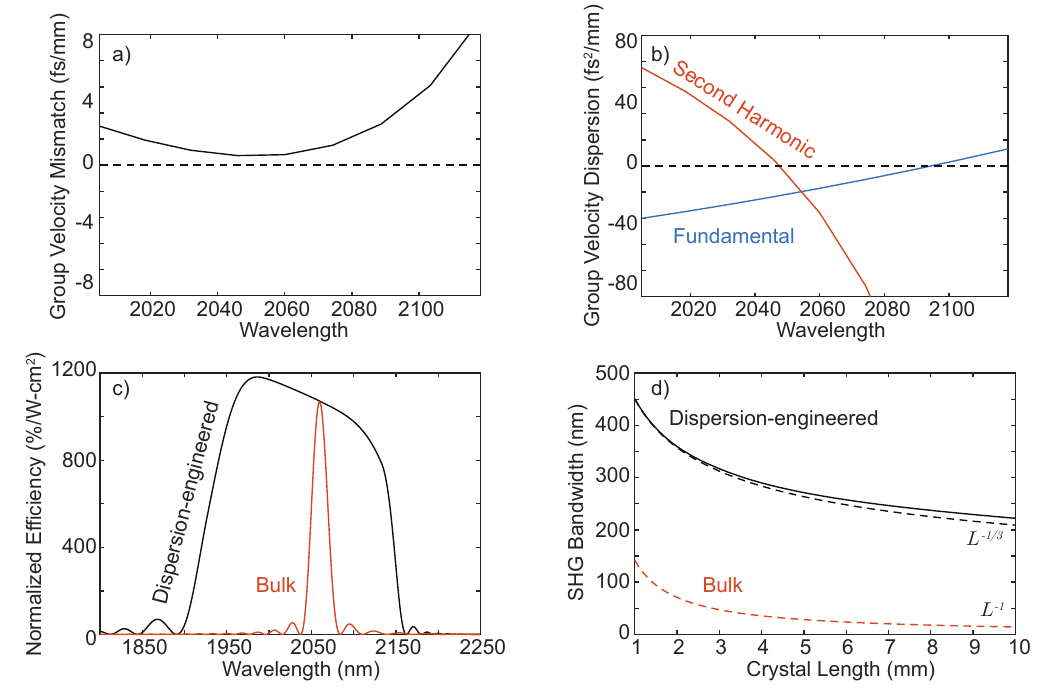}
    \caption{a,b) The calculated group velocity mismatch and group velocity dispersion as a function of wavelength for a chosen geometry corresponding to a film thickness of 700 nm, a top width of 1650 nm, and an etch depth of 364 nm. c) The SHG transfer function for a 5-mm-long device. Solid black - dispersion engineered, solid red - bulk lithium niobate. d) The scaling of SHG bandwidth with increasing device length. For the case considered here, the dispersion engineered device has a bandwidth that scales as $L^{-1/3}$ (dashed black), rather than the conventional $L^{-1}$ scaling associated with temporal walk-off (dashed red).}
    \label{fig:SHGDesign2}
\end{figure}

For a top width of $\sim 1650$ nm and an etch depth of $\sim 360$ nm, we may achieve $\Delta k'\sim 0$ and $k_\omega''\sim 2 k_{2\omega}''$, where $k_\omega''\sim -20$ fs$^2$/mm. The resulting group velocity mismatch and group velocity dispersion as a function of wavelength are shown in Figures \ref{fig:SHGDesign2}(a-b), and confirm that the waveguide maintains a small group-velocity mismatch across a large bandwidth. Figure \ref{fig:SHGDesign2}(c) shows the SHG transfer function for a 5-mm-long device, compared to the transfer function calculated using the dispersion of bulk lithium niobate. The SHG bandwidth for this length of device is 270 nm, nearly a full order of magnitude larger than the $\sim$30-nm-wide transfer function associated with the bulk dispersion relations.  Finally, we note that for the special case of $\Delta k'=0$ and $k_\omega''\sim 2k_{2\omega}''$, the leading order term in the series expansion of $\Delta k(\Omega)$ given by (\ref{eqn:SHGBW}) are of order $\Omega^3$, and therefore the bandwidth should exhibit an $L^{-1/3}$ scaling with device length. The calculated SHG bandwidth is plotted in Figure \ref{fig:SHGDesign2}(d) as a function of device length, showing good agreement with the $L^{-1/3}$ scaling (dashed orange line). Conventional devices relying on the dispersion of bulk lithium niobate exhibit an $L^{-1}$ scaling with device length; for a 1-cm-long waveguide, we expect these devices to have a 14-nm-wide transfer function, whereas dispersion-engineered devices exhibit a 220-nm-wide transfer function.

We close this section by noting that these devices exhibit broad bandwidth by having a phase-mismatch that varies slowly with wavelength. Conversely, the phase-matched wavelength of these devices exhibits an extremely rapid tuning with respect to small changes in phase-mismatch. This rapid tuning behavior and the corresponding implications for the fabrication tolerance are disussed in \ref{sec:tolerance}.

\subsection{The bandwidth of three-wave mixing}

We now generalize the treatment above to consider the bandwidths of arbitrary three-wave interactions. In this case, the three frequencies $\omega_3$, $\omega_2$, and $\omega_1$ are each detuned by $\Omega_3$, $\Omega_2$ and $\Omega_1$ respectively, where $\Omega_3 = \Omega_2+\Omega_1$. It's often convenient to parameterize the frequency detuning of the interacting waves using a common-mode detuning, $\Omega$, and an anti-symmetric detuning, $\Omega'$, such that $\Omega_3 = 2\Omega$, $\Omega_2 = \Omega + \Omega'$, and $\Omega_1 = \Omega - \Omega'$. In this case, the phase mismatch is given by
\begin{eqnarray}
    \Delta k(\Omega,\Omega') = & k_{\omega_3}(\omega_3 + 2\Omega) -  k_{\omega_2}(\omega_2 + \Omega + \Omega')\label{eqn:dkTWM}\\ 
    &- k_{\omega_1}(\omega_1 + \Omega - \Omega') - 2\pi/\Lambda_G.\nonumber 
\end{eqnarray}
For either DFG or OPA with a fixed pump at $\omega_3$ ($\Omega=0$), $\Omega'$ corresponds to the decrease of the generated idler frequency as the frequency of a seeded signal around $\omega_2$ is increased. 

We can again analyze (\ref{eqn:dkTWM}) by series expanding each of the propagation constants $k_{\omega_j}$ around $\omega_j$. To first order in $\Omega$ and $\Omega'$, the tuning of $\Delta k$ with respect to $\Omega$ and $\Omega'$ is again determined by the group velocities of the interacting waves,
\begin{eqnarray}
    \Delta k(\Omega,\Omega') = & k_0 + (\Delta k_{3-2}'+\Delta k_{3-1}')\Omega-\Delta k_{2-1}'\Omega',
\end{eqnarray}
where $\Delta k'_{3-2}=k_{\omega_3}'-k_{\omega_2}'$. When $\Delta k_{3-2}'=-\Delta k_{3-1}'$, corresponding to symmetric temporal walk-off of $\omega_1$ and $\omega_2$ relative to $\omega_3$, the phase-mismatch becomes is a function only of $\Omega'$, $\Delta k(\Omega') = k_0 - \Delta k_{2-1}'\Omega'$. This case has been used in the design of ultrafast optical parametric amplifiers~\cite{LutherDavies2017,Trapani1995,Marchese:2005,Cerullo2003,Manzoni2016}, where it is desirable for the phase-mismatch to be a weak function of the pump bandwidth. In the context of quantum optics, ultrafast parametric amplifiers that achieve symmetric walk-off have been studied both as a source of multimode squeezing~\cite{Rodriguez2020} and as a sources of separable photons. We address the latter case in \sref{sec:SPDC}.

In practice, most experimentally relevant cases impose constraints on $\Omega$ and $\Omega'$ that simplify this expansion and provide clearer insights about the role of higher order dispersion. When the frequency of any one of the three waves is held constant, $\Delta k(\Omega,\Omega')$ becomes a function only of $\Omega$ or $\Omega'$, and the tuning behavior of $\Delta k$ can be understood using the previous analysis for SHG. As an example, when $\Omega=0$,
\begin{eqnarray}
    \Delta k(\Omega,\Omega') = & \Delta k_0 -  \Delta k_{2-1}'\Omega'-\frac{k_{\omega_2}'' + k_{\omega_1}''}{2}(\Omega')^2\label{eqn:dkTWM2}
\end{eqnarray}
In this case, we find that the available bandwidth for TWM is determined by the group velocity mismatch between $\omega_1$ and $\omega_2$, $\Delta\Omega'_\mathrm{TWM}=4\pi/\Delta k_{2-1}' L$, where $\Delta k_{2-1}'=k_{\omega_2}'-k_{\omega_1}'$. The scaling of bandwidth with respect to device length again becomes $\Delta\Omega_\mathrm{TWM}'\propto L^{-1/2}$ when $\Delta k_{2-1}'=0$. Similar results occur for $\Omega+\Omega'=0$ and $\Omega-\Omega'=0$.

In the degenerate case, where $\omega_1=\omega_2=\omega$, the phase-mismatch is given by
\begin{eqnarray}
\Delta k(\Omega,\Omega') = &\Delta k_0 + 2\Delta k'\Omega+\left(2k_{2\omega}''-k_\omega''\right)\Omega^2-k_\omega''(\Omega')^2.\label{eqn:dkSFG}
\end{eqnarray}
When $\Omega=0$, (\ref{eqn:dkSFG}) can be used to calculate the bandwidth around $\omega$ that can be summed to $2\Omega$, $\Delta \Omega'_\mathrm{SFG}=2\sqrt{2\pi/k_\omega''L}$. Similarly, (\ref{eqn:dkSFG}) also describes the amount of bandwidth generated by degenerate OPA. In this case, the amplification bandwidth is determined by $\Delta k(\Omega') = 2\gamma$, rather than the usual $\Delta k(\Omega')L=2\pi$, resulting in $\Delta\Omega_\mathrm{OPA}'= 2\sqrt{2\gamma/k_\omega''}$. Therefore, the bandwidth around $\omega$ usable for SFG and the bandwidth generated by OPA around $\omega$ are both determined to leading order by $k_\omega''$ rather than the temporal walk-off. We note here that when expanding (\ref{eqn:dkSFG}) to arbitrary order in $\Omega'$ all of the odd dispersion orders cancel (e.g. $\Delta k_\omega''' (\Omega')^3 + \Delta k_\omega'''(-\Omega')^3 = 0$). As a result, when a nonlinear waveguide is dispersion-engineered to have $k_\omega''=0$ the $\Omega'$ dependence of $\Delta k$ is dominated by fourth and sixth order dispersion, which enables phase-matching bandwidths on the order of 10's of THz.

\begin{figure}[t]
\centering
\includegraphics[width=\columnwidth]{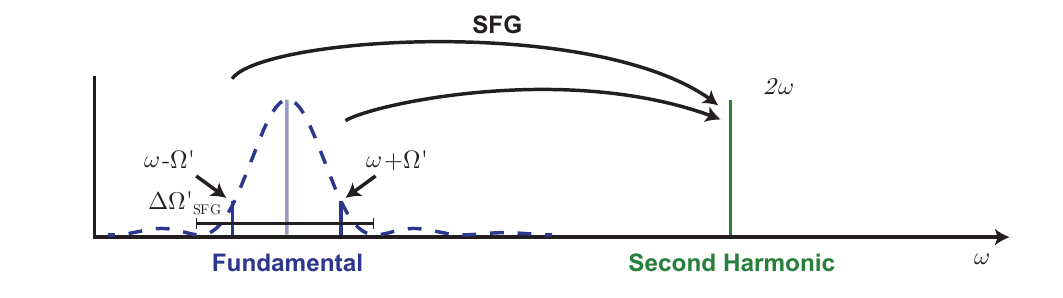}
\caption{\label{fig:TWMBW}The amount of fundamental bandwidth around $\omega$ that can contribute to SHG, $\Delta\Omega'_\mathrm{SFG}$, is limited by the group velocity dispersion of the fundamental, $k_\omega''$. Here, the height of the $\mathrm{sinc}^2$ transfer function denotes a reduced response at $2\omega$ to SFG of the two frequencies at $\omega\pm\Omega'$.}
\end{figure}

\section{Pulsed second-harmonic generation in dispersion-engineered waveguides}\label{sec:pulsednlo}

For nonlinear processes driven using short optical pulses, $A(z,t)$, both $\Omega$ and $\Omega'$ contribute to $\Delta k$, and therefore multiple dispersion orders (e.g. $\Delta k'$ and $k_\omega''$) contribute to the bandwidth generated by a nonlinear process. In bulk media, this behavior imposes a trade-off between the pulse duration of the driving field and the interaction length in the nonlinear crystal; short pulses with large instantaneous power can only be used to achieve efficient interactions in short nonlinear crystals. In this section, we generalize the CWEs for SHG to describe short pulses, and consider two limits: undepleted SHG, and quasi-static SHG. The latter case is a new regime for femtosecond pulses that can only be accessed in dispersion-engineered nonlinear waveguides by simultaneously suppressing the temporal walk-off of the interacting harmonics and the GVD of fundamental. In this limit, short pulses can interact over length scales orders of magnitude longer than in bulk media, which can reduce the energy requirements for efficient frequency conversion to the femtojoule scale in traveling-wave devices. The implications of quasi-static operation for synchronously-pumped nonlinear resonators are discussed in \sref{sec:SPND}.

\subsection{The coupled-wave equations for short pulses}

In the absence of any nonlinearity, the field envelopes for the fundamental, $A_\omega(z,t)$, and second harmonic, $A_{2\omega}(z,t)$, evolve as
\begin{eqnarray}
\partial_z A_\omega &= \hat{D}_\omega A_\omega,\label{eqn:D01}\\
\partial_z A_{2\omega} &= -\Delta k' \partial_t A_{2\omega}+\hat{D}_{2\omega} A_{2\omega},\label{eqn:D02}
\end{eqnarray}
where we have chosen our phase reference such that the pulse envelopes are in a reference frame co-moving with the fundamental pulse envelope, $A_\omega$. The dispersion operators, $\hat{D}$, describe the evolution of the field envelopes due to GVD and higher-order dispersion, $\hat{D}_\omega\sum_{m=2}^{\infty}(-i)^{(m+1)}\left(k_\omega^{(m)}\partial_t^m\right)/m!$, where $k_\omega^{(m)}$ represents the m$^{th}$ derivative of propagation constant $k$ around frequency $\omega$.

We may add nonlinearity by assuming that $\chi^{(2)}$ is sufficiently dispersionless that the nonlinear polarization given by $P_\mathrm{NL}(2\omega)=\epsilon_0 \int \chi^{(2)}_\mathrm{eff}(2\omega;\omega',2\omega-\omega')E(\omega')E(2\omega-\omega')d\omega'$ can be evaluated in the time domain, \textit{i.e.} $P_{\mathrm{NL},2\omega}(t)=2\epsilon_0 d_\mathrm{eff} E_\omega^2(t)$, where $d_\mathrm{eff}=\chi^{(2)}_\mathrm{eff}/2$. In this case, we may add the contributions to $\partial_z A_\omega(z,t)$ from the dispersion operator and the nonlinear coupling in (\ref{eqn:CWE01}-\ref{eqn:CWE02}) to find
\begin{eqnarray}
\partial_z A_\omega &= -i\kappa A_{2\omega}A_\omega^*\exp(-i\Delta k z) + \hat{D}_\omega A_\omega,\label{eqn:CWE21}\\
\partial_z A_{2\omega} &= -i\kappa A_\omega^2\exp(i\Delta k z)-\Delta k' \partial_t A_{2\omega}+\hat{D}_{2\omega} A_{2\omega}.\label{eqn:CWE22}
\end{eqnarray}
We note here that (\ref{eqn:CWE21}-\ref{eqn:CWE22}) are scale invariant with respect to the following transformation,
\begin{eqnarray}
D\rightarrow D/s_1, \Delta k\rightarrow \Delta k/s_1,\Delta k'\rightarrow \Delta k'/s_1,\\
L\rightarrow s_1 L,\mathrm{P_{in}}(t)\rightarrow\mathrm{P_{in}}(t)/s_1^2,
\end{eqnarray}
where $\mathrm{P_{in}}(t)=|A_\omega(0,t)|^2+|A_{2\omega}(0,t)|^2$. For any reduction of the dispersion, temporal walk-off, and phase-mismatch by a factor $s_1$, the interaction length $L$ can be increased by $s_1$, thereby facilitating a quadratic reduction of the power requirements for SHG.

In general, numerical split-step Fourier methods are required to solve (\ref{eqn:CWE21}-\ref{eqn:CWE22}) to account for the evolution of the field envelopes due to dispersion and nonlinearity. However, (\ref{eqn:CWE21}-\ref{eqn:CWE22}) can be solved analytically in two limits: in the limit of undepleted SHG, and in the quasi-static limit, where dispersion is negligible. We consider these cases below.

\subsection{Undepleted second-harmonic generation}

For the case of undepleted SHG, one may solve (\ref{eqn:CWE21}-\ref{eqn:CWE22}) using a transfer function approach~\cite{Imeshev2000a,Imeshev2000b}. In this case, the evolution of $A_\omega(z,t)$ is given by (\ref{eqn:D01}) and is readily solved in the frequency domain, $\hat{A}(z,\Omega)=\hat{A}(0,\Omega)\exp(\hat{D}_\omega(\Omega)z)$. Similarly, in the frequency domain (\ref{eqn:CWE22}) becomes
\begin{eqnarray}
\partial_z \hat{A}_{2\omega}(z,2\Omega) = &-i\kappa \int \hat{A}_{\omega,+}\hat{A}_{\omega,-}\exp(i\Delta k z)d\Omega'\label{eqn:SH_Undep}\\
&-i2\Omega\Delta k' \hat{A}_{2\omega}(z,2\Omega)\nonumber\\
&+\hat{D}_{2\omega}(2\Omega)\hat{A}_{2\omega}(z,2\Omega),\nonumber
\end{eqnarray}
where $\hat{A}_{\omega,\pm}=\hat{A}_\omega(z,\Omega\pm\Omega')$. Defining $\tilde{A}_{2\omega}(z,2\Omega)=\hat{A}_{2\omega}(z,2\Omega)\exp\left(i \Delta k'_\omega 2\Omega z-\hat{D}_{2\omega}(2\Omega)z\right)$, and multiplying both sides of (\ref{eqn:SH_Undep}) by $\exp\left(i \Delta k'_\omega 2\Omega z-\hat{D}_{2\omega}(2\Omega)z\right)$, we have
\begin{equation}
\partial_z \tilde{A}_{2\omega}(z,2\Omega) = -i\kappa \int \hat{A}_{\omega,+}\hat{A}_{\omega,-}\exp(i\Delta k (\Omega,\Omega')z)d\Omega'\label{eqn:pulseSHG}
\end{equation}
where the phase mismatch is given, as before, by $\Delta k(\Omega,\Omega') = k(2\omega + 2\Omega)-k(\omega+\Omega + \Omega')-k(\omega + \Omega-\Omega')$. As with CW undepleted SHG, (\ref{eqn:pulseSHG}) can be integrated to yield the resulting second harmonic,
\begin{eqnarray}
\tilde{A}_{2\omega}(z,2\Omega) = &-i\kappa z\int \hat{A}_\omega(0,\Omega+\Omega')\hat{A}_\omega(0,\Omega-\Omega')\\
&\exp(i\Delta k(\Omega,\Omega') z/2)\mathrm{sinc}(\Delta k(\Omega,\Omega') z/2)d\Omega'.\label{eqn:pulseSH}\nonumber
\end{eqnarray}

The response of the second harmonic to the input fundamental can still be understood in terms of a transfer function, but in this case we need to know $\mathrm{sinc}(\Delta k(\Omega,\Omega') z/2)$ for every pair of interacting frequencies. The dependence of $\Delta k(\Omega,\Omega')$ on the detuning of the fundamental and second harmonic was studied in \sref{sec:BW}, and we reproduce the result here. To second order in $\Omega$ and $\Omega'$, $\Delta k(\Omega,\Omega')$ is given by
\begin{equation}
\Delta k(\Omega,\Omega') = \Delta k_0 + 2\Delta k'\Omega + \left(2k_{2\omega}''-k_\omega''\right)\Omega^2-k_\omega''(\Omega')^2.\label{eqn:dkpulse}
\end{equation}
The main feature of (\ref{eqn:dkpulse}) is that $\Omega$ and $\Omega'$ enter independently, which allows us to interpret the resulting dynamics using our insights from the CW case. We see that the amount of bandwidth around the fundamental that can contribute to SHG is given by $\Delta \Omega'_\mathrm{SFG}=2\sqrt{2\pi/k_\omega''L}$ and determined to leading order by $k_\omega''$. Similarly, we see that the amount of bandwidth generated at the second harmonic $\Delta \Omega_\mathrm{SHG}$ is limited to leading order by the temporal walk-off between the waves $\Delta k'$. Finally we note that for narrowband pulses, or materials with small $k_\omega''$, we can neglect the $\Omega'$ term. This renders $\Delta k(\Omega,\Omega')$ a function only of $\Omega$, and (\ref{eqn:pulseSH}) reduces to
\begin{eqnarray}
\tilde{A}_{2\omega}(z,2\Omega) = &-i\kappa z\exp(i\Delta k(\Omega) z/2)\mathrm{sinc}(\Delta k(\Omega) z/2)\label{eqn:filteredSH}\\
&\int \hat{A}_\omega(0,\Omega+\Omega')\hat{A}_\omega(0,\Omega-\Omega')d\Omega'.\nonumber
\end{eqnarray}
In essence, \eref{eqn:filteredSH} suggests that the response of the generated second harmonic to the nonlinear polarization generated by the fundamental is simply filtered by the CW SHG transfer function. Therefore, a semi-analytical calculation of the second harmonic may be achieved in two steps. First we calculate the second harmonic envelope that would be generated in the absence of dispersion in the time domain, $A_{2\omega}^\mathrm{ND}(z,t)=-i\kappa A_\omega^2(0,t)z$. Then, the power spectral density associated with this envelope is filtered in the frequency domain by the CW transfer function for SHG,
\begin{equation}
|\hat{A}_{2\omega}(z,2\Omega)|^2=\mathrm{sinc^2}(\Delta k(\Omega) z/2)|\hat{A}_{2\omega}^\mathrm{ND}(z,\Omega)|^2.\label{eqn:SHfilter}
\end{equation}

As with CW interactions, the treatment used here is readily extended to TWM by adding the dispersion operators $\hat{D}_{\omega_j}$ and temporal walk-off to the coupled-wave equations. In the undepleted limit, the analysis leading to (\ref{eqn:pulseSH}) is essentially unchanged. We forgo this derivation here and simply note that, as with SHG, the analysis of the transfer function $\mathrm{sinc}(\Delta k(\Omega,\Omega')L/2)$ in terms of the dispersion orders of the interacting waves is sufficient to determine the generated harmonic during SFG, or signal and idler during DFG and SPDC.

\subsection{Quasi-static Interactions}\label{sec:quasi-static}

In the previous section, we found that the bandwidth around the fundamental that can contribute to SHG is determined by $\Delta k_\omega''$, and that the bandwidth generated around the second harmonic is determined by $\Delta k'$. This suggests that in a dispersion-engineered waveguide where these two terms are simultaneously zero we may neglect the temporal walk-off and dispersion operators entirely. In this quasi-static limit we can now solve the CWEs for SHG and account for an arbitrary amount of pump depletion. In this case, the CWEs for the pulse envelopes $A_\omega(z,t)$ and $A_{2\omega}(z,t)$ are given by the CWEs for CW SHG, with each temporal slice of the pulses undergoing conversion independently
\begin{eqnarray}
\partial_z A_\omega(z,t) &= -i\kappa A_{2\omega}(z,t)A_\omega^*(z,t)\exp(-i\Delta k z),\label{eqn:CWE31}\\
\partial_z A_{2\omega}(z,t) &= -i\kappa A_\omega^2(z,t)\exp(i\Delta k z).\label{eqn:CWE32}
\end{eqnarray}
This heuristic model enables us to develop an intuitive understanding of SHG, OPA, and optical parametric oscillation (OPO) since (\ref{eqn:CWE31}-\ref{eqn:CWE32}) can be solved exactly for most cases of interest. This quasi-static heuristic has been shown to be accurate even for pulses with octave-spanning power spectra in the presence of weak dispersion~\cite{jankowski2021supercontinuum}. For phase-matched SHG, the second harmonic envelope is given by
\begin{equation}
A_{2\omega}(z,t) = -i A_\omega(0,t)\mathrm{tanh}(\kappa A_\omega(0,t)z),\label{eqn:QSSH}
\end{equation}
which exhibits an instantaneous conversion efficiency given by $\mathrm{P}_{2\omega}(z,t)/\mathrm{P}_\omega(z,t) = \mathrm{tanh}^2(\kappa A_\omega(0,t)z)$. Conversion occurs rapidly around $t=0$, where $A_\omega(0,t)$ is the largest. In the $t\rightarrow\pm \infty$ tails of $A_\omega$, we recover the conventional $\mathrm{P}_{2\omega}(z,t)/\mathrm{P}_\omega(z,t) = \eta_0 P_\omega(0,t)z^2$ scaling associated with undepleted SHG.

We note here that while in general numerical split-step Fourier methods are required to solve (\ref{eqn:CWE21}-\ref{eqn:CWE22}), in the presence of a small amount of dispersion we can perform a single split-step analytically. In this case, we solve for $A_{2\omega}^{\mathrm{ND}}(z,t)$ in the time-domain using (\ref{eqn:QSSH}), and then filter this envelope in the frequency domain using (\ref{eqn:SHfilter}).

\subsection{Experimental demonstration of ultra-broadband second harmonic generation}

\begin{figure}
    \centering
    \includegraphics[width=\columnwidth]{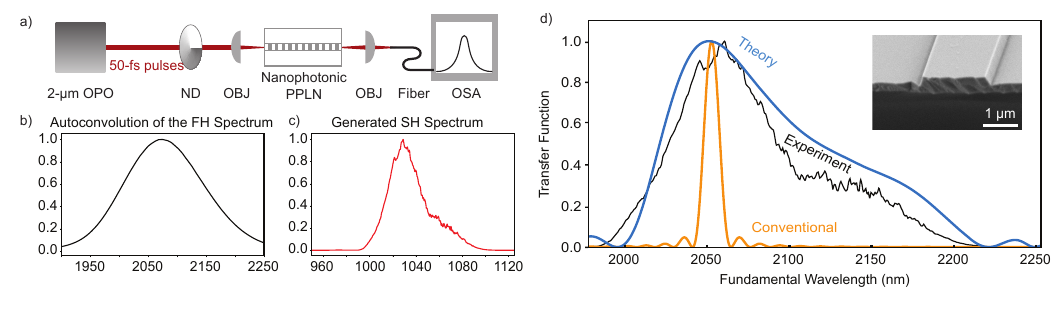}
    \caption{Schematic of experimental setup. The input pulses from an OPO are attenuated using a variable neutral density filter (ND), and are focused into the sample using a reflective objective lens (OBJ). The output harmonics are measured using an optical spectrum analyzer (OSA). b,c) Measured spectrum power spectral density of the driving polarization ($|A_\omega(0,\Omega)*A_\omega(0,\Omega)|^2$) and output second harmonic ($|A_{2\omega}(L,\Omega)|^2$), respectively. d) Measured SHG transfer function (black) for a 6-mm-long nanophotonic waveguide, showing good agreement with theory (blue). The bandwidth of these waveguides exceeds that of bulk PPLN (orange) by more than an order of magnitude. Figure adapted from Jankowski \textit{et al}., Optica \textbf{7}, 40 (2020). Copyright 2020 Authors, licensed under the terms of the OSA Open Access Publishing Agreement.}
    \label{fig:pulsed_TF}
\end{figure}

We now consider an experimental demonstration of SHG in a waveguide that has been designed to achieve quasi-static interactions of femtosecond pulses~\cite{Jankowski2020}. The fabricated waveguides are similar to the designs discussed in \sref{sec:exampleSHG}, except that the waveguide geometry has been chosen to achieve quasi-static SHG around 2050-nm ($\Delta k' \sim 0$ and $k_\omega'' \sim 0$). The fabricated waveguides have a top width of 1850 nm, an etch depth of 340 nm, and a film thickness of 700 nm, corresponding to a temporal walk-off of $\Delta k' = 5$ fs/mm, and group velocity dispersion of $\Delta k_\omega'' = -15$ fs$^2$/mm. We fabricated 6-mm-long waveguides with poling periods ranging from 5.01 $\mu$m to 5.15 $\mu$m in steps of 10 nm, corresponding to a shift of the phase-mismatch by $\Delta k L = 4\pi$ between successive devices.

The experimental setup is shown in Figure \ref{fig:pulsed_TF}(a). 50-fs-long sech$^2$ pulses with a repetition rate of 75-MHz from an optical parametric oscillator (OPO) are focused into, and collected from, the PPLN waveguides using  Thorlabs LMM-40X-P01 reflective objectives (OBJ). This method of focusing ensures that the focused beams are free of chromatic aberrations, and that the in-coupled pulses are free of chirp. The light output from the end-facet of the waveguide is then imaged into a high-NA multi-mode fiber, and the resulting fundamental and second harmonic spectra are captured using two Yokogawa optical spectrum analyzers (OSA). To characterize the SHG transfer function, we record the spectrum input to the waveguide at the fundamental and output from the waveguide at the second harmonic. Then, we estimate $A_{2\omega}^\mathrm{ND}(z,\Omega)\propto A_\omega(z,\Omega) * A_\omega(z,\Omega)$ using the auto-convolution of the spectrum of the fundamental, shown in Figure \ref{fig:pulsed_TF}(b). The ratio of the measured second harmonic spectrum (Figure \ref{fig:pulsed_TF}(c)) with $A_{2\omega}^\mathrm{ND}$ yields the measured SHG transfer function (Figure \ref{fig:pulsed_TF}(d)), showing good agreement between experiment and theory. These devices exhibit a $\Delta\lambda_\mathrm{SHG}$ bandwidth $>$220 nm, which outperforms bulk 2-$\mu$m SHG devices of the same length in PPLN by an order of magnitude, and would have an even greater advantage in longer devices due to the $L^{-1/2}$ scaling of the bandwidth. This broad transfer function confirms that the waveguide achieves nearly-quasi-static interactions of short pulses across the length of the device. The inset of (Figure \ref{fig:pulsed_TF}(d)) shows an SEM image of the end facet of the waveguide, from which we estimated the waveguide geometry. The strong agreement between the measured and theoretical transfer function verifies the waveguide dispersion calculated using these parameters.

\begin{figure}
    \centering
    \includegraphics[width=\columnwidth]{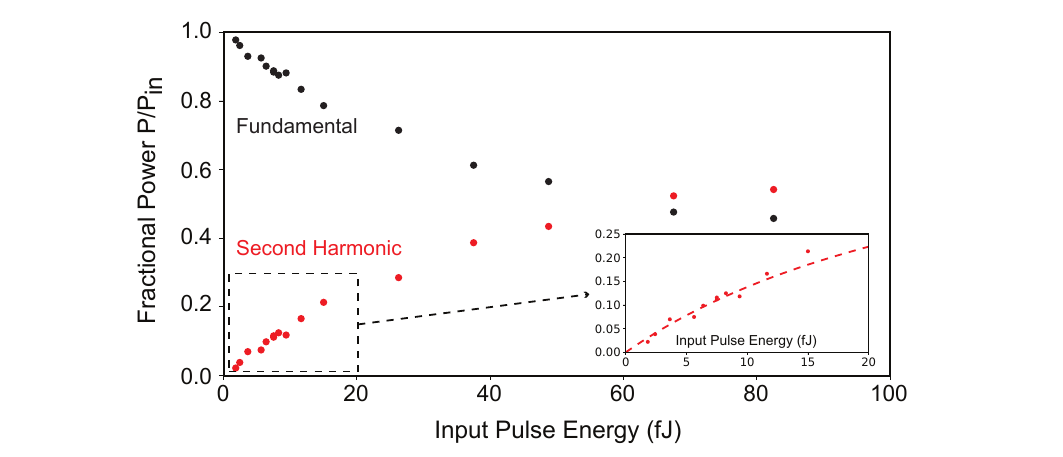}
    \caption{SHG conversion efficiency and pump depletion as a function of input pulse energy, showing 50\% conversion efficiency with an input pulse energy of 60-fJ. Inset: Undepleted regime with fit given by \ref{eqn:SHfilter} and \ref{eqn:QSSH}. Figure reproduced from Jankowski \textit{et al}., Optica \textbf{7}, 40 (2020). Copyright 2020 Authors, licensed under the terms of the OSA Open Access Publishing Agreement.}
    \label{fig:pulsed_conv}
\end{figure}

Having verified that these waveguides achieve phase-matching and ultra-broadband interactions, we measured the conversion efficiency as a function of the pulse energy input to the waveguide. The results are shown in Figure \ref{fig:pulsed_conv}. The inset shows the undepleted regime, denoted by the dotted box, with the dashed line corresponding to a theoretical fit based on a heuristic model for saturation, where the field envelopes calculated using \eref{eqn:QSSH} are then filtered by the measured SHG transfer function using \eref{eqn:SHfilter}. The only fitting parameter used here is a  normalized efficiency of $\eta_0=1000~\%$/W-cm$^2$, which is in good agreement with the theoretically predicted value of $1100~\%$/W-cm$^2$. This normalized conversion efficiency is a 50-fold improvement over the theoretical values for conventional reverse proton-exchanged waveguides due to the small effective area, $\sim 1$ $\mu$m$^2$, of the TFLN waveguide. Here, a normalized efficiency of 20\%/W-cm$^2$ is estimated for doubling of wavelength around 2 $\mu$m in a diffused waveguide, when the quartic scaling of normalized efficiency with wavelength is used to scale from the 1.5-$\mu$m value~\cite{Langrock2006}. When driven with short pulses, these TFLN waveguides achieve a conversion efficiency of 50\% using only 60 fJ of in-coupled pulse energy. The large normalized efficiency reported here combined with a 10-fold increase in interaction length, results in a 5000-fold reduction in the energy requirements needed to achieve saturation when compared to a conventional device with the same bandwidth.

These results represent one of the first examples of dispersion-engineered nonlinear interactions in nanophotonic PPLN devices, and confirm that fabricated devices can achieve the large bandwidth enhancements predicted in \sref{sec:BW}. More recent work has also quasi-static optical parametric amplification in similar devices~\cite{jankowski2021efficient,ledezma2021intense}. These devices demonstrated unsaturated gains as large as 120 dB/cm across nearly a micron of bandwidth using only four picojoules of pump pulse energy, and achieved efficient optical parametric generation with orders of magnitude less pulse energy than previous demonstrations in $\chi^{(2)}$ waveguides. In all of these cases the performance improvements of dispersion-engineered devices when compared to conventional devices are substantial. In the following sections, we consider the role of dispersion engineering in nonlinear devices used to generate non-classical light.

\section{Ultra-broadband squeezed light}\label{sec:OPA}

In the previous sections we established the bandwidths associated with three-wave interactions and verified the behavior of dispersion-engineered nonlinear devices using pulsed SHG. We now consider the design of a dispersion-engineered OPA operating around degeneracy, which can be used both to generate and detect broadband squeezed vacuum~\cite{Shaked2018}. Squeezed states are a critical resource for continuous-variable quantum information processing, and their use in measurement-based quantum computation represent a promising route towards universal, fault-tolerant quantum computation~\cite{Takeda2019,Pfister2019}. In particular, measurement-based quantum computation requires multi-partite entangled states known as cluster states~\cite{Raussendorf2001,Raussendorf2003}. In this scheme, the number of entangled modes is a computational resource, with larger computations requiring more modes.

\begin{figure}
    \centering
    \includegraphics[width=\columnwidth]{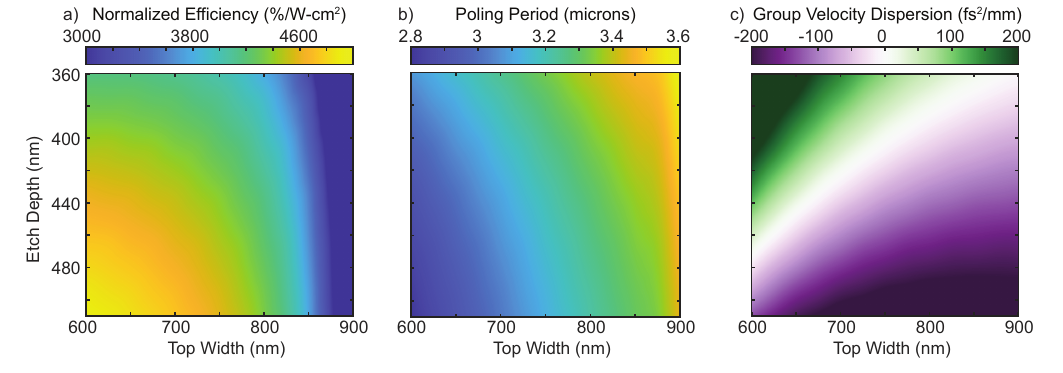}
    \caption{Variation with waveguide geometry of the a) nonlinearity, b) phase-mismatch, and c) group velocity dispersion of a 600-nm thin film for a fundamental centered on 1560 nm.}
    \label{fig:OPADesign}
\end{figure}

Recent work has focused on time domain multiplexed continuous-variable cluster states, which have proven to be extremely scalable~\cite{Yokoyama2013}: these states have been successfully scaled to one million entangled modes~\cite{Yoshikawa2016}, where other schemes typically operate with 10's of modes, and have been used to produce 2-D cluster states~\cite{Asavanant2019}, which are necessary for universal quantum computation. In this approach, broadband squeezed light from two CW-pumped OPAs (or OPOs) is partitioned into time bins of period $T$, where $T\gg 1/\Delta\Omega$ is determined by the bandwidth of the squeezed light. These two spatially separated beams (hereafter referred to as rails) are then combined on a 50:50 beamsplitter to create a series of EPR states separated by $T$. One rail is then delayed by $T$, and the two rails are again interfered on a 50:50 beamsplitter such that each time bin is entangled with two neighboring time bins of the opposite rail, thereby forming a cluster state. This method has three limitations: i) The speed of computation is set by the size of the time bins, $T$, and therefore by the bandwidth of the optical parametric amplifier. Early demonstrations based on OPOs were limited to the bandwidth of a cavity resonance (34 MHz), and therefore utilized time bins of $T$=158 ns~\cite{Yokoyama2013}. ii) The physical size of the computer is set by the delay line, 30 m for $T$=158 ns. iii) The amount of squeezing needed for fault-tolerant quantum computation is $\sim 20$ dB~\cite{Pfister2019}, which exceeds any experimental demonstration to date~\cite{Vahlbruch2016}. Current state of the art devices have focused on using guided-wave OPA, and have achieved 6 dB of squeezing with 2.5 THz of bandwidth~\cite{Kashiwazaki2020}.

\begin{figure}[t]
    \centering
    \includegraphics[width=\columnwidth]{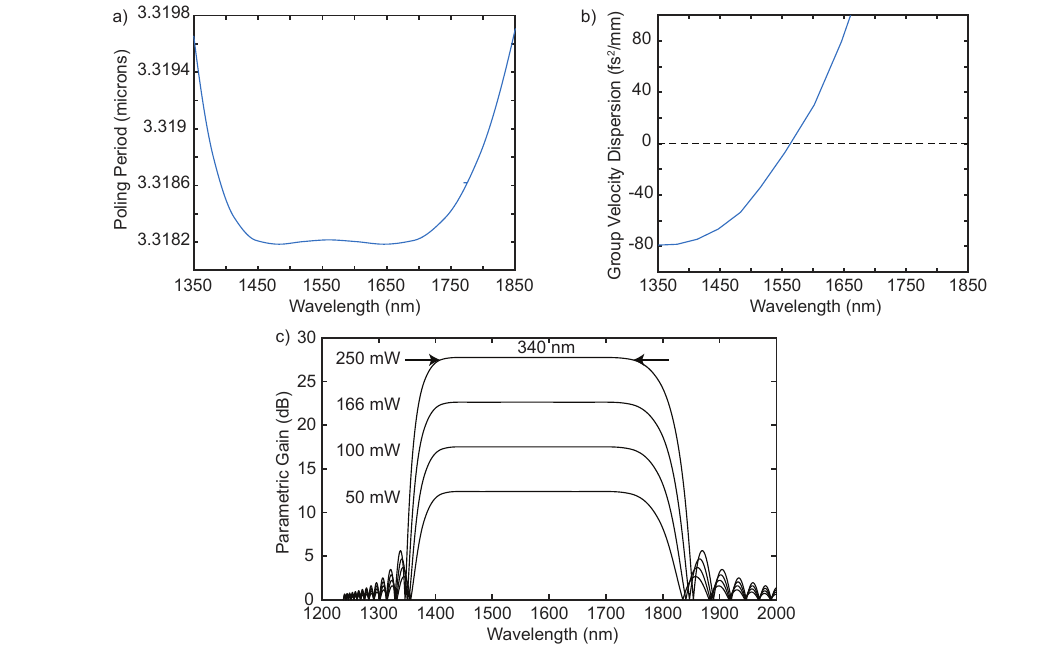}
    \caption{ a) Poling period, b) group velocity dispersion of the fundamental, and c) OPA gain spectrum, $R_+(\omega')$, as a function of signal wavelength. Here we assumed $\eta_0$=4000\%/W-cm$^2$, $\mathrm{P_{in}}=200$ mW, and $\eta_\mathrm{sys}$=1.}
    \label{fig:OPADesign2}
\end{figure}

As discussed in \sref{sec:BW}, the bandwidth of degenerate OPA with a CW pump ($\Omega=0$) is dominated by $k_\omega''$, with a higher order contribution from fourth order dispersion. We therefore focus on $k_\omega''$ and ignore the role of $\Delta k'$, $k_{2\omega}''$, and higher order dispersion. The normalized efficiency, poling period, and $k_\omega''$ are plotted in Figure \ref{fig:OPADesign} as a function of waveguide geometry for a 600-nm thin film. While many waveguide geometries may achieve $k_\omega''=0$, we consider the case where the top width and etch depth are given by 784 nm and 390 nm, respectively, corresponding to  $\eta_0=$4000\%/W-cm$^2$. The poling period and $k_\omega''$ are shown as a function of signal wavelength in Figure \ref{fig:OPADesign2}(a-b). The nominal poling period of 3.3182 $\mu$m remains flat as the signal and idler wavelength are tuned across hundreds of nanometers of bandwidth due to the zero-crossing of $k_\omega''$ around 1560 nm. This observed slow variation of $\Delta k(0,\Omega')$ confirms ultra-broadband operation. It can be shown~\cite{Crouch1988} that the maximum parametric gain and squeezing attainable is given by $G_\pm = |\mu|\pm|\nu|$, where $\mu(z) = \left[\mathrm{cosh}(gz)+\frac{i\Delta k}{2g}\mathrm{sinh}(gz)\right]$, and $\nu(z) = \frac{\gamma}{g}\mathrm{sinh}(gz)$.

\begin{figure}
    \centering
    \includegraphics[width=\columnwidth]{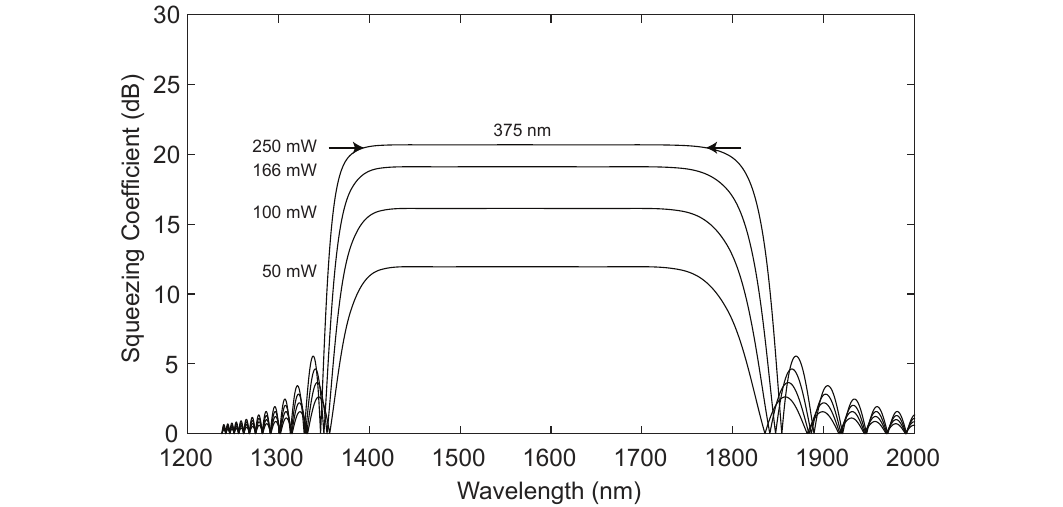}
    \caption{Squeezing spectrum, $R_-(\Omega')$, as a function of signal wavelength. Here we assumed the same parameters as in Figure \ref{fig:OPADesign2}, with $\alpha = 3$ dB/m.}
    \label{fig:OPADesign3}
\end{figure}

The parametric gain, $G_+(\Omega')$ is shown as a function of wavelength and pump power in Figure \ref{fig:OPADesign2}(c) for a 1-cm-long waveguide. These waveguides produce flat parametric gain (to within 1$\%$) across 340 nm of bandwidth ($\Delta \Omega'\sim$40 THz) due to the slow variation of $\Delta k(0,\Omega')$, and may achieve nearly 30-dB of gain for 250 mW of pump power. The amount of detectable squeezing $R_-$ will be limited by the propagation loss $\alpha$ and detection efficiency $\eta_D$,
\begin{equation}
R_\pm = 1 - \eta_\mathrm{sys} + \eta_\mathrm{sys}G_\pm,
\end{equation}
where $\eta_\mathrm{sys}=(1-\exp(-2\alpha L))\eta_D$ is the total detection efficiency~\cite{Schnabel2017}. Assuming values for the loss of $\alpha$=3 dB/m~\cite{Zhang2017}, we find that 20 dB of squeezing is possible for the values considered here (Figure \ref{fig:OPADesign3}). Further increases of $R_-$ are possible by driving short waveguides with more power, provided that pump-induced losses do not become significant. While early SHG experiments pumped with 100's of mW of pump power showed no evidence of such loss mechanisms~\cite{Wang2018}, further experimental study is needed to characterize these effects in the context of squeezing since the waveguide is pumped at shorter wavelengths. In the absence of these detrimental effects, these devices are a promising route to achieve sufficient parametric gain for fault-tolerant quantum computation and the use of dispersion engineering enables OPA with 100's of nanometers of bandwidth, potentially miniaturizing the physical systems used to implement measurement-based quantum computation by another order of magnitude relative to the state-of-the-art~\cite{Kashiwazaki2020}.

\section{Heralded photons from spontaneous parametric downconversion}\label{sec:SPDC}
Having established the role of waveguide dispersion in pulsed nonlinear interactions, we now consider device designs that rely on co-engineering waveguide dispersion with spatially varying poling domains to generate heralded single photons. The generation of indistinguishable (high purity) single photons is essential for numerous applications, including linear optical quantum computation~\cite{Walmsley2005, Kok2007, Humphreys2013}, quantum communication~\cite{Briegel1998}, quantum simulations~\cite{Spring2012}, and quantum metrology~\cite{Motes2015}. Trapped atoms~\cite{Higginbottom2016}, quantum dots~\cite{Dusanowski2019}, single vacancy centers~\cite{Sipahigil2014}, and heralded SPDC~\cite{Chen2017} are some common platforms for generation of pure single photons. Heralded SPDC has advantages of its ease of access, room temperature operation, ease of integration with communication channels, and the ability to engineer emission in precise spatio-temporal modes for high visibility interference.


We begin this section with a brief review of the theory of heralded SPDC and show that the purity of the heralded photon is largely determined by the dispersion of the nonlinear medium. We note here that while high purity heralded single photon generation has been implemented by several means in the past~\cite{Mosley2008, URen2006, Sller2011, Hendrych2007, Zhang2019}, these demonstrations have been limited to nonlinear crystals that achieve the necessary dispersion at desirable wavelengths, or have approximated the desired dispersion through periodic retiming of the pulse envelopes~\cite{URen2006}. In this section, we focus on dispersion-engineered SPDC in TFLN waveguides. There are several advantages to heralded single photon generation in TFLN devices as opposed to traditional LN sources: i) factorable biphoton generation at wavelengths previously not accessible in LN, ii) higher normalized efficiencies, which reduce the necessary pump power by orders of magnitude, and iii) the opportunity for on-chip multiplexing of low-probability, high-fidelity single photon sources for a compact pseudo-deterministic heralded single photon source~\cite{MeyerScott2020}. In principle, the designs discussed here may achieve pair generation rates as large as 1 GHz when pumped with 10 $\mu$W of power, which outperforms state-of-the-art devices by more than an order of magnitude~\cite{Ma2020,Steiner2021}. 


\subsection{Theory of heralded SPDC}

In $\chi^{(2)}$ media heralded SPDC involves pumping a nonlinear crystal at frequency $\omega_{p}$ such that the nonlinear interaction probabilistically annihilates a photon at $\omega_{p}$ and generates a photon at $\omega_{s}$ and at $\omega_{i}$ (with $\omega_{s}$ + $\omega_{i}$ = $\omega_{p}$), one of which can be detected to herald the presence of the other. The generated signal and idler photons are typically entangled in some or all of spatial, spectral, time-bin and polarization degrees of freedom. While time-bin entanglement is essential to the heralding process, the other degrees of entanglement are deleterious to the purity of the heralded photon. We note here that SPDC in well-defined waveguide modes largely eliminates the spatial and polarization entanglements, and we therefore focus on suppressing frequency domain correlations. The discussion here follows that in~\cite{URen2005GenerationOP, Aichele2002, URen04}. 

The SPDC Hamiltonian has the same form as the Hamiltonian for TWM (\ref{eqn:H}), where we assume the nonlinear medium is pumped with a pulsed coherent state and produces photons in a band of frequencies, $\omega_{s,m}$ and $\omega_{i,n}$. \Eref{eqn:H} can thus be written as
\begin{eqnarray}
    \hat{H}_\mathrm{int}/\hbar =  \sum_{m,n} \gamma(\omega_{s,m}+\omega_{i,n})\hat{a}_{s,m}^{\dagger}(\omega_s)\hat{a}_{i,n}^{\dagger }(\omega_i) + h.c.\label{eqn:H_SPDC},
\end{eqnarray}
where $\gamma(\omega_{s,m}+\omega_{i,n})=\kappa v_g \mathrm{sinc}(\Delta k(\omega_{s,m},\omega_{i,n}) L/2)A_p(\omega_{s,m} + \omega_{i,n})$, and $A_p(\omega_p)$ refers to the classical pump amplitude in W$^{1/2}$ at frequency $\omega_p$. The signal and idler frequencies are given by $\omega_{s,m}=\omega_{s} + 2\pi m/T$, where $T$ is the repetition period of the pump pulses used to drive the nonlinear medium. The coupling coefficient is proportional to the classical transfer function for DFG~\eref{eqn:OPA03}, $\gamma(\omega_{s,m}+\omega_{i,n})\propto \mathrm{sinc}(\Delta k(\omega_{s,m},\omega_{i,n})L/2)$, hereafter referred to as the phase-matching function $\Phi(\omega_{s,m},\omega_{i,n})=\mathrm{sinc}(\Delta k(\omega_{s,m},\omega_{i,n}) L/2)$. The signal and idler are seeded only by vacuum fluctuations. In the weak interaction limit, corresponding to small parametric gain, we consider only up to the first order perturbation in the evolution of the signal and idler, thereby ignoring multiple pair emission. Under these assumptions, the state of interest is given by
\begin{equation}
    \ket{\psi } = \ket{0} + \sum_{m,n} f(\omega _{s,m}, \omega _{i,n})\ket{1_\omega} _{s,m}\ket{1_{\omega }} _{i,n},
    \label{eq2}
\end{equation}
where $f(\omega _{s,m}, \omega _{i,n}) = A_p(\omega_{s,m}+ \omega_{i,n}) \Phi(\omega_{s,m}, \omega _{i,n})$ is the joint spectral amplitude (JSA) of the biphoton state. Using the idler as the heralding photon, the density matrix representing the state of the heralded photon (signal) is given by the partial trace over the idler 
\begin{equation}
    \hat{\rho} _s = Tr_i\{\ket{\psi} \bra{\psi }\hat{P}_i\},
    \label{eq3}
\end{equation}
where $\hat{P}_i$ is the measurement operator corresponding to the heralding operation
\begin{equation}
    \hat{P}_i = \sum_n P(\omega ^{\prime })\ket{1_{\omega ^{\prime }}}_{i,n} \bra{1_{\omega ^{\prime }}}_{i,n}.
    \label{eq4}
\end{equation}
Thus we obtain 
\begin{equation}
    \hat{\rho }_s = \sum_n P(\omega ^{\prime }) \ket{\psi^{\prime}(\omega ^{\prime })}_{i,n}\bra{\psi^{\prime}(\omega ^{\prime })}_{i,n},
    \label{eq5}
\end{equation}
where 
\begin{equation}
    \ket{\psi^{\prime}(\omega ^{\prime })}_{i} = \sum_m f(\omega _{s,m}, \omega ^{\prime }_{i,n})\ket{1_{\omega }}_{s,m}.
    \label{eq6}
\end{equation}

\Eref{eq5} is the integral over the ensemble of pure states $\ket{\psi^{\prime}(\omega ^{\prime })}_{i}$ weighted by the heralding probability function ($P(\omega ^{\prime})$), and in general we have $\mathrm{Tr}(\hat{\rho }_s ^{2}) < 1$ indicating a mixed state. For traveling-wave OPA, there are two routes to achieve a $\hat{\rho }_s$ that corresponds to a spectrally separable state with Schmidt coefficient 1: 
\begin{enumerate}
\item $P(\omega ^{\prime }) \rightarrow \delta_{\omega ^{\prime }, \omega_\mathrm{filter}}$, \textit{i.e.} the idler detection bandwidth is restricted to a single spectral component. This spectral filtering method is the most common approach, but suffers from two drawbacks: the resulting reduction in count rate due to filtering of the generated bandwidth, and an inability to resolve the uncertainty in the relative temporal positions of the heralded photons that were produced in a mixed state~\cite{Chen2017}. 
\item The biphoton joint spectrum is factorable, \textit{i.e.} $f(\omega _{s,m}, \omega_{i,n}) = f_{s}(\omega_{s,m}) f_{i}(\omega _{i,m})$~\cite{URen04}. This approach relies on engineering the phase-matching function to achieve separability without filtering.
\end{enumerate}
We focus on the latter approach here, which eliminates both of the problems associated with filtering. For the remainder of this treatment, we approximate both the pump envelope and the phase-matching function to have Gaussian envelopes, and include both temporal walk-off and group velocity dispersion in the phase-mismatch. With these assumptions it can be shown that the following two conditions on the group velocity mismatch and on the pre-chirp of the pump ($\phi_p''$) guarantee a factorable state~\cite{URen2005GenerationOP}:
\begin{eqnarray}
    \Delta k^{\prime }_{p-s}/\Delta k^{\prime }_{p-i} < 0,\label{eqn:fact_cond_01}\\
    2\phi''_{p} + k_{p}''L/2 = 0\label{eqn:fact_cond_02},
\end{eqnarray}
We note here that the second condition (\ref{eqn:fact_cond_02}) is readily satisfied for most realistic $k_{p}''$ using standard pulse-shaping techniques, and therefore focus on satisfying the first condition (\ref{eqn:fact_cond_01}) to achieve a factorable state. 

When the first condition (\ref{eqn:fact_cond_01}) is satisfied, \textit{i.e.} the group velocity of pump lies between that of signal and idler, both the joint spectral amplitude and the joint spectral intensity (JSI; $S(\omega_{s,m}, \omega_{i,n})=|f(\omega_{s,m}, \omega_{i,n})|^2$) become factorable. We use the JSI as a proxy for the JSA since the JSI can be easily visualized by plotting the product of the squared moduli of the pump envelope and the phase-matching function

\begin{equation}
    S(\omega_{s,m}, \omega_{i,n}) = |A_p(\omega_{s,m} + \omega_{i,n})|^{2} |\Phi(\omega_{s,m}, \omega_{i,n})|^{2}\label{eqn:JSI}
\end{equation}
The factorability of the JSI can be determined by the JSI purity evaluated by taking its eigenvalue decomposition and calculating the normalized sum of the squared eigenvalues. The heralded state purity is given by the equivalent formulation in terms of Schmidt coefficients of the biphoton state~\cite{BenDixon2013}. Having noted the conditions for the JSA factorability, we focus on modeling the JSI as it clearly illustrates the dispersion engineering aspects of this problem. 


\subsection{Example designs}

\begin{figure}[ht]
  \centering
  \includegraphics[width=\columnwidth]{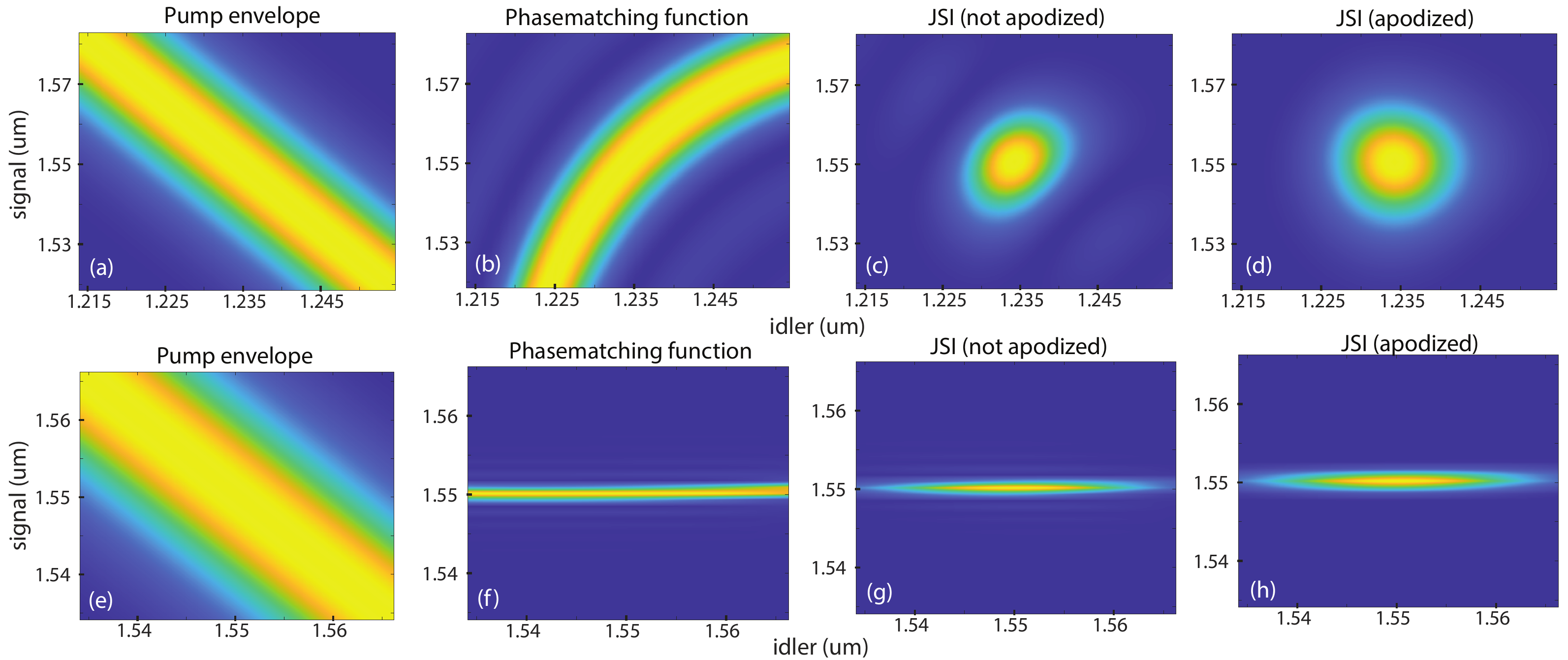}
\caption{\label{fig:JSI}a) The pump envelope, $|A_p(\omega_{s,m}+\omega_{i,n})|^2$, b) phase-matching function, $|\Phi(\omega_{i,n},\omega_{s,m})|^2$, and c) the resulting JSI for the case of symmetric group velocity matching. d) The resulting JSI for a  $\Phi(\omega_{i,n},\omega_{s,m})$ that has been apodized to have a Gaussian dependence on frequency. e)-h) The same as a)-d), for the case of asymmetric group velocity matching.}
\end{figure}

In this section we consider design examples for high purity heralded single photon generation in TFLN waveguides on silica, using the waveguide geometry for dispersion engineering. We consider two waveguide geometries that achieve separability using rather different dispersion relations: symmetric temporal walk-off ($\Delta k^{\prime }_{p-s}/\Delta k^{\prime }_{p-i} \approx -1$), where the group velocity of the pump is midway between the group velocities of the signal and idler, and asymmetric temporal walk-off ($\Delta k^{\prime }_{p-s}/\Delta k^{\prime }_{p-i} \rightarrow 0_{-}$ or $\ll -1$) where the phase-mismatch is dominated by either pump-signal or pump-idler walk-off. In both examples, we plot the power spectrum of the Gaussian pump envelope $|A_p(\omega_{s,m} + \omega_{i,n})|^{2}$ (Figure \ref{fig:JSI} (a,e)) followed by the modulus square of the phase-matching function $|\Phi(\omega_{s,m}, \omega_{i,n})|^{2}$ (Figure \ref{fig:JSI} (b,f)). We initially consider phase-matching functions generated by a uniform grating (constant $\Delta k$), which exhibit a sinc$^2$ spectrum, and compute the JSI using \eref{eqn:JSI} (Figure \ref{fig:JSI} (c,g)). Then, we repeat this calculation using a Gaussian phase-matching function. In the latter case, the shape of the phase-matching function is matched to the shape of the pump spectrum by apodizing the nonlinear interaction~\cite{Fejer1992, Chen2017, BenDixon2013, Phillips2013, Braczyk2010}. 

Gaussian phase-matching functions may be achieved in two ways: changing the poling period along the nonlinear region~\cite{Braczyk2010}, or keeping the poling period constant but varying the strength of the nonlinear coupling along the waveguide $\kappa \rightarrow \kappa(z)$. The latter can be accomplished by either spatially varying the duty cycle directly~\cite{Chen2017}, or selectively deleting some domains in order to reduce the effective duty cycle of the grating when averaged over many periods~\cite{Phillips2013, Huang2006}. For apodized gratings with $\kappa(z) = \kappa \exp{(-(z-L/2)^2/L_\mathrm{apod}^2)}$ the phase-matching functions $\Phi(\omega_{s,m},\omega_{i,n}) \propto \int_0^L \kappa(z)\exp(-i\Delta k(\omega_{s,m},\omega_{i,n})z)dz$ can be calculated by assuming that $L_\mathrm{apod}$ is sufficiently large to take the limits of integration to $\pm \infty$. The transfer function is given by $\Phi(\omega_{s,m},\omega_{i,n}) \propto \exp{(-\Delta k(\omega_{s,m},\omega_{i,n})^2L_\mathrm{apod}^2/4)}$. We may find the $L_\mathrm{apod}$ needed to achieve separability by  Taylor series expanding $\Delta k$ and retaining terms up to first order in detuning around the signal and idler ($\Omega_{s,m}$ and $\Omega_{i,n}$ respectively, where $\Omega_{s,m}=2\pi m/T$), we find $\Phi(\Omega_{s,m},\Omega_{i,n}) \propto \exp{(-L_\mathrm{apod}^2(\Delta k_{p-s}^{\prime}\Omega_{s,m} + \Delta k_{p-i}^{\prime}\Omega_{i,n})^2/4)}$. For a Gaussian pump envelope,  $A_p(\omega_{s,m}+\omega_{i,n}) \propto \exp(-\pi^2 \tau^2(m+n)^2/T^2)$, separability occurs when $\Delta k'_{p-s}\Delta k'_{p-i} L_\mathrm{apod}^2 = -\tau^2$.

The pair generation rate of a given waveguide design is readily calculated using $\sum_m \langle 0|\hat{a}_{s,m}^\dagger \hat{a}_{s,m}|0\rangle/T$, where
\begin{equation}
\hat{a}_{s,m} = -i A_\mathrm{pk} \sum_n c_{m+n}\hat{a}_{i,n}^\dagger \int_0^L \kappa(z)\exp\left(-i\Delta k(\omega_{s,m},\omega_{i,n}) z\right) dz.\label{eqn:PGR}
\end{equation}
Here $A_\mathrm{pk}$ is the peak pump amplitude and the pump envelope is given by $c_{\ell}=\frac{\tau}{T}\sqrt{\pi}\exp (-\ell^2 \pi^2\tau^2/T^2)$. Intuitive analytical expressions for the pair generation rate can be found for several of the cases considered here. In the case of asymmetric walk-off ($\Delta k_{p-s}'\gg\Delta k'_{p-i}$) we find
\begin{equation}
    \frac{\mathrm{pairs}}{\mathrm{s}}=\sqrt{\frac{\pi}{2}}\frac{|\kappa A_\mathrm{pk}L|^2}{T}\frac{\tau}{\Delta k'_{p-s} L },
\end{equation}
where $\tau$ is the transform-limited pulse duration of the Gaussian pump input to the waveguide. We note here that $\Delta k'_{p-s}$ must be large to ensure separability. This large walk-off effectively filters the pair generation rate, thereby increasing the power requirements of these devices. For apodized gratings with either dispersion relation, the pair generation rate is given by
\begin{equation}
    \frac{\mathrm{pairs}}{\mathrm{s}}=\frac{\pi|\kappa A_\mathrm{pk}L_\mathrm{apod}|^2(\Delta k')^2}{2T\sqrt{(\Delta k')^2-(\Delta k'_{p-i})^2}\sqrt{(\Delta k')^2-\Delta (k'_{p-s})^2}},
\end{equation}
where $(\Delta k')^2 = \Delta k'_{p-s}\Delta k'_{p-i}$.

With the conditions for separability (\ref{eqn:fact_cond_01}) and the design rules for apodized gratings given above, we now consider two example designs. The pump envelope, phase-matching function, and resulting JSI for the case of symmetric temporal walk-off ($\Delta k^{\prime }_{p-s}/\Delta k^{\prime }_{p-i} \approx -1$) is shown in \fref{fig:JSI}((a) - (d)). The waveguide geometry is given by a 525 nm etch depth, a 500 nm top-width, a 550-nm thin film, and a side-wall angle of 23 degrees. Phase-matching occurs for a poling period of 2.34 $\mu$m. When a 1-cm-long waveguide is pumped with pulses centered around 687-nm  (3.3-THz bandwidth) in the fundamental TE$_{00}$ mode, signal and idler photons are produced in the TE$_{00}$ mode at 1234 nm and at 1550 nm, respectively. The same parameters are shown for the case of asymmetric temporal walk-off ($\Delta k^{\prime }_{p-s}/\Delta k^{\prime }_{p-i} \rightarrow 0_{-}$), in \fref{fig:JSI}((e) - (f)). In this case, we consider an etch depth of 400 nm, a top-width of 1200 nm, and a film thickness of 900 nm, with a corresponding poling period of 3.85 $\mu$m. When a 1-cm-long waveguide is pumped with 775-nm pulses (2.5-THz bandwidth) in the fundamental TM mode, signal and idler photons are generated at 1550 nm in the waveguide's fundamental TE and TM modes respectively. Without apodization only the design with strongly asymmetric walk-off exhibits high ($>95\%$) purity. For intermediate values of $\Delta k^{\prime }_{p-s}/\Delta k^{\prime }_{p-i}$ around $-1$ the JSI will typically have a purity $\approx$ 70-90$\%$ due to the side-lobes of the sinc$^2$ transfer function seen in \fref{fig:JSI}((b) - (c)). Once  apodized with a Gaussian phase-matching function, the symmetric case  is estimated to have a JSI purity $>95\%$ (Figure \ref{fig:JSI} (d)), while the purity of the asymmetric case is further improved to $>98\%$ (Figure \ref{fig:JSI} (h)).

Both of these devices have exceptionally low energy requirements. An estimated pulse energy of 300 fJ is required for generation of 0.1 photons per pulse for asymmetric walk-off, and an estimated pulse energy of 1 fJ is required for generation of 0.1 photons per pulse for the symmetric walk-off design. In principle, driving this device with a 10 GHz source would produce photon pairs at a rate of 1 GHz with 10 $\mu$W of average power. We note here that at this time considerable effort is focused on heralded SPDC using micro-ring resonators~\cite{Raymer2005,Ma2017,Lu2019NIST,Luo2015,Guo2016,Ma2020,Steiner2021}. While the spectral purity in these devices is expected to be high owing to the very narrow bandwidth of the resonant modes, possible issues regarding low heralding efficiencies remain to be addressed~\cite{Vernon2016}. At this time, state-of-the-art integrated photonic devices based on microresonators achieve pair generation rates of 30 MHz with 10 $\mu$W of pump power~\cite{Ma2020,Steiner2021}, which suggests that the pulsed travelling-wave devices considered here may be a promising alternative. We further note that other recent demonstrations in the TFLN platform include densely integrated and re-configurable linear optical circuit elements ~\cite{Desiatov2019,Wang2019}, highly efficient electro-optic modulators~\cite{Wang2018EOM, Zhang2019EOM}, and integrated single photon detectors~\cite{Sayem2020}. Thus, alongside source engineering, this platform has many of the components needed to implement an on-chip linear optical quantum computer~\cite{Thompson2011, VanMeter2016}.

\section{Routes toward single-photon nonlinear devices}\label{sec:SPND}

An outstanding challenge in the field of nonlinear optics has been the realization of devices that achieve efficient nonlinear interactions at the single-photon level by embedding a highly nonlinear medium in a low-loss resonator~\cite{Mabuchi2012}. These nonlinear resonators may exhibit effects similar to strongly coupled cavity or circuit QED systems, namely, vacuum Rabi splitting~\cite{Mabuchi2012,Agarwal1994}, photon blockading~\cite{Majumdar2013}, and the formation of Schr\"{o}dinger cat states~\cite{Reid1993,onodera2019nonlinear}. The multi-mode behavior of these systems also enables new operating regimes, such as deterministic parametric downconversion~\cite{yanagimoto2020broadband}. Early demonstrations in PPLN microresonators showed normalized efficiencies of 250,000\%/W~\cite{Chen2019,Lu2019,mckenna2021ultralowpower}, and achieved saturation with 100's of $\mu$W of optical power. Remarkably, recent work has begun to approach single-photon nonlinearities using CW-pumped microresonators~\cite{Lu2020}. In this section, we first review this recent work, and we then discuss microresonators synchronously pumped by short pulses. The approach taken here is based on the quasi-static heuristics presented in \sref{sec:quasi-static}. In this approach, we first consider the classical behavior of CW-pumped resonators, and establish simple intuitive relationships between coupling rate $g_\mathrm{cw}$ and the conditions for saturation. Then, we consider the conditions for saturation in synchronously-pumped resonators to gain insights about the effective enhancement of the coupling rate due to multimode operation. These quasi-static heuristics suggest that synchronously-pumped nonlinear resonators can achieve saturated nonlinear interactions with attojoules or even zeptojoules of pulse energy, and therefore present a promising route towards singe-photon nonlinear interactions. We note here that the development of full quantum models of such systems is a topic of ongoing research~\cite{onodera2019nonlinear,yanagimoto2020broadband}, and briefly summarize these results at the end of~\sref{sec:SPOPOs}. We close this section by comparing a number of promising material systems.


\subsection{Continuous-wave interactions in $\chi^{(2)}$ microresonators}

Throughout this section we will consider two kinds of resonators: doubly-resonant (DRO), where the long-wavelengths ($\omega_1$ and $\omega_2$) are resonant in the cavity, and triply-resonant (TRO), where all three interacting waves are resonant. We retain this naming convention even for degenerate operation ($\omega_1=\omega_2$). Both of these configurations are assumed to be ring resonators with nonlinear interactions occurring in one continuous section of the resonator. The interaction Hamiltonian is given by
\begin{equation}
    \hat{H}_\mathrm{int}/\hbar = g_\mathrm{cw}(\hat{a}_{2\omega}\hat{a}_{\omega}^{\dagger }\hat{a}_{\omega}^{\dagger} + h.c.) 
    \label{eqn:HSHG},
\end{equation}
where the coupling rate $g_\mathrm{cw}$ is given by
\begin{eqnarray}
g_\mathrm{cw,TRO}=\sqrt{\hbar\omega\eta_0 L_\mathrm{QPM}^2\Delta f_\mathrm{FSR}^3/2},\\
g_\mathrm{cw,DRO}=\hbar\omega\eta_0 L_\mathrm{QPM}^2\Delta f_\mathrm{FSR}^2/2,
\end{eqnarray}
for a TRO and a DRO~\cite{onodera2019nonlinear}, respectively, where $L_\mathrm{QPM}$ is the length of the QPM grating and $\Delta f_\mathrm{FSR}=v_g/L$ is the free spectral range (FSR) of the resonator. The typical figure of merit for a nonlinear resonator is $g_\mathrm{cw}/\kappa_{\ell,\omega}$, where $\kappa_{\ell,\omega}=\ell_\omega \Delta f_\mathrm{FSR}$ is the loss rate of the cavity and $2\ell_\omega=T_\omega+A_\omega$, with outcoupling $T_\omega$ and dissipative loss $A_\omega$, is the power loss per round trip. In many cases, we will assume that the cavity is critically coupled ($T_\omega=A_\omega$) to achieve the largest possible field enhancement.

\begin{figure}[t]
    \centering
    \includegraphics[width=\columnwidth]{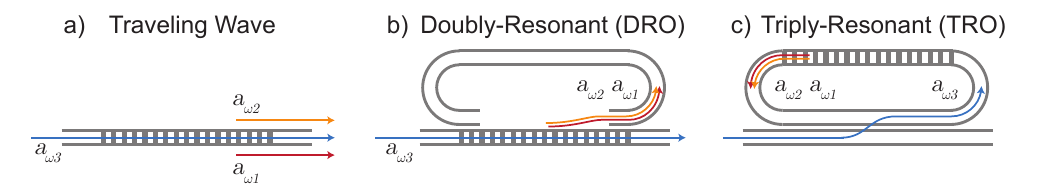}
    \caption{Example geometries for a) traveling-wave OPA, b) doubly-resonant OPO (DRO), with a resonant signal ($\omega_2$) and idler ($\omega_1$), and c) triply-resonant OPO (TRO) with a resonant pump ($\omega_3$), signal, and idler. We retain this naming convention when considering degenerate operation ($\omega_1 = \omega_2$) and SHG.}
    \label{fig:resonators}
\end{figure}

We emphasize here that the coupling rate of DROs and TROs exhibit different scaling with respect to the physical size of a resonator. In DROs the coupling rate is invariant with respect to a simultaneous rescaling of $L_\mathrm{QPM}$ and $\Delta f_\mathrm{FSR}$. When the dominant loss mechanism of a resonator is propagation loss, such that $\ell \propto \alpha L$, the loss rate $\kappa_\ell$ (and therefore $g_\mathrm{DRO}/\kappa_{\ell,\omega}$) is invariant with respect to the size of the resonator. In contrast, when a TRO is uniformly poled ($L_\mathrm{QPM}=L$) the coupling rate (and therefore $g_\mathrm{TRO}/\kappa_{\ell,\omega}$) can be made arbitrarily large by rescaling the dimensions of the resonator. In principle, nonlinear resonators can operate at scales approaching an optical wavelength~\cite{Jahani2021} if care is taken to mitigate thermorefractive noise~\cite{Panuski2020}. In many cases, such as in X-cut thin films, resonators based on quasi-phasematching may contain unpoled regions of fixed size where no nonlinear interactions take place, $L_\mathrm{QPM}+L_\mathrm{lin}=L$, and the coupling rate exhibits a local maximum with respect to $L_\mathrm{QPM}$ due to the competition between the nonlinear length and the cavity FSR. These scaling laws will change when we consider pulsed interactions in resonators.

Classically, there are two cases of interest: resonant SHG, where a resonator pumped at $\omega$ generates a second harmonic at $2\omega$, and optical parametric oscillation, where a resonator pumped at $2\omega$ generates a fundamental at $\omega$. In the case of triply-resonant SHG with an undepleted pump, the conversion efficiency is given by
\begin{equation}
\left(\mathrm{P_{2\omega,out}}/\mathrm{P_{\omega,in}}\right)_\mathrm{TRO}=\frac{\eta_0 L^2 \mathrm{P_{\omega,in}} T_\omega^2 T_{2\omega}}{\ell_\omega^4\ell_{2\omega}^2},
\end{equation}
or for a critically coupled resonator, $\left(\mathrm{P_{2\omega,out}}/\mathrm{P_{\omega,in}}\right)_\mathrm{TRO} = \eta_0 L^2 \mathrm{P_{\omega,in}}/\ell_{\omega}^2\ell_{2\omega}$. We may therefore determine $g_\mathrm{cw}$ by measuring $\mathrm{P_{2\omega,out}}/\mathrm{P_{\omega,in}}$ as a function of $\mathrm{P_{\omega,in}}$, e.g. $\left(\mathrm{P_{2\omega,out}}/\mathrm{P_{\omega,in}}\right)_\mathrm{TRO} = 2g_\mathrm{cw}^2 \mathrm{P_{\omega,in}}/(\hbar\omega\kappa_{\ell,\omega}^2\kappa_{\ell,2\omega})$ for a critically-coupled resonator. Similar relationships can be found for doubly-resonant SHG,
\begin{equation}
\left(\mathrm{P_{2\omega,out}}/\mathrm{P_{\omega,in}}\right)_\mathrm{DRO}=\frac{\eta_0 L^2 \mathrm{P_{\omega,in}} T_\omega^2}{\ell_{\omega}^4},
\end{equation}
or $\left(\mathrm{P_{2\omega,out}}/\mathrm{P_{\omega,in}}\right)_\mathrm{DRO} = 2g_\mathrm{cw} \mathrm{P_{\omega,in}}/(\hbar\omega\kappa_{\ell,\omega}^2)$ for a critically-coupled resonator. In either case, we see that as $g_\mathrm{cw}/\kappa_{\ell,\omega}$ approaches unity saturation occurs when the mean photon number in the cavity is one-half, $\mathrm{P_{\omega,in}}/(\hbar\omega\kappa_{\ell,\omega})=1/2$. Corrections to this undepleted theory to account for saturation can be obtained using Picard iteration~\cite{onodera2019nonlinear}.

Similar behavior occurs in OPOs as the coupling rate approaches the loss rate. In the case of a DRO, the power generated at the fundamental is given by
\begin{equation}
\left(\mathrm{P_{out,\omega}}\right)_\mathrm{DRO}=\frac{2T_\omega \mathrm{P_{th,DRO}}}{\ell_\omega}\left(\sqrt{\frac{\mathrm{P_{in,2\omega}}}{\mathrm{P_{th,DRO}}}}-1\right),\label{eqn:PDRO}
\end{equation}
where $\mathrm{P_{th,DRO}} = \ell_\omega^2/(\eta_0 L^2)$, and full pump depletion occurs when the resonator is driven by $\mathrm{P_{sat,2\omega}}=4\mathrm{P_{th,DRO}}$. The threshold power may be expressed in terms of the coupling rate as $\mathrm{P_{th,DRO}} = \hbar\omega\kappa_\omega^2/(2g_\mathrm{cw})$, and therefore the condition $g_\mathrm{cw}/\kappa_\omega=1$ corresponds to pump depletion occurring with a single pump photon (or two signal photons) present in the cavity  $\mathrm{P_{sat,2\omega}}/(2\hbar\omega \kappa_\omega)=1$. Triply resonant OPOs exhibit nearly identical behavior, 
\begin{equation}
\left(\mathrm{P_{out,\omega}}\right)_\mathrm{TRO}=\frac{T_{2\omega}T_\omega \mathrm{P_{th,TRO}}}{\ell_\omega\ell_{2\omega}}\left(\sqrt{\frac{\mathrm{P_{in,2\omega}}}{\mathrm{P_{th,TRO}}}}-1\right),\label{eqn:PTRO}
\end{equation}
with a threshold reduced by the field enhancement of the resonant second harmonic, $\mathrm{P_{th,TRO}}=\ell_\omega^2\ell_{2\omega}^2/(\eta_0 L^2 T_{2\omega})$.

\begin{figure}
    \centering
    \includegraphics[width=\columnwidth]{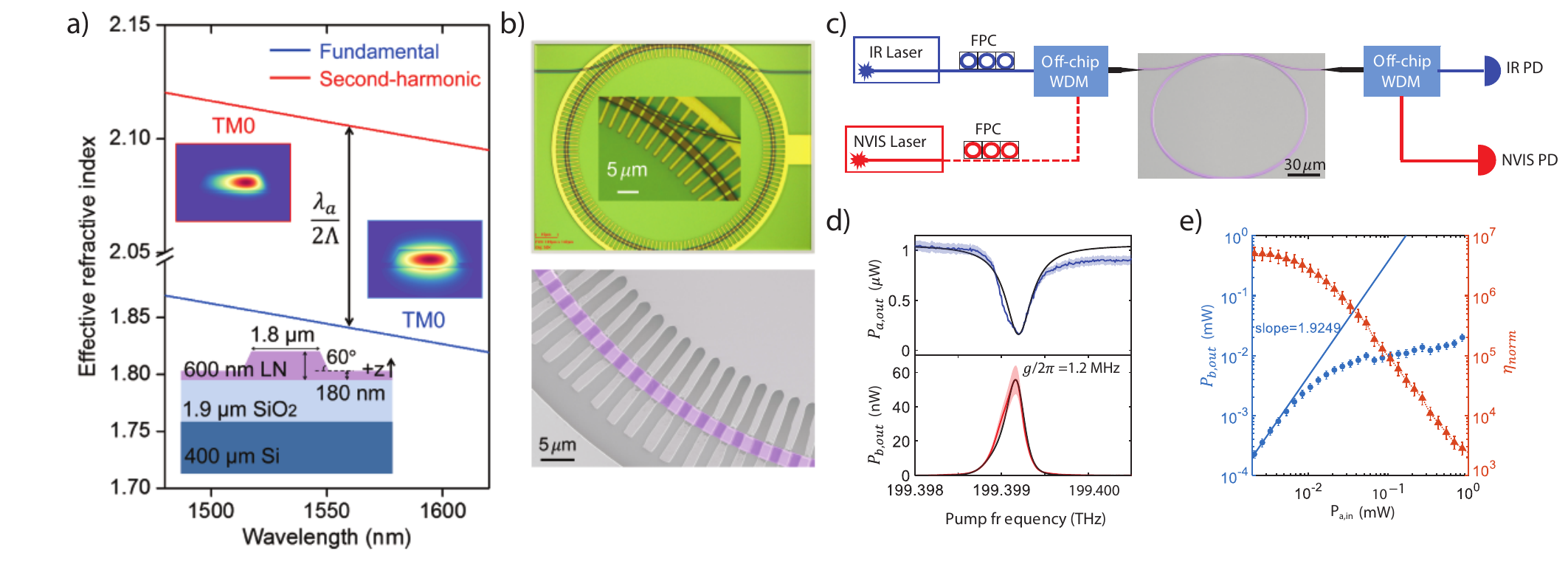}
    \caption{a) Schematic of the waveguide cross-section with associated dispersion relations for the fundamental and second harmonic. Inset: Simulated TM$_{00}$ modes at each wavelength. b) (Top) Optical microscope image of etched ring with radial poling electrode. (Bottom) False-color SEM image of the waveguide after etching in hydrofluoric acid to visualize the poled domains. c) Experimental setup. An 800-nm (NVIS) and 1560-nm laser are combined using wavelength division multiplexing (WDM) to separately measure the resonator Q. d) Depleted fundamental (blue) and generated second harmonic (red) as a function of pump frequency. e) Measured conversion efficiency as a function of pump power. A normalized efficiency of 5,000,000\%/W is extracted by a linear fit to $\mathrm{P_{out}}/\mathrm{P_{in}}$ in the undepleted limit. Figure adapted with permission Lu \textit{et al}., Optica \textbf{7}, 1654 (2020). Copyright 2020 Authors, licensed under the terms of the OSA Open Access Publishing Agreement.}
    \label{fig:SPNL}
\end{figure}

To date, the largest $g_\mathrm{TRO}/\kappa_{\ell,\omega}$ has been demonstrated using triply-resonant SHG in periodically-poled Z-cut lithium niobate thin films. In this case, Z-cut poling enabled QPM of TM$_{00}$ modes throughout the resonator (Figure \ref{fig:SPNL}(a-b))~\cite{Lu2019,Lu2020}, which allowed the authors to take advantage of the favorable scaling of $g_\mathrm{TRO}$ found in fully-poled resonators. These resonators were characterized in three steps. First, the authors measured the quality factor of the resonator at both the fundamental (1560 nm) and second harmonic (780 nm) to determine the loss rates, $\kappa_{\ell,\omega}$ and $\kappa_{\ell,2\omega}$ (Figure \ref{fig:SPNL}(c)). Then, the resonator was temperature tuned such that both the fundamental and second harmonic are resonant in the cavity (Figure \ref{fig:SPNL}(d)). The generated second harmonic power was measured as a function of input power at the fundamental to determine the normalized efficiency, $\left(\mathrm{P_{2\omega,out}}/\mathrm{P_{\omega,in}^2}\right)$, which was then used to determine the coupling rate (Figure \ref{fig:SPNL}(e)). The authors reported loss rates as low as $\kappa_{\ell,\omega}/2\pi = 184.6$ MHz at the fundamental, and a coupling rate as large as $g_\mathrm{cw}/2\pi=1.2$ MHz, corresponding to $g_\mathrm{cw}/\kappa_{\ell,\omega}\sim 0.007$. We note here that the propagation loss of the fundamental, $\alpha_\omega\sim 23$ dB/m, is nearly an order of magnitude larger than the state-of-the-art. Therefore, further reductions in the propagation loss may yield devices with $g_\mathrm{cw}/\kappa_{\ell,\omega}\sim0.1$. Similarly, the coupling rate of these resonators may be further increased by reducing the radius of curvature. The coupling rate of such resonators is ultimately limited by trade-offs between the FSR, bending loss, and bend-induced reductions of the mode overlap.

\subsection{\label{sec:SPOPOs}Pulsed interactions in $\chi^{(2)}$ microresonators}

Having established the relationship between the coupling rate $g_\mathrm{cw}$ and the conditions for saturation in CW-pumped resonators, we now consider a simplified model for saturation in synchronously-pumped resonators as an estimate for an effective enhancement of the coupling rate in these systems. At this time, reduced quantum models for these highly multimode systems are a topic of ongoing research, and we briefly compare the simple insights found here with a number of recent results in this field at the end of this section.

We consider quasi-static nonlinear resonators where the group velocity dispersion of the fundamental and the temporal walk-off may be eliminated using designs similar to those discussed in \sref{sec:exampleSHG}. Therefore, we may model the behavior of nonlinear oscillators driven by short pulses using the heuristics developed in \sref{sec:quasi-static}. In this limit, we may solve the coupled-wave equations using the CW equations of motion for each time slice of the pulses. For an OPO, the power generated at the fundamental is given by
\begin{equation}
\mathrm{P_{\omega,out}}(t)\propto \mathrm{P_{th}}\left(\sqrt{\frac{\mathrm{P_{in,2\omega}}(t)}{\mathrm{P_{th}}}}-1\right),\label{eqn:qs_opo}
\end{equation}
for $t$ that satisfy $\mathrm{P_{in,2\omega}}(t)>\mathrm{P_{th}}$. We have neglected the pre-factors of (\ref{eqn:PDRO}-\ref{eqn:PTRO}) in \ref{eqn:qs_opo} since the behavior of these devices in saturation is largely determined by $\mathrm{P_{in,2\omega}}(t)/\mathrm{P_{th}}$.

For quasi-static devices, threshold and saturation are determined by the peak power of the pump pulses used to drive the resonator, $\mathrm{P_{in,2\omega}}(0) = U/2\tau$ for a sech$^2$ pulse, where $U=\mathrm{P_{av}}/\Delta f_\mathrm{FSR}$ is the pulse energy. The intracavity photon number needed to achieve threshold or saturation is therefore reduced by a factor of $(2\tau\Delta f_\mathrm{FSR})$, which suggests that the coupling rate can be substantially enhanced by using short pulses,
\begin{eqnarray}
g_\mathrm{pulse,DRO}\sim g_\mathrm{cw,DRO}/(2\tau\Delta f_\mathrm{FSR}),\label{eqn:gDRO}\\
g_\mathrm{pulse,TRO}^2\sim g_\mathrm{cw,TRO}^2/(2\tau\Delta f_\mathrm{FSR}).\label{eqn:gTRO}
\end{eqnarray}
As an example, we first consider a realistic case using the designs realized in \sref{sec:pulsednlo}, where $\Delta f_\mathrm{FSR}=10$ GHz, $\eta_0=1000\%$/W-cm$^2$, $L=7$ mm, $\tau=28$ fs (50 fs 3-dB), and we assume $\ell =$ 10\% loss per round trip ($Q\sim 10^6$). The doubly-resonant OPO is pumped at 1030-nm. In this case, threshold occurs for a pump pulse energy of 32 aJ, or $\sim$170 photons. Similar quasi-static designs can be found for fundamental wavelengths around 1560 nm with $\eta_0 = 5000\%$/W-cm$^2$. Assuming best-case numbers ($\eta_0=5000\%$/W-cm$^2$, and $\ell =$ 1\% loss per round trip), threshold occurs at 59 zeptojoules, or 0.23 pump photons. Further reductions to the energy requirements of these systems may be achieved with a resonant second harmonic, which would again reduce the threshold of oscillation by a factor of $\sim T_{2\omega}$. The thresholds for the two cases considered here would be reduced 10-fold and 100-fold, respectively. We emphasize here that (\ref{eqn:gDRO}-\ref{eqn:gTRO}) suggest that synchronously pumped resonators exhibit rather different scaling laws than their CW-pumped counterparts; synchronously pumped TROs exhibit coupling rates that are invariant with respect to rescaling of the dimensions of the resonator, $L$ and $\Delta f_\mathrm{FSR}$, and the coupling rate of synchronously pumped DROs can be made arbitrarily large by increasing the size of the resonator.

We close this section by noting that the non-classical behavior of such highly-nonlinear pulsed interactions must be understood using multimode quantum models. These models are generally intractable due to the large size of the Hilbert space, and the development of reduced models is the subject of ongoing work. A promising first step was taken in~\cite{onodera2019nonlinear}, where the authors considered the role of intracavity dispersion on the behavior of a synchronously pumped DRO. Remarkably, when $\Delta k'=0$ and $k_\omega''=k_{2\omega}''$ the authors found that the system could be described using a single pulsed supermode with an effective coupling enhanced by $1/(\Delta f_\mathrm{FSR}\tau)$, consistent with (\ref{eqn:gDRO}). Similarly, in~\cite{yanagimoto2020broadband} the authors studied parametric downconversion of a single photon input to a traveling-wave OPA and found an effective coupling rate between a single-mode pump and a continuum of signal and idler modes that is enhanced by the available OPA bandwidth. Both of these results confirm that dispersion engineering is useful tool for developing few-photon nonlinear devices, and the emergence of design rules that reduce these systems to effectively single mode behavior with large enhancements of the nonlinear coupling represents a promising direction for the field. At this time there is no reduced model for triply-resonant oscillators or traveling-wave devices driven by short pulses equivalent to the model for DROs found in~\cite{onodera2019nonlinear}.

\subsection{Choice of nonlinear materials}\label{sec:materials}

We close this section by comparing the relative nonlinearity of a number of emerging platforms for nonlinear photonics. At this time, only a few of these materials have demonstrated quasi-phasematched interactions in tightly-confining waveguides, however many of these materials have been used to realize either QPM in bulk media or tightly-confining photonics in direct-etched thin films separately. For classical devices the figure of merit is $\eta_0$, which determines the power required to achieve efficient nonlinear interactions. For non-classical devices we use the coupling rate as a point of comparison, which scales as $g_\mathrm{cw,DRO}\sim \omega\eta_0$ for DROs and $g_\mathrm{cw,TRO}\sim \sqrt{\omega\eta_0}$ for TROs. We note here that $\eta_0$ exhibits a quartic ($\omega^4$) scaling with frequency due to the scale invariance of Maxwell's equations, where an explicit $\omega^2$ dependence occurs in the expression for $\kappa^2$, and an implicit $\omega^2$ comes from rescaling all of the relevant waveguide dimensions with wavelength, $A_\mathrm{eff}\propto\omega^{-2}$. In practice, this scaling is slightly faster than $\omega^4$ due to the dispersion of $d_\mathrm{eff}$.

\begin{figure}
    \centering
    \includegraphics[width=\columnwidth]{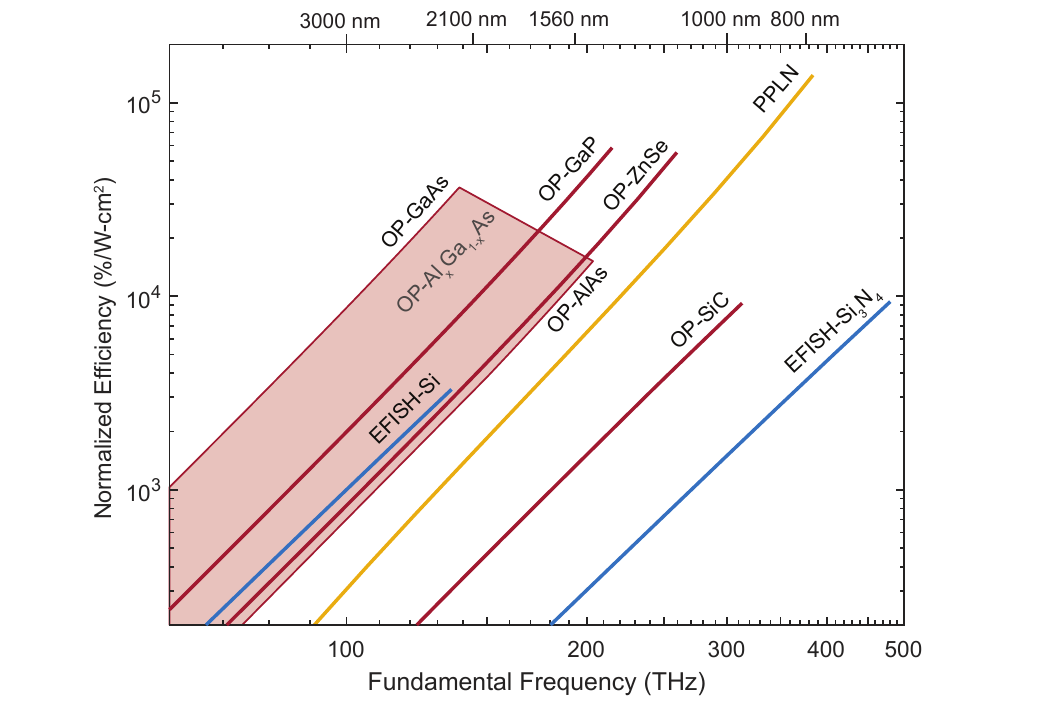}
    \caption{Normalized efficiency as a function of fundamental wavelength for a $(\lambda/n)\times(\lambda/n)$ suspended ridge waveguide. Orange: ferroelectric materials, red: orientation patterned semiconductors, blue: $\chi^{(3)}$ waveguides with electric field induced nonlinearities. Materials with intrinsic $\chi^{(2)}$ nonlinearities are plotted for fundamental frequencies as large as 40\% of the bandgap. The $\eta_0$ associated with materials with induced $\chi^{(2)}$ nonlinearities are plotted for frequencies as large as as 50\% of the bandgap.}
    \label{fig:materials}
\end{figure}

Figure \ref{fig:materials} shows $\eta_0$ for a number of materials of interest as a function of wavelength. Here we have a assumed a suspended square ridge with cross-sectional area $(\lambda/n_\omega)^2$, such that the relative normalized efficiency of a waveguide scales as $\eta_0 \sim \omega^4 d_\mathrm{eff}^2/n_{2\omega}$. All of the materials, with the exception of silicon nitride (Si$_3$N$_4$), are assumed to be periodically poled or orientation patterned with a 50\% duty cycle, and therefore the effective nonlinear coefficient is reduced by a factor of $2/\pi$. Both Si$_3$N$_4$ and silicon rely on electric-field induced nonlinearities, where an effective second order nonlinear susceptibility is given by $\chi^{(2)}_{ZZZ}=3\chi^{(3)}_{ZZZZ}E_\mathrm{DC}$~\cite{Timurdogan2017,Billat2017,Porcel2017}. In the case of Si$_3$N$_4$, the material forms a self-organized nonlinear grating with a Fourier coefficient given by $\chi^{(2)}/2$ rather than $2 \chi^{(2)}/\pi$~\cite{Billat2017,Porcel2017,Hickstein2019}. For orientation-patterned semiconductors with four-fold rotational symmetry, such as GaP, GaAs, AlAs, and ZnSe, we assume that propagation occurs along the $<$011$>$ direction in an X-cut film such that $d_{14}$ can be used to couple between TE modes of the fundamental and TM modes of the second harmonic. Materials with with intrinsic $\chi^{(2)}$ nonlinearities are plotted for fundamental frequencies as large as 40\% of the bandgap, rather than 50\%, to avoid large errors in Miller's delta scaling and linear absorption of the second harmonic. In many cases, two-photon absorption of the second harmonic may limit operation to $0.25 E_g$, rather than $0.4 E_g$.

The maximum $\eta_0$ that may be achieved in a particular material system is determined both by $d_\mathrm{eff}^2/n_{2\omega}$ and the bandgap of the material, with large bandgap materials able to take advantage of the  $\omega^4$ scaling of $\eta_0$. GaAs exhibits the largest $d_\mathrm{eff}$ of any materials shown here, but is constrained to moderate normalized efficiencies by a bandgap of 1.44 eV. Similarly, Si$_3$N$_4$ has the largest bandgap of any material considered here (5 eV), but is constrained to relatively low normalized efficiencies by the induced $\chi^{(2)}$ of $\sim 1$ pm/V. Remarkably large nonlinearities have been demonstrated in Al$_x$Ga$_{1-x}$As recently~\cite{Chang2018,Chiles2019,Stanton2020}. This material system can take advantage of the tunable bandgap associated with ternary alloys; the aluminium concentration can be tuned to optimize the nonlinearity for a fundamental wavelength of 1560 nm. Pure AlAs has a $d_{14}$ roughly four times smaller than GaAs~\cite{Ohashi1993}, and therefore the Al concentration can be decreased to increase $d_{14}$, provided that the bandgap does not become sufficiently small to introduce loss at the second harmonic or two-photon loss at the fundamental. 

SiC, GaP, and ZnSe have all seen rapid development in the past decade. Both SiC and GaP have been developed into low-loss platforms for nonlinear nanophotonics~\cite{Lukin2019,Lukin2020,Wilson2019}, with key results including SHG~\cite{Rivoire2011,Song:19}, OPO~\cite{Guidry:20}, and comb formation~\cite{Wilson2019,Guidry:20}. GaP and ZnSe have both been grown as orientation patterned thin films~\cite{Maidment2016,Schunemann2014,Tassev:19,Vangala:19,Schunemann2019}. Both of these materials appear to be particularly promising in the near-infrared, with OP-GaP having the largest possible normalized efficiency of any material considered here at 1560 nm ($\sim 50,000$\%/W-cm$^2$). However, to date, none of these systems have combined low-loss nanophotonics with quasi-phasematching.

Lithium niobate may access one of the largest normalized efficiencies ($\eta_0\sim 130,000\%/$W-cm$^2$) as the second harmonic approaches the bandgap, which is two orders of magnitude larger than the current state-of-the-art. Therefore, a yet unexplored route towards single-photon nonlinearities would be to use tightly-confining waveguides designed for frequency doubling Ti:Sapphire wavelengths. We note, however, that care must be taken when operating at such short wavelengths. Frequency doubling these wavelengths requires sub-micron poling periods and therefore such devices have extremely stringent fabrication tolerances, as discussed in \ref{sec:tolerance}. Furthermore, while there exist geometries that achieve $k_\omega''=0$, group velocity matching between an 800-nm TE$_{00}$ fundamental and a 400-nm TE$_{00}$ second harmonic cannot be achieved using the simple ridge waveguides considered here. Finally, we note that when short wavelengths are used as a pump, both photorefractive damage and pump-induced absorption may become significant. These effects may be mitigated by using 5\% MgO-doped lithium niobate, which has been used in a number of recent demonstrations~\cite{Wang2018,Jankowski2020,mckenna2021ultralowpower,jankowski2021supercontinuum,jankowski2021efficient, ledezma2021intense}. In congruent lithium niobate microresonators without MgO doping, photorefractive effects may substantially alter the observed dynamics~\cite{He:19}.

\begin{table}
\caption{\label{tabone}The maximum normalized efficiency attainable in each material system. The wavelength, $\lambda_\mathrm{ref}$, corresponds to highest fundamental frequency potted in figure \ref{fig:materials}. These values may be rescaled to other wavelengths using the power law $\omega^x$, where $x$ is determined by the dispersion relations of a given material and the Miller's delta estimate of the scaling of the nonlinear coefficients.}

\begin{indented}
\lineup
\item[]\begin{tabular}{@{}*{6}{l}}
\br                              
Material & $\lambda_\mathrm{ref}$ & $n$ & $d_\mathrm{iJ}^{\rm a}(\lambda_\mathrm{ref})$ & \0\0\0$\eta_{0} $& Scaling\\ 
\ns
& (nm) & & (pm/V) & (\%/W-cm$^2$) & \0\0$\omega^x$\\
\mr
PPLN &\0 780  & 2.17~\cite{Gayer2008} & 28.4~\cite{Shoji1997,Choy1976} &\0138,700 &4.66\cr
OP-GaAs &2,165 & 3.33~\cite{Skauli2003} & 107~\cite{Shoji1997,Skauli2002} &\0 36,500&4.24 \cr 
OP-AlAs &1,472 & 2.90~\cite{Fern1971} & 31.6~\cite{Ohashi1993} &\0 15,200&4.48  \cr 
OP-SiC &\0 957  & 2.63~\cite{Wang2013} & -11.7~\cite{Sato2009} &\0\0 9,200 &4.25   \cr
OP-GaP &1,393  & 3.07~\cite{Shoji1997} & 57.0~\cite{Wei2018} &\0 58,500 &4.28   \cr
OP-ZnSe &1,156  & 2.47~\cite{Tatian1984,Connolly1979} & 36.2~\cite{Shoji1997} &\0 55,300 &4.28   \cr
EFISH-Si &2,214  & 3.44~\cite{Frey2006} & 41~\cite{Timurdogan2017} &\0\0 3,300 &4   \cr
EFISH-Si$_3$N$_4$ &\0 624  & 2.04~\cite{Luke2015} & 3.7~\cite{Porcel2017} &\0\0 9,400 &4   \cr 
\br
\end{tabular}
\item[] $^{\rm a}$ $d_{33}$ for materials with 3-fold or 6-fold rotational symmetry,
\item[] $^{\rm\,\,}$ $d_{14}$ for materials with 4-fold rotational symmetry.
\end{indented}
\end{table}

\section{Summary}\label{sec:conclusion}

Dispersion-engineered nonlinear waveguides with quasi-phasematched $\chi^{(2)}$ interactions are a promising platform for integrated quantum photonic devices. A key feature of this platform is that quasi-phasematching frees up the waveguide geometry as a design parameter, which allows for simultaneous engineering of the group velocities and group velocity dispersion of the interacting waves. Using extra this degree of freedom we are able to engineer the bandwidth of continuous-wave interactions as well as the interaction lengths associated with pulsed interactions. In both of these cases, we have been able to demonstrate designs that achieve at least an order of magnitude better performance than the state of the art, in terms of bandwidth and interaction length.

We then applied these design rules to a number of quantum photonic technologies. First, we considered the design of degenerate optical parametric amplifiers for producing time domain multiplexed cluster states, and demonstrated that in principle this platform enables orders of magnitude larger parametric gain and bandwidth than current state of the art traveling-wave OPAs. Second, we considered sources of heralded photons using SPDC, and demonstrated several designs that could achieve high purity photons at telecomm wavelengths without any filtering. Third, we demonstrated quasi-static nonlinear interactions, where femtosecond pulses can interact over long lengths without accumulating substantial distortions of their pulse envelopes. In the context of synchronously-pumped nonlinear resonators, we showed that quasi-static interactions could enable saturated nonlinear interactions at the single-photon level.

\subsection{Future directions}

We close this review by noting that there is a rather large amount of work left to be done, and summarize some of the possible directions for future work here.

\textit{Platform development---} At this time, quasi-phasematching has only been demonstrated in three platforms for nonlinear nanophotonics: silicon, silicon nitride, and lithium niobate. As a result, the development of devices with dispersion-engineered $\chi^{(2)}$ interactions has been limited to these materials~\cite{Hickstein2019,Singh2020,Jankowski2020}. Similar designs can be realized in thin films of stoichiometric lithium tantalate (SLT), which is commercially available, has a larger bandgap than lithium niobate, and is less susceptible to damage. These films may be poled and etched using recipes similar to those discussed in \sref{sec:fab}. An extremely promising direction for the field is to realize dispersion engineered QPM devices in both AlGaAs and GaP. Both of these materials have been grown as orientation-patterned thin films and have been used to realize low-loss nonlinear photonics. These systems are particularly interesting at longer wavelenghts; both AlGaAs and GaP have broad transparency windows extending into the mid-infrared, and exhibit the largest nonlinearities of any material considered here. Other potential material systems include ZnSe, SiC, AlN, and KTP. ZnSe has been grown as an orientation patterned thin film, but there have been no demonstrations of low loss ZnSe nanophotonics. Similarly, while bulk KTP is regularly periodically poled and used for nonlinear optics, low-loss nanophotonics have not yet been demonstrated in KTP thin films. SiC and AlN have both been used to realize a number of nonlinear photonic devices, but neither of these materials have been grown as orientation patterned thin films. Dispersion engineered nonlinear interactions may become possible in these systems when more complicated geometries are used to achieve both phase-velocity matching and higher-order dispersion engineering.

\textit{Advanced approaches to dispersion engineering---} While the ridge geometries considered here are the simplest approach to dispersion engineering, this limits the number of dispersion orders that can be engineered simultaneously and the range of wavelengths that can realize $\Delta k'=0$. In the case of X-cut lithium niobate, group-velocity-matched designs are limited to wavelengths longer than 1300 nm, which prevents quasi-static devices from taking advantage of the large normalized efficiencies found at short wavelengths. More flexibility may be found by incorporating additional degrees of freedom, such as a side ridge~\cite{Guo2020} or a multilayer cladding. Designs with many degrees of freedom become difficult to engineer using heuristics and parameter sweeps, and new approaches to dispersion engineering such as photonic inverse design~\cite{Vercruysse2020} may resolve these limitations of the current generation of devices.

\textit{Device demonstrations---} Many of the devices proposed here have not yet been realized, and at this time of writing it is likely that a number of technical challenges need to be resolved. Ultrabroadband OPAs have recently been demonstrated at wavelengths around 2-$\mu$m~\cite{jankowski2021efficient,ledezma2021intense}, but there have been no demonstrations in the C-band where low-loss optical fibers and high quantum efficiency detectors are readily available. Similarly, there have not yet been any demonstrations of squeezing using TFLN waveguides, and the ability to detect the squeezing will likely be limited by the relatively low collection efficiencies found in these waveguides. This may be resolved by integrating couplers, modulators, and detectors directly into the waveguides, or by developing extremely low-loss chip-to-fiber interfaces~\cite{He:lowloss19,Hu:21}. In many cases the degree of squeezing attainable in waveguides is limited by pump-induced absorption, which has not yet been characterized in nanophotonic devices with the level of accuracy needed in squeezing experiments. Successful realizations of these devices may enable chip-scale sources of cluster states for measurement-based quantum computation.

There have not yet been any experimental demonstrations of separable biphotons in integrated photonics platforms using designs similar to those proposed here, and it is unclear what technical problems still need to be resolved. One potential hurdle may be inhomogeneities in the waveguide dimensions, such as variations of the film thickness over the length of the nonlinear section, which may degrade the separability of the generated photon pairs by distorting the phase-matching function. In principle, successful realizations of these devices may achieve photon generation rates approaching GHz with 10 $\mu$W of pump power.

Considerable efforts have been focused on realizing single photon nonlinearities, but current implementations are still several orders of magnitude below the nonlinear couplings needed to achieve strong coupling. Possible routes towards such devices include operating with an 800-nm fundamental to take advantage of the large nonlinearities available at short wavelengths, achieving higher quality factors, or by using resonator geometries that enable smaller cavities and larger free spectral ranges~\cite{Jahani2021}. We note, however, that recent work suggests care must be taken to mitigate thermorefractive noise in the latter approach~\cite{Panuski2020}.

\textit{System-level integration---} Integrated systems for quantum optics typically require many linear and nonlinear components such as squeezers, couplers, and modulators to be functioning together with low loss~\cite{Lenzini2018}. At this time, most of the work in nonlinear nanophotonic devices has focused on the demonstration and characterization of new devices, and there have been few realizations of integrated systems with multiple linear and nonlinear components~\cite{Wang2019}. One of the largest obstacles to realizing photonic integrate circuits with the complexity needed for quantum optics in these platforms is the stringent fabrication tolerances of the nonlinear devices. In most cases electro-optic tuning and temperature tuning alone are not sufficient to compensate the phase-mismatch due to small fabrication errors of the waveguide. These problems can be solved by developing new tuning methods or by developing new waveguide geometries that achieve non-critical phasematching, where the gradient of the poling period with respect to one or more waveguide dimensions is zero.

\textit{Reduced models for multimode quantum optics---} The heuristic approach for calculating the coupling rate of synchronously pumped resonators taken here suggests that short pulses can be used to realize much larger coupling rates in synchronously pumped microresonators. In practice, a more accurate calculation of the effective coupling rate between short pulses must rely on reduced quantum models. The coupling rate between two pulsed supermodes in a reduced model will ultimately depend on the model, and care must be taken to ensure few-mode operation. In the case of a synchronously pumped DRO, recent results have found a reduced quantum model that achieves both few-mode operation and enhanced coupling rates comparable to the heuristic approach used here~\cite{onodera2019nonlinear}. Conversely, recent work studying traveling-wave parametric down-conversion and TROs found an enhancement of the coupling rate due to the number of modes available for parametric down-conversion, but no enhancement due to the multi-mode nature of the pump~\cite{yanagimoto2020broadband}. At this time there are likely undiscovered reduced models for nonlinear waveguides and resonators driven by short pulses that will yield clearer insights and new design rules. Solving these problems will open up new routes towards single-photon nonlinear interactions in integrated photonic devices.

\section*{Acknowledgements}
The authors wish to thank NTT Research for their financial and technical support. The results reported in~\cite{Wang2018,Jankowski2020,jankowski2021supercontinuum} were the result of a collaboration between the Fejer group at Stanford University and the Lon\v{c}ar group at Harvard University. For these results, electrode patterning and poling was performed at the Stanford Nanofabrication Facility, the Stanford Nano Shared Facilities (NSF award ECCS-2026822), and the Cell Sciences Imaging Facility (NCRR award S10RR02557401). Patterning and dry etching was performed at the Harvard University Center for Nanoscale Systems (CNS), a member of the National Nanotechnology Coordinated Infrastructure (NNCI) supported by the National Science Foundation. The results reported in~\cite{mckenna2021ultralowpower} were the result of a collaboration between the Fejer and Safavi-Naeini groups at Stanford University. For these results, electrode patterning and poling, as well as patterning and dry etching of waveguides, was performed at the Stanford Nanofabrication Facility, the Stanford Nano Shared Facilities (NSF award ECCS-2026822), and the Cell Sciences Imaging Facility (NCRR award S10RR02557401).

\section*{Funding}
NTT Research Inc. Physics and Informatics Labs (146395); National Science Foundation (NSF) (ECCS-1609688, EFMA-1741651, CCF-1918549); Department of Energy (DoE) (DE-AC02-76SF00515); Army Research Laboratory (ARL) (W911NF-15-2-0060, 48635-Z8401006).

\appendix

\section{Derivation of the coupled-wave equations in nonlinear waveguides}\label{sec:CWEs}

We first briefly summarize the relevant aspects of nonlinear optical waveguide theory for the devices considered throughout this review, such as the definition of waveguide modes and their associated dispersion relations~\cite{Snyder1984}. We then establish a convenient mode normalization to derive the nonlinear coupling between two waveguide modes. This inter-modal coupling, when combined with the linear dispersion relations of each waveguide mode, enables an accurate description of the behavior of nonlinear devices.

\subsection{Waveguide Modes}

\begin{figure}
\centering
\includegraphics[width=\columnwidth]{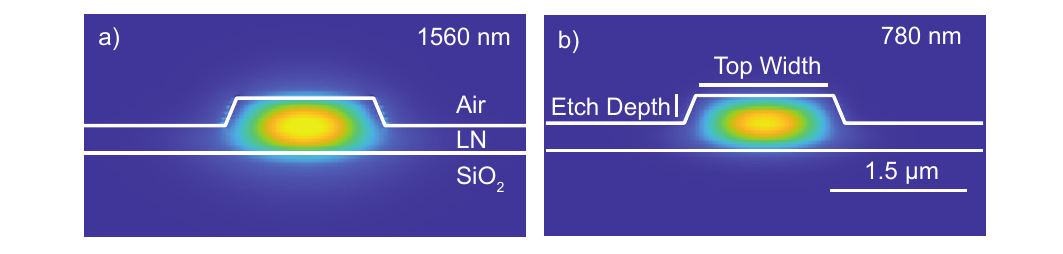}
\caption{\label{fig:wgmodes}Schematic of typical ridge waveguide, with the associated electric field $E_{x,\mu}$ of the waveguide mode for both the fundamental (a) and second harmonic (b). The top cladding is air, the etched thin film is X-cut MgO:LN, and we approximate the 2-$\mu$m-thick silica adhesion layer as extending to infinity.}
\end{figure}

A typical nonlinear waveguide considered here comprises an LN ridge, an air top cladding, and a silica substrate as shown in figure \ref{fig:wgmodes} with the associated $E_x$ field distribution of the TE$_{00}$ mode. Cross sections with more degrees of freedom can be used to better tailor the dispersion of the interacting modes~\cite{Guo2020}, but the simple geometry shown here already allows for engineering of a wide variety of new devices. Waveguide modes arise as the solution to Maxwell's equations in the absence of a nonlinear polarization, with a dielectric constant $\bar{\epsilon}(x,y,\omega)$ that varies in two spatial dimensions,

\begin{eqnarray}
\nabla\cdot [\bar{\epsilon}(x,y,\omega)\mathbf{E}(x,y,z,\omega)] = 0,\\
\nabla\cdot \mathbf{H}(x,y,z,\omega)= 0,\\
\nabla\times \mathbf{H}(x,y,z,\omega)=i\omega\bar{\epsilon}(x,y,\omega)\mathbf{E}(x,y,z,\omega),\\
\nabla\times \mathbf{E}(x,y,z,\omega)=-i\omega\mu_0 \mathbf{H}(x,y,z,\omega).
\end{eqnarray}
The media considered here are uniaxial, with the crystal coordinates aligned to the waveguide coordinates such that $\bar{\epsilon}(x,y,\omega)$ is a second rank diagonal tensor
\begin{equation}
\bar{\epsilon}(x,y,\omega)=\left[
\begin{array}{ccc}
     \epsilon_{xx}(x,y,\omega) & 0 & 0\\
     0 & \epsilon_{yy}(x,y,\omega) & 0\\
     0 & 0 & \epsilon_{zz}(x,y,\omega).
\end{array}{}
\right]
\end{equation}
Throughout this review, we will use capital letters $(X,Y,Z)$ to denote crystal coordinates, and lower-case letters $(x,y,z)$ to denote waveguide coordinates. In typical straight waveguides, the direction of propagation $(z)$ is taken along the crystalline Y-axis of the lithium niobate. There are two cases of interest: X-cut lithium niobate films, which have their crystalline Z-axis aligned with the waveguide x-axis ($\epsilon_{xx} = \epsilon_{ZZ}$, $\epsilon_{yy} = \epsilon_{XX}$, $\epsilon_{zz} = \epsilon_{XX}$) and Z-cut lithium niobate films, which have their crystalline Z-axis aligned with the waveguide y-axis ($\epsilon_{xx} = \epsilon_{XX}$, $\epsilon_{yy} = \epsilon_{ZZ}$, $\epsilon_{zz} = \epsilon_{XX}$). This section will focus predominantly on TE$_{00}$ modes in X-cut films, which exhibit both large nonlinearities and allow for dispersion engineering at many wavelengths of interest.
Since $\bar{\epsilon}(x,y,\omega)$ is translationally invariant in $z$, we may solve Maxwell's equations by expanding the fields in a series of guided modes

\begin{eqnarray}
\mathbf{E}(x,y,z,\omega) &= \sum_\mu a_{\mu}(\omega)\mathbf{E}_{\mu}(x,y,\omega)e^{-ik_\mu(\omega) z},\\    
\mathbf{H}(x,y,z,\omega) &= \sum_\mu a_{\mu}(\omega)\mathbf{H}_{\mu}(x,y,\omega)e^{-ik_\mu(\omega) z},
\end{eqnarray}
where $a_{\mu}$ represents the component of $\mathbf{E}$ contained in mode $\mu$ around frequency $\omega$. The transverse mode profiles, $\mathbf{E}_\mu$ and $\mathbf{H}_\mu$, and their associated propagation constant, $k_\mu$, arise as solutions to an eigenvalue problem, and may be found using the methods described in~\cite{Fallahkhair2008}. The dispersion relations given by $k_\mu(\omega)$ are determined both by the materials that comprise the waveguide and, for tightly confining structures, the geometry of the waveguide. It can be shown that the propagation constant of mode $\mu$ is given by~\cite{Snyder1984}
\begin{equation}
k_\mu(\omega) = \frac{\int_{A_\infty}\mathbf{H}_\mu \cdot \left(\omega\mu_0\right)\mathbf{H}_\mu + \mathbf{E}_\mu\cdot \left(\omega\bar{\epsilon}\right)\mathbf{E}_\mu dA}{2\int_{A_\infty} \left(\mathbf{E}\times \mathbf{H}^*\right)\cdot \hat{z}dA},\label{eqn:disp1}
\end{equation}
where $\int_{A_\infty} f(x,y) dA=\int_{-\infty}^{\infty}\int_{-\infty}^{\infty} f(x,y) dx dy$ denotes the integral of $f(x,y)$ over the waveguide cross section. The inverse group velocity of mode $\mu$ is given by a similar expression
\begin{equation}
k'_\mu(\omega) = \frac{\int_{A_\infty} \mathbf{H}_\mu \cdot \partial_\omega(\mu_0 \omega)\mathbf{H}_\mu^* + \mathbf{E}_\mu\cdot \partial_\omega(\omega\bar{\epsilon})\mathbf{E}_\mu^* dA}{2\int_{A_\infty} \left(\mathbf{E}\times \mathbf{H}^*\right)\cdot \hat{z}dA}.\label{eqn:disp2}
\end{equation}
A casual comparison of \eref{eqn:disp1} and \eref{eqn:disp2} might suggest that n$^{th}$ derivatives of the propagation constant can be found simply by replacing the partial derivatives $\partial_\omega$ in \eref{eqn:disp2} with $\partial_\omega^n$. This is not the case since the inner products in \eref{eqn:disp1} have the form $\mathbf{u}\cdot \bar{L} \mathbf{u}$, whereas the inner products in \eref{eqn:disp2} have the form $\mathbf{u}\cdot \bar{L} \mathbf{u}^*$, where $\bar{L}$ is an arbitrary linear operator.

A recurring theme throughout this review is that the choice of waveguide geometry and the resulting $k_\mu(\omega)$ are crucial in determining the performance of nonlinear devices. This section will focus predominantly on TE$_{00}$ modes in X-cut LN films, which exhibit both large nonlinearities and allow for dispersion engineering at many wavelengths of interest. In this context there are largely two contributions that modify $k_\mu(\omega)$ and $k'_\mu(\omega)$. When the wavelength of a mode is larger than the dimensions of the waveguide, the fields expand into the cladding layers around the waveguide, which modifies the $\mathbf{E}_\mu\cdot (\omega\bar{\epsilon})\mathbf{E}_\mu$ and $\mathbf{E}_\mu\cdot \partial_\omega(\omega\bar{\epsilon})\mathbf{E}_\mu^*$ overlap integrals. This typically reduces the effective index associated with the mode $n_{\mu}=c k_\mu/\omega$ due to the increasing overlap of the fields with the low-index cladding, and often increases the group index $n_{g,\mu} = c k'_\mu$ due to a more rapid variation of the propagation constant with frequency. The second contribution to $k_\mu(\omega)$ is avoided crossings of different spatial modes. In this case, two modes $\mathbf{e}_\mu$ and $\mathbf{e}_\nu$ that share the same parity (e.g. TE$_{00}$ and TM$_{10}$) may approach the same propagation constant at a wavelength $\lambda_{AC}$ due to their different group velocities. However since two modes of the same parity cannot have the same propagation constant, the modes will hybridize and $k_{\mu,\nu}$ will exhibit an anti-crossing around $\lambda_{AC}$. This strong variation of $k_\mu(\omega)$ around $\lambda_{AC}$ can modify dispersion orders hundreds of nanometers away from the anti-crossing.

We conclude our discussion of linear waveguide theory by summarizing a few properties of waveguide modes that will be convenient for deriving the nonlinear coupling between two modes. First, we note that all of these eigenmodes satisfy an orthogonality relation,
\begin{eqnarray}
\int_{A}\frac{1}{2}\mathrm{Re}\left(\left[\mathbf{E}_\mu\times \mathbf{H}_\nu^*\right] \cdot \hat{z}\right)dxdy =\mathrm{P}\delta_{\mu,\nu},\label{eqn:orth}
\end{eqnarray}
where the fields are normalized such that $\mathrm{P}=1$~W, and therefore the power contained in mode $\mu$ is $\mathrm{P}|a_\mu|^2$. Second, we note that it is convenient to express these mode profiles using dimensionless functions $\mathbf{e}(x,y)$ and $\mathbf{h}(x,y)$
\begin{eqnarray}
\mathbf{E}_\mu(x,y) = \sqrt{\frac{2Z_0 \mathrm{P}}{n_\mu A_{\mathrm{mode},\mu}}}\mathbf{e}_\mu(x,y)\label{eqn:E_mode},\\
\mathbf{H}_\mu(x,y) = \sqrt{\frac{2n_\mu \mathrm{P}}{Z_0 A_{\mathrm{mode},\mu}}}\mathbf{h}_\mu(x,y)\label{eqn:H_mode},
\end{eqnarray}
where $Z_0$ is the impedance of free space. The dimensionless field distributions $\mathbf{e}(x,y)$ and $\mathbf{h}(x,y)$ are normalized such that the peak value of $\mathrm{Re}(\mathbf{e}_\mu\times \mathbf{h}_\mu^*) \cdot \hat{z}$ is unity. As a consequence of (\ref{eqn:orth}), the area of mode $\mu$ is given by $A_{\mathrm{mode},\mu}=\int\mathrm{Re}(\mathbf{e}_\mu\times \mathbf{h}_\mu^*)\cdot \hat{z}dx dy$. The modal area is a measure of how tightly confined a mode is and largely determines the strength of nonlinear interactions, with more tightly confined modes producing stronger nonlinear couplings. The definitions used here are chosen to establish an intuitive correspondence between nonlinear interactions in nanowaveguides, which require a fully-vectorial description, and nonlinear interactions between the conventional transverse modes that occur in loosely-guiding waveguides and in a bulk medium. As an example, for an x-polarized Gaussian beam propagating in free space,  $\mathbf{e}(x,y)=\exp(-(x^2+y^2)/w^2)\hat{\mathbf{x}}$, $\mathbf{h}(x,y)=\exp(-(x^2+y^2)/w^2)\hat{\mathbf{y}}$, and $A_\mathrm{mode}=\pi w^2/2$.

Having established the waveguide modes, their dispersion relations, and their normalization, we now consider nonlinear interactions between waveguide modes. The treatment used in the following sections accounts for the fully-vectorial nature of the modes~\cite{Fejer1986,Kolesik2004}, with each field component of $\mathbf{E}_\mu$ coupled together by the full nonlinear tensor, $d_{ijk}$, of the media that comprise the waveguide.

\subsection{Nonlinear coupling}

The presence of a nonlinear polarization at frequency $\omega$ gives rise to driving terms that cause the content of each mode, $a_\mu$, to evolve in $z$. The derivation of the coupled-wave equations under the influence of a nonlinear polarization is similar to that of the orthogonality relations, where Maxwell's equations now include the nonlinear polarization,
\begin{equation}
\nabla\times \mathbf{H}(x,y,z,\omega)=i\omega\bar{\epsilon}(x,y,\omega)\mathbf{E}(x,y,z,\omega)+i\omega \mathbf{P}_\mathrm{NL}(x,y,\omega).
\end{equation}
We consider a pair of modes $\mathbf{E}_1 = a_\mu(z) \mathbf{E}_\mu(x,y)\exp(-i k_\mu z)$, $\mathbf{E}_2 = a_\nu \mathbf{E}_\nu(x,y)\exp(-i k_\nu z)$, such that such that $\mathbf{H}_\nu$ and $\mathbf{H}_\mu$ independently satisfy Maxwell's equations. Substituting $\mathbf{E}_1$ and $\mathbf{H}_1$ into Maxwell's curl equations, taking the dot product with $-\mathbf{H}_2^*$ and $\mathbf{E}_2^*$, respectively, and adding them together yields
\begin{eqnarray}
-\mathbf{H}_2^* \cdot (\nabla\times \mathbf{E}_1) + \mathbf{E}_2^* \cdot (\nabla\times \mathbf{H}_1)\\
=i\omega \mu_0 \mathbf{H}_2^*\cdot \mathbf{H}_1 + i\omega\epsilon_0 \mathbf{E}_2^* \cdot \epsilon \mathbf{E}_1+i\omega \mathbf{E}_2^*\cdot \mathbf{P}_\mathrm{NL}(x,y,\omega)\label{Poynting}.
\end{eqnarray}
Taking the complex conjugate of (\ref{Poynting}), interchanging the indices, and adding them together (assuming real $\bar{\epsilon}$) yields
\begin{eqnarray}
-\mathbf{H}_2^* \cdot (\nabla\times \mathbf{E}_1) + \mathbf{E}_2^* \cdot (\nabla\times \mathbf{H}_1)-\\
\mathbf{H}_1 \cdot (\nabla\times \mathbf{E}_2^*) + \mathbf{E}_1 \cdot (\nabla\times \mathbf{H}_2^*)=\\
-\left(i\omega \mathbf{E}_2^*\cdot \mathbf{P}_\mathrm{NL,1}-i\omega \mathbf{E}_1^*\mathbf{P}_\mathrm{NL,2}\right)\label{Poynting2}
\end{eqnarray}
We rewrite (\ref{Poynting2}) using $\nabla\cdot (\mathbf{A}\times \mathbf{B}) = \mathbf{B}\cdot\nabla\times \mathbf{A} - \mathbf{A}\cdot \nabla\times \mathbf{B}$ to arrive at
\begin{equation}
\nabla\cdot\left(\mathbf{E}_1\times \mathbf{H}_2^* + \mathbf{E}_2^*\times \mathbf{H}_1\right)=-\left(i\omega \mathbf{E}_2^*\cdot \mathbf{P}_\mathrm{NL,1}-i\omega \mathbf{E}_1^*\mathbf{P}_\mathrm{NL,2}\right),\label{PoyntingNL}
\end{equation}
and integrate over all space. Using the orthogonality relations, we find that $a_\mu$ evolves as
\begin{equation}
\partial_z a_{\mu}(z,\omega) = \frac{-i\omega}{4\mathrm{P}}e^{i k_{\mu}z}\int \mathbf{E}_{1}^* \cdot \mathbf{P}_\mathrm{NL,2} dx dy. \label{CWE}
\end{equation}
For second-harmonic generation in the limit where one pair of modes is close to phasematching, we calculate $P_\mathrm{NL,2}$ using one mode for the fundamental at frequency $\omega$ and for the second harmonic at frequency $2\omega$ without loss of generality. For the remainder of this section, the modes under consideration will be referred to as $a_{\omega}$ and $a_{2\omega}$ for the fundamental and second harmonic, respectively. In this case, the nonlinear polarization is given by
\begin{eqnarray}
\mathbf{P}_\mathrm{NL,\omega} = 2\epsilon_0 d_\mathrm{eff} a_{2\omega}a_{\omega}^*\sum_{jk}\bar{d}_{ijk}{E}_{j,2\omega}{E}_{k,\omega}^*e^{-i(k_{2\omega}-k_\omega)z}\label{PNL1}\\
\mathbf{P}_\mathrm{NL,2\omega} = \epsilon_0  d_\mathrm{eff} a_{\omega}^2\sum_{jk}\bar{d}_{ijk}{E}_{j,\omega}{E}_{k,\omega}e^{-2i k_\omega z}\label{PNL2}
\end{eqnarray}
where $i,j,k\in\lbrace x,y,z\rbrace$. For nonlinear interactions between modes polarized predominantly along the crystalline Z-axis in lithium niobate, $d_\mathrm{eff} = \frac{2}{\pi}d_{33}$ is the effective nonlinear coefficient for a 50\% duty cycle periodically poled waveguide, and $\bar{d}_{ijk}$ is the normalized $\chi^{(2)}$ tensor. Assuming Kleinman symmetry, this is expressed using contracted notation~\cite{Nye1985} in the coordinates of the crystal as
\begin{equation*}
\bar{d}_{iJ}=\frac{1}{d_{33}}\left[
\begin{array}{c c c c c c}
0 & 0 & 0 & 0 & d_{15} & d_{16}\\
d_{16} & -d_{16} & 0 & d_{15} & 0 & 0\\
d_{15} & d_{15} & d_{33} & 0 & 0 & 0
\end{array}
\right]
\end{equation*}
where $d_{15}=3.67$ pm/V, $d_{16}=1.78$ pm/V, and $d_{33}=20.5$ pm/V for SHG of 2-$\mu$m light. These values are found using a least squares fit of Miller's delta scaling to the values reported in~\cite{Shoji1997,Choy1976}, and have relative uncertainties of $\pm 5\%$. We therefore expect a relative uncertainty in any calculated normalized efficiency to be $\pm 10\%$.

We arrive at the coupled-wave equations for SHG by substituting Eqns. (\ref{PNL1}-\ref{PNL2}) into (\ref{CWE}) and defining $A_\omega=\sqrt{\mathrm{P}}a_\omega$
\begin{eqnarray}
\partial_z A_\omega = -i\kappa A_{2\omega}A_{\omega}^* e^{-i\Delta k z},\label{NanoCWE1}\\
\partial_z A_{2\omega} = -i\kappa^* A_\omega^2 e^{i\Delta k z}.\label{NanoCWE2}
\end{eqnarray}
The nonlinear coupling, $\kappa$, and the associated effective area are given by 
\begin{eqnarray}
\kappa = \frac{\sqrt{2 Z_0} \omega d_\mathrm{eff}}{c n_\omega \sqrt{A_{\mathrm{eff}}n_{2\omega}}}\exp(-i\phi_\kappa),\\
A_\mathrm{eff}=\frac{A_{\mathrm{mode,}\omega}^2 A_{\mathrm{mode,}2\omega}}{\left|\int \sum_{i,j,k} \bar{d}_{ijk}e^*_{i,2\omega}e_{j,\omega}e_{k,\omega} dx dy\right|^2}.\label{Aeff}
\end{eqnarray}
We remark here that the coupling coefficient $\kappa$ is complex in a nanophotonic waveguide, due to coupling between the purely real transverse components of the fields associated with the waveguide mode with the purely imaginary z-component of the fields. The phase of $\kappa$, $\phi_\kappa$, is given by the phase of the overlap integral in (\ref{Aeff}), and can be neglected without loss of generality. When $\phi_\kappa$ is nonzero, the nonlinear coupling imparts a small phase shift between each of the interacting envelopes, but does not contribute any meaningful change in the resulting nonlinear dynamics. We can remove this phase from the coupled-wave equations by shifting phase reference of the second harmonic, $A_{2\omega}(z,t)\rightarrow A_{2\omega}(z,t)\exp(-i\phi_k)$.

The usual figure of merit for a nonlinear waveguide is the normalized efficiency, $\eta_0=\kappa^2$, which determines the power and device length needed to achieve efficient conversion; devices with larger $\eta_0$ can operate with either less power or shorter devices. The smallest possible effective area for a given wavelength is comparable to $A_\mathrm{eff}\sim (\lambda/n)^2$. Given the scale invariance of Maxwell's equations, the $A_\mathrm{eff}$ of any given device scales as $\lambda^2$, provided that all of the dimensions of the waveguide are rescaled. Therefore, we expect $\eta_0$ to exhibit a quartic scaling with frequency as given designs are rescaled to shorter wavelengths, with a factor of $\omega^2$ coming from the explicit $\omega$-dependence of $\kappa$, and another factor of $\omega^2$ coming from $A_\mathrm{eff}$. In practice, the scaling of $\eta_0$ for a given waveguide is slightly greater than $\omega^4$ due to the dispersion of $d_\mathrm{eff}$.

\section{Tuning and tolerance}\label{sec:tolerance}

Given the relationship between $\Delta k$ and the frequency detuning of the interacting waves, $\Omega$ and $\Omega'$ for SHG and three-wave interactions (\sref{sec:BW}), we are now equipped to discuss the role of tuning mechanisms, such as temperature and fabrication errors. This treatment will yield a set of rules for calculating the fabrication tolerance of a particular device. For the particular case of ultra-broadband interactions, such as those discussed above, we will see that these devices exhibit extremely rapid wavelength tuning behavior and therefore have stringent fabrication requirements.

\subsection{Wavelength tuning due to phase-mismatch}\label{sec:wl_tuning}

We begin by considering the tuning of the phase-matched wavelength with respect to small changes in the phase-mismatch. As an example, we consider CW SHG where $\Delta k_0 = k_{2\omega}-2k_\omega-2\pi/\Lambda_G$ and define $\Lambda_\mathrm{QPM} = 2\pi/(k_{2\omega}-2k_\omega)$ as the period needed to achieve phase-matching at the desired wavelengths. Small deviations between $\Lambda_\mathrm{QPM}$ and the fabricated poling period, $\Lambda_G$, result in a residual phase-mismatch,
\begin{eqnarray}
\delta k &= \frac{2\pi}{\Lambda_\mathrm{QPM}}-\frac{2\pi}{\Lambda_G}\nonumber\\
&\approx \frac{2\pi \Delta \Lambda}{\Lambda_G^2},
\end{eqnarray}
where $\Delta \Lambda = \Lambda_G - \Lambda_\mathrm{QPM}$ is assumed to be much smaller than $\Lambda_G$. Given $\delta k$, the shift in phase-matched wavelength is determined by the GVM, $\Delta k'$, of the interacting waves, 
\begin{equation}
\Delta k(\Omega_\mathrm{PM}) = 0 = \delta k + 2\Delta k'\Omega_\mathrm{PM},\nonumber
\end{equation}
or $\Omega_\mathrm{PM} = -\delta k/2\Delta k'$. A simple rule-of-thumb for calculating the tuning of frequency or wavelength with respect to $\delta k$ is to note that peak of the transfer function shifts to the position of the first zero (e.g. $\Delta \Omega_\mathrm{SHG}/2$ for SHG) when $\delta k L = 2\pi$. For a phase-matching error $\delta k L = 2m\pi$, corresponding to
\begin{equation}
m = \frac{\Delta \Lambda L}{\Lambda_G^2},\label{eqn:m}   
\end{equation}
the shift in the peak phasematching wavelength $\Delta \lambda$ obeys $\Delta\lambda=m\Delta\lambda_\mathrm{SHG}/2$, where $\Delta\lambda_\mathrm{SHG}$ is the full-width of the measured SHG transfer function. This tuning behavior can be expressed as
\begin{equation}
\frac{\Delta\lambda}{\Delta\lambda_\mathrm{SHG}}=\frac{\Delta\Omega}{\Delta\Omega_\mathrm{SHG}}=\frac{m}{2}.\label{eqn:bwtuning}\\
\end{equation}
Equations (\ref{eqn:m}-\ref{eqn:bwtuning}) gives us an intuitive picture of tolerance with respect to $\Delta \Lambda$. Longer devices are more sensitive, with $m$ growing linearly with $L$. Surprisingly, $m$ exhibits a quadratic scaling with poling period; devices with short periods tolerate less fractional error, $\Delta \Lambda/\Lambda_G$, than devices with long periods.

For dispersion-engineered devices with $\Delta k'= 0$ the tuning of wavelength becomes a nonlinear function of $\delta k$, with $\Omega_\mathrm{PM}$ varying rapidly for frequencies around $\omega$,
\begin{equation}
\partial_{\delta k}\Omega_\mathrm{PM} = -\frac{1}{\sqrt{4\delta k(2k_{2\omega}''-k_\omega'')}}.\label{eqn:dom_ddk}
\end{equation}
We need to retain dispersion terms beyond second order in our series expansion for $\Delta k(\Omega)$ to accurately describe this tuning behavior since (\ref{eqn:dom_ddk}) diverges for small errors ($\delta k\sim 0$). A simpler approach is to use (\ref{eqn:m}-\ref{eqn:bwtuning}), which gives the average tuning of $\Delta \Omega$ with respect to $\delta k$ for $\delta k L\in\left[0,2\pi\right]$ and is a reasonable approximation for arbitrary $\Delta k(\omega)$. Taking the device in \sref{sec:exampleSHG} as an example, we have $\Delta \lambda_\mathrm{SHG}=300$ nm for a 5-mm-long device and a period of $\sim 5$ $\mu$m. Therefore, $m=1$ when $\Delta \Lambda = 5$ nm and the SHG peak shifts by 150 nm, or $\Delta \lambda/\Delta\Lambda = 30$.

Given this strong dependence of the phase-matched wavelength on the phase-mismatch in dispersion-engineered devices, care needs to be taken to ensure that the phase-mismatch of fabricated devices can be controlled with sufficient precision. The remainder of this section addresses typical fabrication errors and tuning mechanisms.

\subsection{Waveguide geometry errors}

Small deviations in the geometry of a fabricated waveguide from a nominal design shift $k_\omega$ and $k_{2\omega}$, thereby changing the poling period needed to achieve phase-matching, $\Lambda_\mathrm{QPM}$. For a typical waveguide, the geometry can be  parameterized by top width, $w$, etch depth, $h$, and film thickness, $y$. Small errors in these parameters shift $\Lambda_\mathrm{QPM}$ by
\begin{equation}
\Delta\Lambda_\mathrm{QPM}=dw\partial_w \Lambda_\mathrm{QPM} + dy\partial_y \Lambda_\mathrm{QPM} + dh\partial_h \Lambda_\mathrm{QPM},
\end{equation}
or $\Delta\Lambda_\mathrm{QPM}=d\mathbf{x}\cdot \nabla \Lambda_\mathrm{QPM}$ for an arbitrary parameterization given by $\mathbf{x}$. Higher order contributions to $\Delta k(\Omega,\Omega')$ such as $\Delta k'$ and $k_\omega''$ typically exhibit more tolerance to waveguide geometry errors than the phase-mismatch, and are neglected here. Therefore, to reasonable approximation, we can quantify the role of geometry errors entirely using $\Delta \Lambda_\mathrm{QPM}$ and the resulting tuning of the phase-matched wavelength.

Typical values for $\nabla \Lambda_\mathrm{QPM}$ are given by $\partial_w \Lambda_\mathrm{QPM}=2$ nm/nm, $\partial_h \Lambda_\mathrm{QPM}=-2$ nm/nm, and $\partial_y \Lambda_\mathrm{QPM}=5$ nm/nm. As the waveguide dimensions are made larger and the guided modes become more loosely confined, $\Lambda_\mathrm{QPM}$ becomes larger until it asymptotes to the values found in bulk media. In practice, using the fabrication methods described in \sref{sec:fab}, variations in $dw$, $dh$, and $dy$ between fabrication runs may be as large as $\pm 50$ nm, $\pm 10$ nm, and $\pm 3$ nm, respectively. These geometry errors correspond to an upper and lower bound of $\Delta \Lambda_\mathrm{QPM}\sim \pm140$ nm and a variation in the phase-matched wavelength, $\Delta \lambda$, by \emph{microns}. This large uncertainty in $\Delta \lambda$ illustrates the difficulty of producing dispersion-engineered nonlinear devices using precise fabrication alone; for the numbers considered here, the precision in $dy$ required to fabricate a waveguide with $|\Delta \lambda|<100$ nm is 6.7 Angstroms. While these difficulties may be overcome by using non-critical designs, where $\partial_y \Lambda_\mathrm{QPM}$ or $\partial_w \Lambda_\mathrm{QPM}$ vanishes, such waveguide geometries rarely have the desired dispersion. Instead, fine tuning of the phase-mismatch may be achieved by fabricating many identical waveguides with slightly different poling periods, and by using temperature tuning. Phase-matching can be found reliably when the range of fabricated poling periods spans the bounds of $\Delta\Lambda_\mathrm{QPM}$.

\subsection{Choice of poling periods}

The lithographically patterned poling period $\Lambda_G$ can be written with a precision much finer than resolution ($dx$) of the lithography tools used to define the grating. As a result, optical lithography can be used to define $\Lambda_G$ with better control than $\Lambda_\mathrm{QPM}$, even when the waveguide geometry is patterned using electron-beam lithography. In this section, we briefly derive the resolution needed to pattern a period  $\Lambda_G$ without substantial degradation to device performance. When $dx/\Lambda_G$ is sufficiently small, we make choose an arbitrarily fine step between successive $\Lambda_G$, which enables us to tune the phase-matched wavelength of dispersion-engineered devices in small steps. Having established this tuning behavior, we then discuss commonly used heuristics for fabricating dispersion-engineered devices. The treatment presented here follows~\cite{Fejer1992}.

For grating patterned with resolution $dx$, the desired poling period is given by
\begin{equation}
\Lambda_{G} = (Q+\epsilon)dx = \Lambda_Q + \Delta\Lambda,
\end{equation}
where $Q$ is an integer ($\Lambda_Q = Q dx$) and $-\frac{1}{2}<\epsilon<\frac{1}{2}$ represents the quantization error between the desired poling period $\Lambda_G$ and the patterned period $\Lambda_Q$. The phase drift of the nonlinear polarization due to this small error in poling period can be corrected by patterning an inverted domain of width $\Lambda/2+dx$ instead of $\Lambda/2$ once every $P$ periods, where
\begin{equation}
P = \frac{1}{2\epsilon} = \frac{dx}{2\Delta \Lambda}.
\end{equation}
It can be shown that the degradation of the nonlinear coupling $\kappa$ for SHG or three-wave mixing is given by~\cite{Fejer1992}
\begin{equation}
\bar{\kappa} = \kappa\left(\frac{1 + 2P\mathrm{sinc}(\phi_\mathrm{max})}{1 + 2P}\right)
\end{equation}
provided that $\phi_\mathrm{max} \ll 2P$, where $\phi_\mathrm{max} = \frac{\pi dx}{\Lambda_G}$ is the largest phase error incurred by the nonlinear polarization. The nonlinear coupling $\bar{\kappa}$ will be degraded by less than 1\% if $\phi_\mathrm{max} < 0.08 \pi$, and therefore the resolution needed to achieve high-fidelity poling is
\begin{equation}
dx < 0.08\Lambda_G.\label{eqn:res_limit}
\end{equation} 
\Eref{eqn:res_limit} is remarkably lax; most experimentally relevant poling periods are on the order of microns, whereas many optical lithography tools can pattern poling electrodes with a resolution of $\leq 100$~nm. When \eref{eqn:res_limit} is violated, the fidelity scales as $\bar{\kappa}/\kappa\sim\mathrm{sinc}(\phi_\mathrm{max})\sim 1-\phi_\mathrm{max}^2/6$. As a result, $\phi_\mathrm{max}=0.25 \pi$ corresponds to $\bar{\kappa}/\kappa = 0.9$.

When \eref{eqn:res_limit} is satisfied, there is negligible degradation of device operation due to quantization error, and the desired poling period $\Lambda_G$ can be patterned in arbitrarily fine steps. For a dispersion-engineered waveguide driven by a pulsed laser without wavelength tuning, the fabricated poling periods must have i) a sufficiently fine spacing to efficiently sample the transfer function of a desired nonlinear interaction, and ii) a sufficiently coarse spacing to account for fabrication errors. A natural choice of the spacing in $\Lambda_G$ to satisfy condition (i) is given by adding or removing a single domain, corresponding to a change in $m$ by 1/2, which bounds the total accumulated phase-error by $\delta k L = \pi/2$ for the waveguide closest to phase-matching. In this case, the change between successive periods is given by $\Delta\Lambda_{G}=\Lambda_G^2/(2L)$, where $L$ is the length of the nonlinear waveguide. For our previous example of a 5-mm-long device with a nominal period of 5-$\mu$m the poling period would change in steps of 2.5 nm, which tunes the peak of the SHG transfer function in discrete steps of 75 nm. We note here that achieving the same tuning using waveguide dimensions would require successive waveguides to be patterned with widths that change by $\sim 1$~nm. While this spacing of period still produces large steps of the phase-matched wavelength, it is often too fine to span the bounds of $\Delta\Lambda_\mathrm{QPM}$ for a realistic number of fabricated devices. For the $\pm 140$ nm bounds of $\Delta\Lambda_\mathrm{QPM}$ in the example above, we would need 112 poling periods.

These limitations can be overcome by using multiple tuning mechanisms in parallel. Varying the temperature $T$ of a waveguide can be used to fine-tune the phase-mismatch since $\partial_T k_{2\omega}\neq \partial_T k_{\omega}$ and typical devices, such as those considered in \sref{sec:exampleSHG}, exhibit a tuning of $\partial_T\Lambda_\mathrm{QPM}\sim 0.2-0.3$~nm/C. This allows for continuous tuning of the phase-mismatch between successive poling periods, and the range of available temperatures can enable coarser spacing of the poling period, e.g. $\Delta \Lambda_G \sim 10$~nm for a temperature range of 30 C. In this case, only 28 poled waveguides are necessary to span the  bounds of $\Delta\Lambda_\mathrm{QPM}$. Further improvements in fabrication tolerance are possible by patterning multiple waveguides per poling period. This effectively bounds the width error $dw$ (and therefore $\Delta \Lambda_\mathrm{QPM}$), which in turn reduces the number of necessary poling periods.

\section*{References}

\bibliographystyle{iopart-num}
\bibliography{biblio}

\providecommand{\newblock}{}
\begin{thebibliography}{100}
\expandafter\ifx\csname url\endcsname\relax
  \def\url#1{{\tt #1}}\fi
\expandafter\ifx\csname urlprefix\endcsname\relax\def\urlprefix{URL }\fi
\providecommand{\eprint}[2][]{\url{#2}}

\bibitem{URen2005GenerationOP}
U'Ren A, Silberhorn C, Erdmann R, Banaszek K, Grice W, Walmsley I~A and Raymer
  M 2005 {\em arXiv: Quantum Physics\/}

\bibitem{Yokoyama2013}
Yokoyama S, Ukai R, Armstrong S~C, Sornphiphatphong C, Kaji T, Suzuki S,
  Yoshikawa J, Yonezawa H, Menicucci N~C and Furusawa A 2013 {\em Nature
  Photonics\/} {\bf 7} 982--986
  \urlprefix\url{https://doi.org/10.1038/nphoton.2013.287}

\bibitem{Wakui:07}
Wakui K, Takahashi H, Furusawa A and Sasaki M 2007 {\em Opt. Express\/} {\bf
  15} 3568--3574
  \urlprefix\url{http://www.opticsexpress.org/abstract.cfm?URI=oe-15-6-3568}

\bibitem{DeGreve2012}
Greve K~D, Yu L, McMahon P~L, Pelc J~S, Natarajan C~M, Kim N~Y, Abe E, Maier S,
  Schneider C, Kamp M, H\"{o}fling S, Hadfield R~H, Forchel A, Fejer M~M and
  Yamamoto Y 2012 {\em Nature\/} {\bf 491} 421--425
  \urlprefix\url{https://doi.org/10.1038/nature11577}

\bibitem{Liao2017}
Liao S~K, Yong H~L, Liu C, Shentu G~L, Li D~D, Lin J, Dai H, Zhao S~Q, Li B,
  Guan J~Y, Chen W, Gong Y~H, Li Y, Lin Z~H, Pan G~S, Pelc J~S, Fejer M~M,
  Zhang W~Z, Liu W~Y, Yin J, Ren J~G, Wang X~B, Zhang Q, Peng C~Z and Pan J~W
  2017 {\em Nature Photonics\/} {\bf 11} 509--513
  \urlprefix\url{https://doi.org/10.1038/nphoton.2017.116}

\bibitem{Pelc:11}
Pelc J~S, Ma L, Phillips C~R, Zhang Q, Langrock C, Slattery O, Tang X and Fejer
  M~M 2011 {\em Opt. Express\/} {\bf 19} 21445--21456
  \urlprefix\url{http://www.opticsexpress.org/abstract.cfm?URI=oe-19-22-21445}

\bibitem{Shaked2018}
Shaked Y, Michael Y, Vered R~Z, Bello L, Rosenbluh M and Pe'er A 2018 {\em
  Nature Communications\/} {\bf 9}
  \urlprefix\url{https://doi.org/10.1038/s41467-018-03083-5}

\bibitem{Armstrong1962}
Armstrong J~A, Bloembergen N, Ducuing J and Pershan P~S 1962 {\em Physical
  Review\/} {\bf 127} 1918--1939
  \urlprefix\url{https://doi.org/10.1103/physrev.127.1918}

\bibitem{Franken1963}
Franken P~A and Ward J~F 1963 {\em Reviews of Modern Physics\/} {\bf 35} 23--39
  \urlprefix\url{https://doi.org/10.1103/revmodphys.35.23}

\bibitem{Langrock2006}
Langrock C, Kumar S, McGeehan J, Willner A and Fejer M 2006 {\em Journal of
  Lightwave Technology\/} {\bf 24} 2579--2592
  \urlprefix\url{https://doi.org/10.1109/jlt.2006.874605}

\bibitem{Chen2018}
Chen J~Y, Sua Y~M, Fan H and Huang Y~P 2018 {\em {OSA} Continuum\/} {\bf 1} 229
  \urlprefix\url{https://doi.org/10.1364/osac.1.000229}

\bibitem{Luo2019}
Luo R, He Y, Liang H, Li M, Ling J and Lin Q 2019 {\em Physical Review
  Applied\/} {\bf 11}
  \urlprefix\url{https://doi.org/10.1103/physrevapplied.11.034026}

\bibitem{Bruch2018}
Bruch A~W, Liu X, Guo X, Surya J~B, Gong Z, Zhang L, Wang J, Yan J and Tang H~X
  2018 {\em Applied Physics Letters\/} {\bf 113} 131102
  \urlprefix\url{https://doi.org/10.1063/1.5042506}

\bibitem{Bruch2019}
Bruch A~W, Liu X, Surya J~B, Zou C~L and Tang H~X 2019 {\em Optica\/} {\bf 6}
  1361 \urlprefix\url{https://doi.org/10.1364/optica.6.001361}

\bibitem{Chang2018}
Chang L, Boes A, Guo X, Spencer D~T, Kennedy M~J, Peters J~D, Volet N, Chiles
  J, Kowligy A, Nader N, Hickstein D~D, Stanton E~J, Diddams S~A, Papp S~B and
  Bowers J~E 2018 {\em Laser {\&} Photonics Reviews\/} {\bf 12} 1800149
  \urlprefix\url{https://doi.org/10.1002/lpor.201800149}

\bibitem{Chiles2019}
Chiles J, Nader N, Stanton E~J, Herman D, Moody G, Zhu J, Skehan J~C, Guha B,
  Kowligy A, Gopinath J~T, Srinivasan K, Diddams S~A, Coddington I, Newbury
  N~R, Shainline J~M, Nam S~W and Mirin R~P 2019 {\em Optica\/} {\bf 6} 1246
  \urlprefix\url{https://doi.org/10.1364/optica.6.001246}

\bibitem{Stanton2020}
Stanton E~J, Chiles J, Nader N, Moody G, Volet N, Chang L, Bowers J~E, Nam S~W
  and Mirin R~P 2020 {\em Optics Express\/} {\bf 28} 9521
  \urlprefix\url{https://doi.org/10.1364/oe.389423}

\bibitem{Guidry:20}
Guidry M~A, Yang K~Y, Lukin D~M, Markosyan A, Yang J, Fejer M~M and
  Vu\v{c}kovi\'{c} J 2020 {\em Optica\/} {\bf 7} 1139--1142
  \urlprefix\url{http://www.osapublishing.org/optica/abstract.cfm?URI=optica-7-9-1139}

\bibitem{Wang2018}
Wang C, Langrock C, Marandi A, Jankowski M, Zhang M, Desiatov B, Fejer M~M and
  Lon{\v{c}}ar M 2018 {\em Optica\/} {\bf 5} 1438
  \urlprefix\url{https://doi.org/10.1364/optica.5.001438}

\bibitem{Chang2016}
Chang L, Li Y, Volet N, Wang L, Peters J and Bowers J~E 2016 {\em Optica\/}
  {\bf 3} 531 \urlprefix\url{https://doi.org/10.1364/optica.3.000531}

\bibitem{Timurdogan2017}
Timurdogan E, Poulton C~V, Byrd M~J and Watts M~R 2017 {\em Nature Photonics\/}
  {\bf 11} 200--206 \urlprefix\url{https://doi.org/10.1038/nphoton.2017.14}

\bibitem{Billat2017}
Billat A, Grassani D, Pfeiffer M~H~P, Kharitonov S, Kippenberg T~J and
  Br{\`{e}}s C~S 2017 {\em Nature Communications\/} {\bf 8}
  \urlprefix\url{https://doi.org/10.1038/s41467-017-01110-5}

\bibitem{Porcel2017}
Porcel M~A, Mak J, Taballione C, Schermerhorn V~K, Epping J~P, van~der Slot P~J
  and Boller K~J 2017 {\em Optics Express\/} {\bf 25} 33143
  \urlprefix\url{https://doi.org/10.1364/oe.25.033143}

\bibitem{XLu2020}
Lu X, Moille G, Rao A, Westly D~A and Srinivasan K 2020 {\em Nature
  Photonics\/} \urlprefix\url{https://doi.org/10.1038/s41566-020-00708-4}

\bibitem{Chen2020}
Chen J~Y, Tang C, Ma Z~H, Li Z, Sua Y~M and Huang Y~P 2020 {\em Optics
  Letters\/} {\bf 45} 3789 \urlprefix\url{https://doi.org/10.1364/ol.393445}

\bibitem{Nagy2019}
Nagy J~T and Reano R~M 2019 {\em Optical Materials Express\/} {\bf 9} 3146
  \urlprefix\url{https://doi.org/10.1364/ome.9.003146}

\bibitem{Nagy2020}
Nagy J~T and Reano R~M 2020 {\em Optical Materials Express\/} {\bf 10} 1911
  \urlprefix\url{https://doi.org/10.1364/ome.394724}

\bibitem{Hickstein2019}
Hickstein D~D, Carlson D~R, Mundoor H, Khurgin J~B, Srinivasan K, Westly D,
  Kowligy A, Smalyukh I~I, Diddams S~A and Papp S~B 2019 {\em Nature
  Photonics\/} {\bf 13} 494--499
  \urlprefix\url{https://doi.org/10.1038/s41566-019-0449-8}

\bibitem{Singh2020}
Singh N, Raval M, Ruocco A and Watts M~R 2020 {\em Light: Science {\&}
  Applications\/} {\bf 9}
  \urlprefix\url{https://doi.org/10.1038/s41377-020-0254-7}

\bibitem{Jankowski2020}
Jankowski M, Langrock C, Desiatov B, Marandi A, Wang C, Zhang M, Phillips C~R,
  Lon{\v{c}}ar M and Fejer M~M 2020 {\em Optica\/} {\bf 7} 40
  \urlprefix\url{https://doi.org/10.1364/optica.7.000040}

\bibitem{OBrien2007}
O'Brien J~L 2007 {\em Science\/} {\bf 318} 1567--1570
  \urlprefix\url{https://doi.org/10.1126/science.1142892}

\bibitem{OBrien2009}
O'Brien J~L, Furusawa A and Vu{\v{c}}kovi{\'{c}} J 2009 {\em Nature
  Photonics\/} {\bf 3} 687--695
  \urlprefix\url{https://doi.org/10.1038/nphoton.2009.229}

\bibitem{zhu2021integrated}
Zhu D, Shao L, Yu M, Cheng R, Desiatov B, Xin C~J, Hu Y, Holzgrafe J, Ghosh S,
  Shams-Ansari A, Puma E, Sinclair N, Reimer C, Zhang M and Lon\v{c}ar M 2021
  Integrated photonics on thin-film lithium niobate (\textit{Preprint}
  \eprint{2102.11956})

\bibitem{Wang2020Q}
Wang C, Zhang M and Lon\v{c}ar M 2020 High-q lithium niobate microcavities and
  their applications {\em Ultra-High-Q Optical Microcavities\/} ({WORLD}
  {SCIENTIFIC}) pp 1--35

\bibitem{Honardoost2020}
Honardoost A, Abdelsalam K and Fathpour S 2020 {\em Laser {\&} Photonics
  Reviews\/} {\bf 14} 2000088
  \urlprefix\url{https://doi.org/10.1002/lpor.202000088}

\bibitem{Qi2020}
Qi Y and Li Y 2020 {\em Nanophotonics\/} {\bf 9} 1287--1320
  \urlprefix\url{https://doi.org/10.1515/nanoph-2020-0013}

\bibitem{Boes2018}
Boes A, Corcoran B, Chang L, Bowers J and Mitchell A 2018 {\em Laser {\&}
  Photonics Reviews\/} {\bf 12} 1700256
  \urlprefix\url{https://doi.org/10.1002/lpor.201700256}

\bibitem{Sun2020}
Sun D, Zhang Y, Wang D, Song W, Liu X, Pang J, Geng D, Sang Y and Liu H 2020
  {\em Light: Science {\&} Applications\/} {\bf 9}
  \urlprefix\url{https://doi.org/10.1038/s41377-020-00434-0}

\bibitem{Moody2020}
Moody G, Chang L, Steiner T~J and Bowers J~E 2020 {\em {AVS} Quantum Science\/}
  {\bf 2} 041702 \urlprefix\url{https://doi.org/10.1116/5.0020684}

\bibitem{Lukin2019}
Lukin D~M, Dory C, Guidry M~A, Yang K~Y, Mishra S~D, Trivedi R, Radulaski M,
  Sun S, Vercruysse D, Ahn G~H and Vu{\v{c}}kovi{\'{c}} J 2019 {\em Nature
  Photonics\/} {\bf 14} 330--334
  \urlprefix\url{https://doi.org/10.1038/s41566-019-0556-6}

\bibitem{Lukin2020}
Lukin D~M, Guidry M~A and Vu{\v{c}}kovi{\'{c}} J 2020 {\em {PRX} Quantum\/}
  {\bf 1} \urlprefix\url{https://doi.org/10.1103/prxquantum.1.020102}

\bibitem{Wang2019quantum}
Wang J, Sciarrino F, Laing A and Thompson M~G 2019 {\em Nature Photonics\/}
  {\bf 14} 273--284 \urlprefix\url{https://doi.org/10.1038/s41566-019-0532-1}

\bibitem{mckenna2021ultralowpower}
McKenna T~P, Stokowski H~S, Ansari V, Mishra J, Jankowski M, Sarabalis C~J,
  Herrmann J~F, Langrock C, Fejer M~M and Safavi-Naeini A~H 2021
  Ultra-low-power second-order nonlinear optics on a chip (\textit{Preprint}
  \eprint{2102.05617})

\bibitem{Zhao2020}
Zhao J, Ma C, R\"using M and Mookherjea S 2020 {\em Phys. Rev. Lett.\/} {\bf
  124}(16) 163603
  \urlprefix\url{https://link.aps.org/doi/10.1103/PhysRevLett.124.163603}

\bibitem{Rao2019}
Rao A, Abdelsalam K, Sjaardema T, Honardoost A, Camacho-Gonzalez G~F and
  Fathpour S 2019 {\em Optics Express\/} {\bf 27} 25920
  \urlprefix\url{https://doi.org/10.1364/oe.27.025920}

\bibitem{Zhang2017}
Zhang M, Wang C, Cheng R, Shams-Ansari A and Lon{\v{c}}ar M 2017 {\em Optica\/}
  {\bf 4} 1536 \urlprefix\url{https://doi.org/10.1364/optica.4.001536}

\bibitem{Fejer1992}
{Fejer} M~M, {Magel} G~A, {Jundt} D~H and {Byer} R~L 1992 {\em IEEE Journal of
  Quantum Electronics\/} {\bf 28} 2631--2654

\bibitem{Bortz1994}
Bortz M, Field S, Fejer M, Nam D, Waarts R and Welch D 1994 {\em {IEEE} Journal
  of Quantum Electronics\/} {\bf 30} 2953--2960
  \urlprefix\url{https://doi.org/10.1109/3.362710}

\bibitem{Hum2007}
Hum D~S and Fejer M~M 2007 {\em Comptes Rendus Physique\/} {\bf 8} 180--198
  \urlprefix\url{https://doi.org/10.1016/j.crhy.2006.10.022}

\bibitem{Boes2019}
Boes A, Chang L, Knoerzer M, Nguyen T~G, Peters J~D, Bowers J~E and Mitchell A
  2019 {\em Optics Express\/} {\bf 27} 23919
  \urlprefix\url{https://doi.org/10.1364/oe.27.023919}

\bibitem{Shoji1997}
Shoji I, Kondo T, Kitamoto A, Shirane M and Ito R 1997 {\em Journal of the
  Optical Society of America B\/} {\bf 14} 2268
  \urlprefix\url{https://doi.org/10.1364/josab.14.002268}

\bibitem{Choy1976}
Choy M~M and Byer R~L 1976 {\em Physical Review B\/} {\bf 14} 1693--1706
  \urlprefix\url{https://doi.org/10.1103/physrevb.14.1693}

\bibitem{Caves1987}
Caves C~M and Crouch D~D 1987 {\em Journal of the Optical Society of America
  B\/} {\bf 4} 1535 \urlprefix\url{https://doi.org/10.1364/josab.4.001535}

\bibitem{Crouch1988}
Crouch D~D 1988 {\em Physical Review A\/} {\bf 38} 508--511
  \urlprefix\url{https://doi.org/10.1103/physreva.38.508}

\bibitem{Harris2007}
Harris S~E 2007 {\em Physical Review Letters\/} {\bf 98}
  \urlprefix\url{https://doi.org/10.1103/physrevlett.98.063602}

\bibitem{LutherDavies2017}
Luther-Davies B and Yu Y 2017 Efficient generation of ultra-short pulses in the
  infrared from a simple {PPLN} optical parametric amplifier {\em Nonlinear
  Optics\/} ({OSA}) \urlprefix\url{https://doi.org/10.1364/nlo.2017.ntu2a.4}

\bibitem{Trapani1995}
Trapani P~D, Andreoni A, Solcia C, Foggi P, Danielius R, Dubietis A and
  Piskarskas A 1995 {\em Journal of the Optical Society of America B\/} {\bf
  12} 2237 \urlprefix\url{https://doi.org/10.1364/josab.12.002237}

\bibitem{Marchese:2005}
Marchese S, Innerhofer E, Paschotta R, Kurimura S, Kitamura K, Arisholm G and
  Keller U 2005 {\em Applied Physics B\/} {\bf 81} 1049--1052
  \urlprefix\url{https://doi.org/10.1007/s00340-005-1964-5}

\bibitem{Cerullo2003}
Cerullo G and Silvestri S~D 2003 {\em Review of Scientific Instruments\/} {\bf
  74} 1--18 \urlprefix\url{https://doi.org/10.1063/1.1523642}

\bibitem{Manzoni2016}
Manzoni C and Cerullo G 2016 {\em Journal of Optics\/} {\bf 18} 103501
  \urlprefix\url{https://doi.org/10.1088/2040-8978/18/10/103501}

\bibitem{Rodriguez2020}
Roman-Rodriguez V, Brecht B, Kaali S, Silberhorn C, Treps N, Diamanti E and
  Parigi V 2020 Continuous variable multimode quantum states via symmetric
  group velocity matching (\textit{Preprint} \eprint{2012.13629})

\bibitem{Imeshev2000a}
Imeshev G, Arbore M~A, Fejer M~M, Galvanauskas A, Fermann M and Harter D 2000
  {\em Journal of the Optical Society of America B\/} {\bf 17} 304
  \urlprefix\url{https://doi.org/10.1364/josab.17.000304}

\bibitem{Imeshev2000b}
Imeshev G, Arbore M~A, Kasriel S and Fejer M~M 2000 {\em Journal of the Optical
  Society of America B\/} {\bf 17} 1420
  \urlprefix\url{https://doi.org/10.1364/josab.17.001420}

\bibitem{jankowski2021supercontinuum}
Jankowski M, Langrock C, Desiatov B, Loncar M and Fejer M~M 2021 Supercontinuum
  generation by saturated $\chi^{(2)}$ interactions (\textit{Preprint}
  \eprint{2102.12856})

\bibitem{jankowski2021efficient}
Jankowski M, Jornod N, Langrock C, Desiatov B, Marandi A, Lončar M and Fejer
  M~M 2021 Efficient octave-spanning parametric down-conversion at the
  picojoule level (\textit{Preprint} \eprint{2104.07928})

\bibitem{ledezma2021intense}
Ledezma L, Sekine R, Guo Q, Nehra R, Jahani S and Marandi A 2021 Intense
  optical parametric amplification in dispersion engineered nanophotonic
  lithium niobate waveguides (\textit{Preprint} \eprint{2104.08262})

\bibitem{Takeda2019}
Takeda S and Furusawa A 2019 {\em {APL} Photonics\/} {\bf 4} 060902
  \urlprefix\url{https://doi.org/10.1063/1.5100160}

\bibitem{Pfister2019}
Pfister O 2019 {\em Journal of Physics B: Atomic, Molecular and Optical
  Physics\/} {\bf 53} 012001
  \urlprefix\url{https://doi.org/10.1088/1361-6455/ab526f}

\bibitem{Raussendorf2001}
Raussendorf R and Briegel H~J 2001 {\em Physical Review Letters\/} {\bf 86}
  5188--5191 \urlprefix\url{https://doi.org/10.1103/physrevlett.86.5188}

\bibitem{Raussendorf2003}
Raussendorf R, Browne D~E and Briegel H~J 2003 {\em Physical Review A\/} {\bf
  68} \urlprefix\url{https://doi.org/10.1103/physreva.68.022312}

\bibitem{Yoshikawa2016}
Yoshikawa J, Yokoyama S, Kaji T, Sornphiphatphong C, Shiozawa Y, Makino K and
  Furusawa A 2016 {\em {APL} Photonics\/} {\bf 1} 060801
  \urlprefix\url{https://doi.org/10.1063/1.4962732}

\bibitem{Asavanant2019}
Asavanant W, Shiozawa Y, Yokoyama S, Charoensombutamon B, Emura H, Alexander
  R~N, Takeda S, Yoshikawa J, Menicucci N~C, Yonezawa H and Furusawa A 2019
  {\em Science\/} {\bf 366} 373--376
  \urlprefix\url{https://doi.org/10.1126/science.aay2645}

\bibitem{Vahlbruch2016}
Vahlbruch H, Mehmet M, Danzmann K and Schnabel R 2016 {\em Physical Review
  Letters\/} {\bf 117}
  \urlprefix\url{https://doi.org/10.1103/physrevlett.117.110801}

\bibitem{Kashiwazaki2020}
Kashiwazaki T, Takanashi N, Yamashima T, Kazama T, Enbutsu K, Kasahara R, Umeki
  T and Furusawa A 2020 {\em {APL} Photonics\/} {\bf 5} 036104
  \urlprefix\url{https://doi.org/10.1063/1.5142437}

\bibitem{Schnabel2017}
Schnabel R 2017 {\em Physics Reports\/} {\bf 684} 1--51
  \urlprefix\url{https://doi.org/10.1016/j.physrep.2017.04.001}

\bibitem{Walmsley2005}
Walmsley I~A 2005 {\em Science\/} {\bf 307} 1733--1734
  \urlprefix\url{https://doi.org/10.1126/science.1107451}

\bibitem{Kok2007}
Kok P, Munro W~J, Nemoto K, Ralph T~C, Dowling J~P and Milburn G~J 2007 {\em
  Reviews of Modern Physics\/} {\bf 79} 135--174
  \urlprefix\url{https://doi.org/10.1103/revmodphys.79.135}

\bibitem{Humphreys2013}
Humphreys P~C, Metcalf B~J, Spring J~B, Moore M, Jin X~M, Barbieri M,
  Kolthammer W~S and Walmsley I~A 2013 {\em Physical Review Letters\/} {\bf
  111} \urlprefix\url{https://doi.org/10.1103/physrevlett.111.150501}

\bibitem{Briegel1998}
Briegel H~J, D\"{u}r W, Cirac J~I and Zoller P 1998 {\em Physical Review
  Letters\/} {\bf 81} 5932--5935
  \urlprefix\url{https://doi.org/10.1103/physrevlett.81.5932}

\bibitem{Spring2012}
Spring J~B, Metcalf B~J, Humphreys P~C, Kolthammer W~S, Jin X~M, Barbieri M,
  Datta A, Thomas-Peter N, Langford N~K, Kundys D, Gates J~C, Smith B~J, Smith
  P~G~R and Walmsley I~A 2012 {\em Science\/} {\bf 339} 798--801
  \urlprefix\url{https://doi.org/10.1126/science.1231692}

\bibitem{Motes2015}
Motes K~R, Olson J~P, Rabeaux E~J, Dowling J~P, Olson S~J and Rohde P~P 2015
  {\em Physical Review Letters\/} {\bf 114}
  \urlprefix\url{https://doi.org/10.1103/physrevlett.114.170802}

\bibitem{Higginbottom2016}
Higginbottom D~B, Slodi{\v{c}}ka L, Araneda G, Lachman L, Filip R, Hennrich M
  and Blatt R 2016 {\em New Journal of Physics\/} {\bf 18} 093038
  \urlprefix\url{https://doi.org/10.1088/1367-2630/18/9/093038}

\bibitem{Dusanowski2019}
Dusanowski {\L}, Kwon S~H, Schneider C and H\"{o}fling S 2019 {\em Physical
  Review Letters\/} {\bf 122}
  \urlprefix\url{https://doi.org/10.1103/physrevlett.122.173602}

\bibitem{Sipahigil2014}
Sipahigil A, Jahnke K, Rogers L, Teraji T, Isoya J, Zibrov A, Jelezko F and
  Lukin M 2014 {\em Physical Review Letters\/} {\bf 113}
  \urlprefix\url{https://doi.org/10.1103/physrevlett.113.113602}

\bibitem{Chen2017}
Chen C, Bo C, Niu M~Y, Xu F, Zhang Z, Shapiro J~H and Wong F~N~C 2017 {\em
  Optics Express\/} {\bf 25} 7300
  \urlprefix\url{https://doi.org/10.1364/oe.25.007300}

\bibitem{Mosley2008}
Mosley P~J, Lundeen J~S, Smith B~J, Wasylczyk P, U'Ren A~B, Silberhorn C and
  Walmsley I~A 2008 {\em Physical Review Letters\/} {\bf 100}
  \urlprefix\url{https://doi.org/10.1103/physrevlett.100.133601}

\bibitem{URen2006}
U'Ren A~B, Erdmann R~K, de~la Cruz-Gutierrez M and Walmsley I~A 2006 {\em
  Physical Review Letters\/} {\bf 97}
  \urlprefix\url{https://doi.org/10.1103/physrevlett.97.223602}

\bibitem{Sller2011}
S\"{o}ller C, Cohen O, Smith B~J, Walmsley I~A and Silberhorn C 2011 {\em
  Physical Review A\/} {\bf 83}
  \urlprefix\url{https://doi.org/10.1103/physreva.83.031806}

\bibitem{Hendrych2007}
Hendrych M, Micuda M and Torres J~P 2007 {\em Optics Letters\/} {\bf 32} 2339
  \urlprefix\url{https://doi.org/10.1364/ol.32.002339}

\bibitem{Zhang2019}
Zhang Y, Spiniolas R, Shinbrough K, Fang B, Cohen O and Lorenz V~O 2019 {\em
  Optics Express\/} {\bf 27} 19050
  \urlprefix\url{https://doi.org/10.1364/oe.27.019050}

\bibitem{MeyerScott2020}
Meyer-Scott E, Silberhorn C and Migdall A 2020 {\em Review of Scientific
  Instruments\/} {\bf 91} 041101
  \urlprefix\url{https://doi.org/10.1063/5.0003320}

\bibitem{Ma2020}
Ma Z, Chen J~Y, Li Z, Tang C, Sua Y~M, Fan H and Huang Y~P 2020 {\em Physical
  Review Letters\/} {\bf 125}
  \urlprefix\url{https://doi.org/10.1103/physrevlett.125.263602}

\bibitem{Steiner2021}
Steiner T~J, Castro J~E, Chang L, Dang Q, Xie W, Norman J, Bowers J~E and Moody
  G 2021 {\em {PRX} Quantum\/} {\bf 2}
  \urlprefix\url{https://doi.org/10.1103/prxquantum.2.010337}

\bibitem{Aichele2002}
Aichele T, Lvovsky A and Schiller S 2002 {\em The European Physical Journal D -
  Atomic, Molecular and Optical Physics\/} {\bf 18} 237--245
  \urlprefix\url{https://doi.org/10.1140/epjd/e20020028}

\bibitem{URen04}
U'Ren A 2004 {\em Multi-photon state engineering for quantum information
  processing applications\/} Ph.D. thesis University of Rochester

\bibitem{BenDixon2013}
Dixon P~B, Shapiro J~H and Wong F~N~C 2013 {\em Optics Express\/} {\bf 21} 5879
  \urlprefix\url{https://doi.org/10.1364/oe.21.005879}

\bibitem{Phillips2013}
Phillips C~R, Langrock C, Chang D, Lin Y~W, Gallmann L and Fejer M~M 2013 {\em
  Journal of the Optical Society of America B\/} {\bf 30} 1551
  \urlprefix\url{https://doi.org/10.1364/josab.30.001551}

\bibitem{Braczyk2010}
Bra{\'{n}}czyk A~M, Fedrizzi A, Stace T~M, Ralph T~C and White A~G 2010 {\em
  Optics Express\/} {\bf 19} 55
  \urlprefix\url{https://doi.org/10.1364/oe.19.000055}

\bibitem{Huang2006}
Huang J, Xie X~P, Langrock C, Roussev R~V, Hum D~S and Fejer M~M 2006 {\em
  Optics Letters\/} {\bf 31} 604
  \urlprefix\url{https://doi.org/10.1364/ol.31.000604}

\bibitem{Raymer2005}
Raymer M~G, Noh J, Banaszek K and Walmsley I~A 2005 {\em Physical Review A\/}
  {\bf 72} \urlprefix\url{https://doi.org/10.1103/physreva.72.023825}

\bibitem{Ma2017}
Ma C, Wang X, Anant V, Beyer A~D, Shaw M~D and Mookherjea S 2017 {\em Optics
  Express\/} {\bf 25} 32995
  \urlprefix\url{https://doi.org/10.1364/oe.25.032995}

\bibitem{Lu2019NIST}
Lu X, Li Q, Westly D~A, Moille G, Singh A, Anant V and Srinivasan K 2019 {\em
  Nature Physics\/} {\bf 15} 373--381
  \urlprefix\url{https://doi.org/10.1038/s41567-018-0394-3}

\bibitem{Luo2015}
Luo K~H, Herrmann H, Krapick S, Brecht B, Ricken R, Quiring V, Suche H, Sohler
  W and Silberhorn C 2015 {\em New Journal of Physics\/} {\bf 17} 073039
  \urlprefix\url{https://doi.org/10.1088/1367-2630/17/7/073039}

\bibitem{Guo2016}
Guo X, ling Zou C, Schuck C, Jung H, Cheng R and Tang H~X 2016 {\em Light:
  Science {\&} Applications\/} {\bf 6} e16249--e16249
  \urlprefix\url{https://doi.org/10.1038/lsa.2016.249}

\bibitem{Vernon2016}
Vernon Z, Liscidini M and Sipe J~E 2016 {\em Optics Letters\/} {\bf 41} 788
  \urlprefix\url{https://doi.org/10.1364/ol.41.000788}

\bibitem{Desiatov2019}
Desiatov B, Shams-Ansari A, Zhang M, Wang C and Lon{\v{c}}ar M 2019 {\em
  Optica\/} {\bf 6} 380 \urlprefix\url{https://doi.org/10.1364/optica.6.000380}

\bibitem{Wang2019}
Wang C, Zhang M, Yu M, Zhu R, Hu H and Loncar M 2019 {\em Nature
  Communications\/} {\bf 10}
  \urlprefix\url{https://doi.org/10.1038/s41467-019-08969-6}

\bibitem{Wang2018EOM}
Wang C, Zhang M, Chen X, Bertrand M, Shams-Ansari A, Chandrasekhar S, Winzer P
  and Lon{\v{c}}ar M 2018 {\em Nature\/} {\bf 562} 101--104
  \urlprefix\url{https://doi.org/10.1038/s41586-018-0551-y}

\bibitem{Zhang2019EOM}
Zhang M, Buscaino B, Wang C, Shams-Ansari A, Reimer C, Zhu R, Kahn J~M and
  Lon{\v{c}}ar M 2019 {\em Nature\/} {\bf 568} 373--377
  \urlprefix\url{https://doi.org/10.1038/s41586-019-1008-7}

\bibitem{Sayem2020}
Sayem A~A, Cheng R, Wang S and Tang H~X 2020 {\em Applied Physics Letters\/}
  {\bf 116} 151102 \urlprefix\url{https://doi.org/10.1063/1.5142852}

\bibitem{Thompson2011}
Thompson M, Politi A, Matthews J and O'Brien J 2011 {\em {IET} Circuits,
  Devices {\&} Systems\/} {\bf 5} 94
  \urlprefix\url{https://doi.org/10.1049/iet-cds.2010.0108}

\bibitem{VanMeter2016}
Meter R~V and Devitt S~J 2016 {\em Computer\/} {\bf 49} 31--42
  \urlprefix\url{https://doi.org/10.1109/mc.2016.291}

\bibitem{Mabuchi2012}
Mabuchi H 2012 {\em Physical Review A\/} {\bf 85}
  \urlprefix\url{https://doi.org/10.1103/physreva.85.015806}

\bibitem{Agarwal1994}
Agarwal G~S 1994 {\em Physical Review Letters\/} {\bf 73} 522--524
  \urlprefix\url{https://doi.org/10.1103/physrevlett.73.522}

\bibitem{Majumdar2013}
Majumdar A and Gerace D 2013 {\em Physical Review B\/} {\bf 87}
  \urlprefix\url{https://doi.org/10.1103/physrevb.87.235319}

\bibitem{Reid1993}
Reid M~D and Krippner L 1993 {\em Physical Review A\/} {\bf 47} 552--555
  \urlprefix\url{https://doi.org/10.1103/physreva.47.552}

\bibitem{onodera2019nonlinear}
Onodera T, Ng E, Lörch N, Yamamura A, Hamerly R, McMahon P~L, Marandi A and
  Mabuchi H 2019 Nonlinear quantum behavior of ultrashort-pulse optical
  parametric oscillators (\textit{Preprint} \eprint{1811.10583})

\bibitem{yanagimoto2020broadband}
Yanagimoto R, Ng E, Jankowski M~P, Onodera T, Fejer M~M and Mabuchi H 2020
  Broadband parametric downconversion as a discrete-continuum fano interaction
  (\textit{Preprint} \eprint{2009.01457})

\bibitem{Chen2019}
Chen J~Y, Ma Z~H, Sua Y~M, Li Z, Tang C and Huang Y~P 2019 {\em Optica\/} {\bf
  6} 1244 \urlprefix\url{https://doi.org/10.1364/optica.6.001244}

\bibitem{Lu2019}
Lu J, Surya J~B, Liu X, Bruch A~W, Gong Z, Xu Y and Tang H~X 2019 {\em
  Optica\/} {\bf 6} 1455
  \urlprefix\url{https://doi.org/10.1364/optica.6.001455}

\bibitem{Lu2020}
Lu J, Li M, Zou C~L, Sayem A~A and Tang H~X 2020 {\em Optica\/} {\bf 7} 1654
  \urlprefix\url{https://doi.org/10.1364/optica.403931}

\bibitem{Jahani2021}
Jahani S, Roy A and Marandi A 2021 {\em Optica\/} {\bf 8} 262
  \urlprefix\url{https://doi.org/10.1364/optica.411708}

\bibitem{Panuski2020}
Panuski C, Englund D and Hamerly R 2020 {\em Physical Review X\/} {\bf 10}
  \urlprefix\url{https://doi.org/10.1103/physrevx.10.041046}

\bibitem{Ohashi1993}
Ohashi M, Kondo T, Ito R, Fukatsu S, Shiraki Y, Kumata K and Kano S~S 1993 {\em
  Journal of Applied Physics\/} {\bf 74} 596--601
  \urlprefix\url{https://doi.org/10.1063/1.355272}

\bibitem{Wilson2019}
Wilson D~J, Schneider K, H\"{o}nl S, Anderson M, Baumgartner Y, Czornomaz L,
  Kippenberg T~J and Seidler P 2019 {\em Nature Photonics\/} {\bf 14} 57--62
  \urlprefix\url{https://doi.org/10.1038/s41566-019-0537-9}

\bibitem{Rivoire2011}
Rivoire K, Buckley S, Hatami F and Vu{\v{c}}kovi{\'{c}} J 2011 {\em Applied
  Physics Letters\/} {\bf 98} 263113
  \urlprefix\url{https://doi.org/10.1063/1.3607288}

\bibitem{Song:19}
Song B~S, Asano T, Jeon S, Kim H, Chen C, Kang D~D and Noda S 2019 {\em
  Optica\/} {\bf 6} 991--995
  \urlprefix\url{http://www.osapublishing.org/optica/abstract.cfm?URI=optica-6-8-991}

\bibitem{Maidment2016}
Maidment L, Schunemann P~G and Reid D~T 2016 {\em Optics Letters\/} {\bf 41}
  4261 \urlprefix\url{https://doi.org/10.1364/ol.41.004261}

\bibitem{Schunemann2014}
{Schunemann} P~G, {Mohnkern} L, {Vera} A, {Yang} X~S, {Lin} A~C, {Harris} J~S,
  {Tassev} V and {Snure} M 2014 Growth of device-quality orientation-patterned
  gallium phosphide (op-gap) by improved hydride vapour phase epitaxy {\em 2014
  Conference on Lasers and Electro-Optics (CLEO) - Laser Science to Photonic
  Applications\/} pp 1--2

\bibitem{Tassev:19}
Tassev V~L and Vangala S~R 2019 New heteroepitaxially grown materials for
  frequency conversion in the mid and longwave infrared {\em Nonlinear Optics
  (NLO)\/} (Optical Society of America) p NTu4A.33
  \urlprefix\url{http://www.osapublishing.org/abstract.cfm?URI=NLO-2019-NTu4A.33}

\bibitem{Vangala:19}
Vangala S, Tassev V and Snure M 2019 Thick heteroepitaxial growth of znse on
  gaas substrates for frequency conversion in the mlwir {\em Nonlinear Optics
  (NLO)\/} (Optical Society of America) p NTu4A.40
  \urlprefix\url{http://www.osapublishing.org/abstract.cfm?URI=NLO-2019-NTu4A.40}

\bibitem{Schunemann2019}
Schunemann P~G and Zawilski K~T 2019 {Vapor transport growth of single crystal
  zinc selenide (Conference Presentation)} {\em Nonlinear Frequency Generation
  and Conversion: Materials and Devices XVIII\/} vol 10902 ed Schunemann P~G
  and Schepler K~L International Society for Optics and Photonics (SPIE)
  \urlprefix\url{https://doi.org/10.1117/12.2514055}

\bibitem{He:19}
He Y, Yang Q~F, Ling J, Luo R, Liang H, Li M, Shen B, Wang H, Vahala K and Lin
  Q 2019 {\em Optica\/} {\bf 6} 1138--1144
  \urlprefix\url{http://www.osapublishing.org/optica/abstract.cfm?URI=optica-6-9-1138}

\bibitem{Gayer2008}
Gayer O, Sacks Z, Galun E and Arie A 2008 {\em Applied Physics B\/} {\bf 91}
  343--348 \urlprefix\url{https://doi.org/10.1007/s00340-008-2998-2}

\bibitem{Skauli2003}
Skauli T, Kuo P~S, Vodopyanov K~L, Pinguet T~J, Levi O, Eyres L~A, Harris J~S,
  Fejer M~M, Gerard B, Becouarn L and Lallier E 2003 {\em Journal of Applied
  Physics\/} {\bf 94} 6447--6455
  \urlprefix\url{https://doi.org/10.1063/1.1621740}

\bibitem{Skauli2002}
Skauli T, Vodopyanov K~L, Pinguet T~J, Schober A, Levi O, Eyres L~A, Fejer M~M,
  Harris J~S, Gerard B, Becouarn L, Lallier E and Arisholm G 2002 {\em Optics
  Letters\/} {\bf 27} 628 \urlprefix\url{https://doi.org/10.1364/ol.27.000628}

\bibitem{Fern1971}
Fern R~E and Onton A 1971 {\em Journal of Applied Physics\/} {\bf 42}
  3499--3500 \urlprefix\url{https://doi.org/10.1063/1.1660760}

\bibitem{Wang2013}
Wang S, Zhan M, Wang G, Xuan H, Zhang W, Liu C, Xu C, Liu Y, Wei Z and Chen X
  2013 {\em Laser {\&} Photonics Reviews\/} {\bf 7} 831--838
  \urlprefix\url{https://doi.org/10.1002/lpor.201300068}

\bibitem{Sato2009}
Sato H, Abe M, Shoji I, Suda J and Kondo T 2009 {\em Journal of the Optical
  Society of America B\/} {\bf 26} 1892
  \urlprefix\url{https://doi.org/10.1364/josab.26.001892}

\bibitem{Wei2018}
Wei J, Murray J~M, Barnes J~O, Krein D~M, Schunemann P~G and Guha S 2018 {\em
  Optical Materials Express\/} {\bf 8} 485
  \urlprefix\url{https://doi.org/10.1364/ome.8.000485}

\bibitem{Tatian1984}
Tatian B 1984 {\em Applied Optics\/} {\bf 23} 4477
  \urlprefix\url{https://doi.org/10.1364/ao.23.004477}

\bibitem{Connolly1979}
Connolly J, diBenedetto B and Donadio R 1979 Specifications of raytran material
  {\em Contemporary Optical Systems and Components Specifications\/} ed Fischer
  R~E ({SPIE}) \urlprefix\url{https://doi.org/10.1117/12.957359}

\bibitem{Frey2006}
Frey B~J, Leviton D~B and Madison T~J 2006 Temperature-dependent refractive
  index of silicon and germanium {\em Optomechanical Technologies for
  Astronomy\/} ed Atad-Ettedgui E, Antebi J and Lemke D ({SPIE})
  \urlprefix\url{https://doi.org/10.1117/12.672850}

\bibitem{Luke2015}
Luke K, Okawachi Y, Lamont M~R~E, Gaeta A~L and Lipson M 2015 {\em Optics
  Letters\/} {\bf 40} 4823 \urlprefix\url{https://doi.org/10.1364/ol.40.004823}

\bibitem{Guo2020}
Guo H, Weng W, Liu J, Yang F, H\"{a}nsel W, Br{\`{e}}s C~S, Th{\'{e}}venaz L,
  Holzwarth R and Kippenberg T~J 2020 {\em Optica\/} {\bf 7} 1181
  \urlprefix\url{https://doi.org/10.1364/optica.396542}

\bibitem{Vercruysse2020}
Vercruysse D, Sapra N~V, Su L and Vuckovic J 2020 {\em {IEEE} Journal of
  Selected Topics in Quantum Electronics\/} {\bf 26} 1--6
  \urlprefix\url{https://doi.org/10.1109/jstqe.2019.2950803}

\bibitem{He:lowloss19}
He L, Zhang M, Shams-Ansari A, Zhu R, Wang C and Marko L 2019 {\em Opt.
  Lett.\/} {\bf 44} 2314--2317
  \urlprefix\url{http://ol.osa.org/abstract.cfm?URI=ol-44-9-2314}

\bibitem{Hu:21}
Hu C, Pan A, Li T, Wang X, Liu Y, Tao S, Zeng C and Xia J 2021 {\em Opt.
  Express\/} {\bf 29} 5397--5406
  \urlprefix\url{http://www.opticsexpress.org/abstract.cfm?URI=oe-29-4-5397}

\bibitem{Lenzini2018}
Lenzini F, Janousek J, Thearle O, Villa M, Haylock B, Kasture S, Cui L, Phan
  H~P, Dao D~V, Yonezawa H, Lam P~K, Huntington E~H and Lobino M 2018 {\em
  Science Advances\/} {\bf 4} eaat9331
  \urlprefix\url{https://doi.org/10.1126/sciadv.aat9331}

\bibitem{Snyder1984}
Snyder A~W and Love J~D 1984 {\em Optical Waveguide Theory\/} (Springer {US})
  \urlprefix\url{https://doi.org/10.1007/978-1-4613-2813-1}

\bibitem{Fallahkhair2008}
Fallahkhair A~B, Li K~S and Murphy T~E 2008 {\em Journal of Lightwave
  Technology\/} {\bf 26} 1423--1431
  \urlprefix\url{https://doi.org/10.1109/jlt.2008.923643}

\bibitem{Fejer1986}
Fejer M 1986 {\em Single Crystal Fibers: Growth Dynamics and Nonlinear Optical
  Interactions\/} Ph.D. thesis Stanford University

\bibitem{Kolesik2004}
Kolesik M and Moloney J~V 2004 {\em Physical Review E\/} {\bf 70}
  \urlprefix\url{https://doi.org/10.1103/physreve.70.036604}

\bibitem{Nye1985}
Nye J~F 1985 {\em Physical Properties of Crystals: Their Representation by
  Tensors and Matrices\/} (Oxford University Press)

\end{thebibliography}

\end{document}